\documentclass[
superscriptaddress,
showpacs,
amsmath,amssymb,
aps,
prb,
twocolumn,
]{revtex4-1}

\usepackage{bibunits}
\usepackage{graphicx}% Include figure files
\usepackage{dcolumn}% Align table columns on decimal point
\usepackage{bm}% bold math
\usepackage[breaklinks,colorlinks = true,linkcolor = red,urlcolor=blue,citecolor=red]{hyperref}
\usepackage{multirow}
\usepackage{array}
\usepackage{booktabs}
\usepackage{ctable}
\usepackage{upgreek}
\usepackage{epsfig}
\usepackage{mathrsfs}
\usepackage{amssymb}
\usepackage{amsbsy}
\usepackage{color}
\usepackage{cancel}
\usepackage{pifont}
\usepackage{marginnote}
\usepackage{float}
\usepackage{verbatim}
\usepackage{subfigure}
\usepackage{bbm, dsfont}
\usepackage{lineno}
\newcommand{\bcen}{\begin{center}}
\newcommand{\ecen}{\end{center}}
\newcommand{\btab}{\begin{tabular}}
\newcommand{\etab}{\end{tabular}}
\newcommand{\bdes}{\begin{description}}
\newcommand{\edes}{\end{description}}
\newcommand{\mc}{\multicolumn}
\newcommand{\ul}{\underline}
\newcommand{\beq}{\begin{equation}}
\newcommand{\eeq}{\end{equation}}
\newcommand{\bea}{\begin{eqnarray}}
\newcommand{\eea}{\end{eqnarray}}
\newcommand{\non}{\nonumber}
\newcommand{\etal}{et.~al.\ }
\newcommand{\half}{\frac{1}{2}}
\newcommand{\bary}{\begin{array}}
\newcommand{\eary}{\end{array}}
\newcommand{\benum}{\begin{enumerate}}
\newcommand{\eenum}{\end{enumerate}}
\newcommand{\bitem}{\begin{itemize}}
\newcommand{\eitem}{\end{itemize}}
\newcommand{\cuup}[1]{c_{#1 \uparrow}}
\newcommand{\cdown}[1]{c_{#1 \downarrow}}
\newcommand{\cdup}[1]{c^\dagger_{#1 \uparrow}}
\newcommand{\cddown}[1]{c^\dagger_{#1 \downarrow}}
\newcommand{\beps}{\mbox{\boldmath $ \epsilon $}}
\newcommand{\bsig}{\mbox{\boldmath $ \sigma $}}
\newcommand{\bpi}{\mbox{\boldmath $ \pi $}}
\newcommand{\bkap}{\mbox{\boldmath $ \kappa $}}
\newcommand{\bgam}{\mbox{\boldmath $ \gamma $}}
\newcommand{\bphi}{\mbox{\boldmath $ \phi $}}
\newcommand{\balp}{\mbox{\boldmath $ \alpha $}}
\newcommand{\beot}{\mbox{\boldmath $ \eta $}}
\newcommand{\btau}{\mbox{\boldmath $ \tau $}}
\newcommand{\blam}{{\boldsymbol{\lambda}}}
\newcommand{\bomg}{\mbox{\boldmath $ \omega $}}
\newcommand{\bOmg}{\mbox{\boldmath $ \Omega $}}
\newcommand{\bxhi}{\mbox{\boldmath $ \xi $}}
\newcommand{\bmu} {\mbox{\boldmath $ \mu $}}
\newcommand{\bnu} {\mbox{\boldmath $ \nu $}}
\newcommand{\bdelta}{{\boldsymbol{\delta}}}
\newcommand{\bDelta}{{\boldsymbol{\Delta}}}
\newcommand{\bPi}{{\boldsymbol{\Pi}}}
\newcommand{\bpsi}{\mbox{\boldmath $ \psi $}}
\newcommand{\brho}{\mbox{\boldmath $ \rho $}}
\newcommand{\bGam}{{\boldsymbol{\Gamma}}}
\newcommand{\bLam}{\mbox{\boldmath $ \Lambda $}}
\newcommand{\bPhi}{\mbox{\boldmath $ \Phi $}}
\newcommand{\bOne}{{\boldsymbol 1}}
\newcommand{\ba} { \bm{a} }
\newcommand{\bb} { \mbox{\boldmath $b$}}
\newcommand{\bc} { {\mathbf c} }
\newcommand{\bd} { \mbox{\boldmath $d$}}
\newcommand{\be} { \mbox{\boldmath $e$}}
\newcommand{\bff}{ \mbox{\boldmath $f$}}
\newcommand{\bg} { \mbox{\boldmath $g$}}
\newcommand{\bh} { \mbox{\boldmath $h$}}
\newcommand{\bi} { \mbox{\boldmath $i$}}
\newcommand{\bj} { \mbox{\boldmath $j$}}
\newcommand{\bk} { \bm{k} }
\newcommand{\bl} { \mbox{\boldmath $l$}} 
\newcommand{\bmm} { \mbox{\boldmath $m$}}
\newcommand{\bn} { \mbox{\boldmath $n$}}
\newcommand{\bo} { \mbox{\boldmath $o$}}
\newcommand{\bp} { \bm{p} }
\newcommand{\bq} { \bm{q} }
\newcommand{\br} { \boldsymbol{r}}
\newcommand{\bs} { \mbox{\boldmath $s$}}
\newcommand{\bt} {\boldsymbol{t}} 
\newcommand{\bu} { \mbox{\boldmath $u$}}
\newcommand{\bv} { \mbox{\boldmath $v$}}
\newcommand{\bw} { \mbox{\boldmath $w$}}
\newcommand{\bx} { \mbox{\boldmath $x$}}
\newcommand{\by} { \mbox{\boldmath $y$}}
\newcommand{\bz} { \mbox{\boldmath $z$}}

\newcommand{\bA} { \mbox{\boldmath $A$}}
\newcommand{\bB} { \mbox{\boldmath $B$}}
\newcommand{\bC} { \mbox{\boldmath $C$}}
\newcommand{\bD} { \mbox{\boldmath $D$}}
\newcommand{\bE} { \mbox{\boldmath $E$}}
\newcommand{\bF} { \mbox{\boldmath $F$}}
\newcommand{\bG} { \mbox{\boldmath $G$}}
\newcommand{\bH} { \mbox{\boldmath $H$}}
\newcommand{\bI} { \mbox{\boldmath $I$}}
\newcommand{\bJ} { \mbox{\boldmath $J$}}
\newcommand{\bK} { \mbox{\boldmath $K$}}
\newcommand{\bL} { \mbox{\boldmath $L$}}
\newcommand{\bM} { \mbox{\boldmath $M$}}
\newcommand{\bN} { \mbox{\boldmath $N$}}
\newcommand{\bO} { \mbox{\boldmath $O$}}
\newcommand{\bP} { \mbox{\boldmath $P$}}
\newcommand{\bQ} { \boldsymbol{Q} }
\newcommand{\bR} { {\mathbf R} }
\newcommand{\bS} { \mbox{\boldmath $S$}}
\newcommand{\bT} { \mbox{\boldmath $T$}}
\newcommand{\bU} { \mbox{\boldmath $U$}}
\newcommand{\bV} { \mbox{\boldmath $V$}}
\newcommand{\bW} { \mbox{\boldmath $W$}}
\newcommand{\bX} { \mbox{\boldmath $X$}}
\newcommand{\bY} { \mbox{\boldmath $Y$}}
\newcommand{\bZ} { \mbox{\boldmath $Z$}}
\newcommand{\bzero} { {\boldsymbol{0}}}
\newcommand{\bfell} {\mbox{\boldmath $ \ell $}}

\newcommand{\dou}{\partial}
\newcommand{\leftjb} {[\![}
\newcommand{\rightjb} {]\!]}
\newcommand{\ju}[1]{ \leftjb #1 \rightjb }
\newcommand{\D}[1]{\mbox{d}{#1}} 
\newcommand{\grad}{\mbox{\boldmath $\nabla$}}
\newcommand{\modulus}[1]{|#1|}
\renewcommand{\div}[1]{\grad \cdot #1}
\newcommand{\curl}[1]{\grad \times #1}
\newcommand{\mean}[1]{\langle #1 \rangle}
\newcommand{\bra}[1]{{\langle #1 |}}
\newcommand{\ket}[1]{| #1 \rangle}
\newcommand{\braket}[2]{\langle #1 | #2 \rangle}
\newcommand{\dbdou}[2]{\frac{\dou #1}{\dou #2}}
\newcommand{\dbdsq}[2]{\frac{\dou^2 #1}{\dou #2^2}}
\newcommand{\Pint}[2]{ P \!\!\!\!\!\!\!\int_{#1}^{#2}}

\newcommand{\PARA}[1]{{\noindent {$\bigstar$} {\sc #1}:~}}

\newcommand{\eqn}[1] {eqn.~(\ref{#1})}
\newcommand{\prn}[1] {(\ref{#1})}
\newcommand{\sect}[1] {Section~\ref{#1}}
\newcommand{\Sect}[1] {Section~\ref{#1}}
\newcommand{\fig}[1]{fig.~\ref{#1}}
\newcommand{\Fig}[1]{Fig.~\ref{#1}}
\newcommand{\corr}{\color{red}{}}

\newcommand{\ci}{\mathbbm{i}}
\makeatletter
\newcommand{\rmnum}[1]{\romannumeral #1}
\newcommand{\Rmnum}[1]{\expandafter\@slowromancap\romannumeral #1@}
\makeatother

\newcommand{\uncon}[1]{\centerline{\epsfysize=#1 \epsfbox{/usr2/shenoy/styles/construction.eps}}}
\newcommand{\checkup}[1]{{(\tt #1)}\typeout{#1}}
\newcommand{\ttd}[1]{{\color[rgb]{1,0,0}{\bf #1}}}
\newcommand{\ttds}[1]{{\color[rgb]{0,0,1}{\bf #1}}}
\newcommand{\red}[1]{{\color[rgb]{1,0,0}{\protect{#1}}}}
\newcommand{\blue}[1]{{\color[rgb]{0,0,1}{#1}}}
\newcommand{\green}[1]{{\color[rgb]{0.0,0.5,0.0}{#1}}}
\newcommand{\citebyname}[1]{\citeauthor{#1}\cite{#1}}
\newcommand{\signum}[0]{\mathop{\mathrm{sign}}}
\newcommand{\skup}{\ket{s \uparrow}}
\newcommand{\skdn}{\ket{s \downarrow}}
\newcommand{\pkup}{\ket{p \uparrow}}
\newcommand{\pkdn}{\ket{p \downarrow}}
\newcommand{\sbup}{\bra{s \uparrow}}
\newcommand{\sbdn}{\bra{s \downarrow}}
\newcommand{\pbup}{\bra{p \uparrow}}
\newcommand{\pbdn}{\bra{p \downarrow}}

\newcommand{\myfigwidth}{0.4\paperwidth}
\newcommand{\myhalffigwidth}{0.2\paperwidth}

\newcommand{\asc}{a_{sc}}
\newcommand{\as}{a_{s}}
\newcommand{\Eb}{E_{b}}
\newcommand{\Ef}{E_F}
\newcommand{\kf}{k_F}
\newcommand{\lambdaT}{{\lambda_T}}
\newcommand{\cF}{{\cal F} }
\newcommand{\ie}{{i.e., } }
\newcommand{\etaT}{{\eta_t}}
\newcommand{\muNI}{{\mu_{NI}}}

\newcommand{\bbc}{\color{blue}}
\newcommand{\sbc}{\color{red}}
\newcommand{\vbc}{\color{violet}}
\newcommand{\mylabel}[1]{\label{#1}} 

\newcommand{\myonlinecite}[1]{[\onlinecite{#1}]}
\newcommand{\mycite}[1]{\cite{#1}}

\newcommand{\titlename}{Statistics-tuned phases of pseudofermions in one dimension}

\begin{document}

%\linenumbers
\relax

%\preprint{}

%\preprint{}

\title{\titlename}

\author{Adhip Agarwala}\email{adhip.agarwala@icts.res.in}
\affiliation{International Centre for Theoretical Sciences, Tata Institute of Fundamental Research, Bengaluru 560089, India}
\author{Gaurav Kumar Gupta}\email{ggaurav@iisc.ac.in}\thanks{Current address: Physics Department, Technion, 32000 Haifa, Israel.}
\affiliation{Centre for Condensed Matter Theory, Department of Physics, Indian Institute of Science, Bangalore 560 012, India}
\author{Vijay B. Shenoy}\email{shenoy@iisc.ac.in}
\affiliation{Centre for Condensed Matter Theory, Department of Physics, Indian Institute of Science, Bangalore 560 012, India}
\author{Subhro Bhattacharjee}\email{subhro@icts.res.in}
\affiliation{International Centre for Theoretical Sciences, Tata Institute of Fundamental Research, Bengaluru 560089, India}

\date{\today}

\begin{abstract}

We show that a quadratic system of {\it pseudofermions}, with tunable fractionalised statistics, can host a rich phase diagram on a one dimensional chain with nearest and next nearest neighbour hopping. Using a combination of numerical and analytical techniques, we show that that by varying the statistical angle and the ratio of the hopping, the system stabilizes two Tomonaga-Luttinger liquids (TLL) with central charges $c=1$ and $2$ respectively along with the inversion symmetry broken bond ordered (BO) insulating phase. Interestingly, the two quantum phase transitions in the system -- (1) between the two TLLs, and, (2) the $c=1$ TLL and BO phase can be engendered by solely tuning the statistics of the pseudofermions. Our analysis shows that both these transition are continuous and novel with the former lacking a local order-parameter based description and the latter of Berezinskii-Kosterlitz-Thouless type. These phases and phase transitions can be of direct experimental relevance in context of recent studies of fermionic cold atoms.

 \end{abstract}

\maketitle
%\begin{bibunit}
%\tableofcontents

\paragraph*{Introduction : }  Advances in the physics of one dimensional systems motivated by, inter alia, the possibility of Majorana zero modes\cite{Kitaev_PU_2001,Alicea_RPP_2012, mourik2012signatures}, have ushered in many new possibilities and opportunities. Particularly remarkable is the prospect of creating quantum entangled phases with fractional quantum numbers and statistics  \cite{Leinaas_NC_1977, Lieb_PR_1963,Kundu_PRL_1999, Pasquier_IMS_1994,  Ha_PRL_1994, Ha_NPB_1995}. Several recent proposals indeed suggest that starting with bosons or fermions%\cite{Keilmann_NatCom_2011,Greschner_PRL_2015,, Cardarelli_PRA_2016, }
, effective local Hamiltonians with degrees of freedom following {\it fractionalized} or {\it intermediate} statistics can be realized, for example, in ultra-cold atomic systems \cite{Keilmann_NatCom_2011,Strater_PRL_2016, Greschner_PRL_2015, Cardarelli_PRA_2016,Greschner_PRA_2018}. Exploring the physics of such system with tunable statistics has hence emerged as an active field of research.

In a one dimensional(D) chain with sites labeled $i,j$ etc., such tunable statistics is captured by the algebra generated by the onsite creation/annihilation operators given by 
\bea
&a_j a_i \pm a_i a_j e^{ \ci \phi~sign(i-j)} =& 0 \nonumber\\
&a_j a^\dagger_i \pm a^\dagger_i a_j e^{- \ci \phi~sign(i-j)}=& \delta_{ij}\nonumber\\
&[N_i,a_j]=-\delta_{ij}a_j;~~~[N_i,a^\dagger_j]&=\delta_{ij}a_j^\dagger
\label{eq_comm}
\eea
(where $N_i=a_i^\dagger a_i$). The underlying physics consistent with $sign(0)=0$  produces an onsite algebra that is bosonic or fermionic depending on the relative sign ($\pm$). Owing to this, we refer the two cases as {\it pseudofermions} ($+$ sign) or {\it pseudobosons} ($-$ sign) respectively even for $\phi\neq 0$. In either case, the off-site algebra can be tuned from fermionic to bosonic by tuning {\it statistical parameter} $\phi\in [0,\pi]$.
Both {\it pseudobosons} and {\it pseudofermions} defined above are generalizations of two dimensional ``anyons" to one spatial dimensions following Leinass and Myrheim\cite{Leinaas_NC_1977}.
% in the sense of generalized boundary condition of the wave-function of two identical quantum particles in one spatial dimension. 

 Subsequent works on exactly solvable one dimensional interacting bosonic \cite{Lieb_PR_1963,Kundu_PRL_1999} and fermionic systems \cite{Pasquier_IMS_1994, Ha_PRL_1994,Ha_NPB_1995} have shown interesting implications on generalized operator algebra \cite{Aneziris_IJMP_1991, Posske_PRB_2017,Frau_arXiv_1994,  Baz_IJMP_2003} as well as understanding of such one dimensional anyons in terms of exclusion statistics\cite{Haldane_PRL_1991} and generalized distribution functions\cite{Wu_PRL_1994,Murthy_PRL_1994}. While much recent work\cite{Batchelor_PRL_2006, Guo_PRA_2009, Eckholt_NJP_2009,Eckholt_PRA_2008, Greschner_PRL_2015, Zhang_PRA_2017, Lange_PRL_2017,Lange_PRA_2017, Forero_PRA_2018, Zuo_PRB_2018}, has concentrated on {\it pseudobosons}, here we show that {\it pseudofermions} can provide natural access to a complementary set of phases and phase transitions.  {\it Pseudofermions}, unlike {\it pseudobosons}, satisfy a hard-core constraint at {\it any} $\phi$ with $\phi=\pi$ limit being the hard-core boson limit. While this constraint may also be accessed as the infinite on-site interaction limit of {\it pseudoboson}s, {\it pseudofermions} are naturally relevant to studies of ultracold fermionic atoms \cite{Ketterle_arXiv_2008, Giorgini_RMP_2008}.  

\begin{figure}
	\includegraphics[width=0.95\columnwidth]{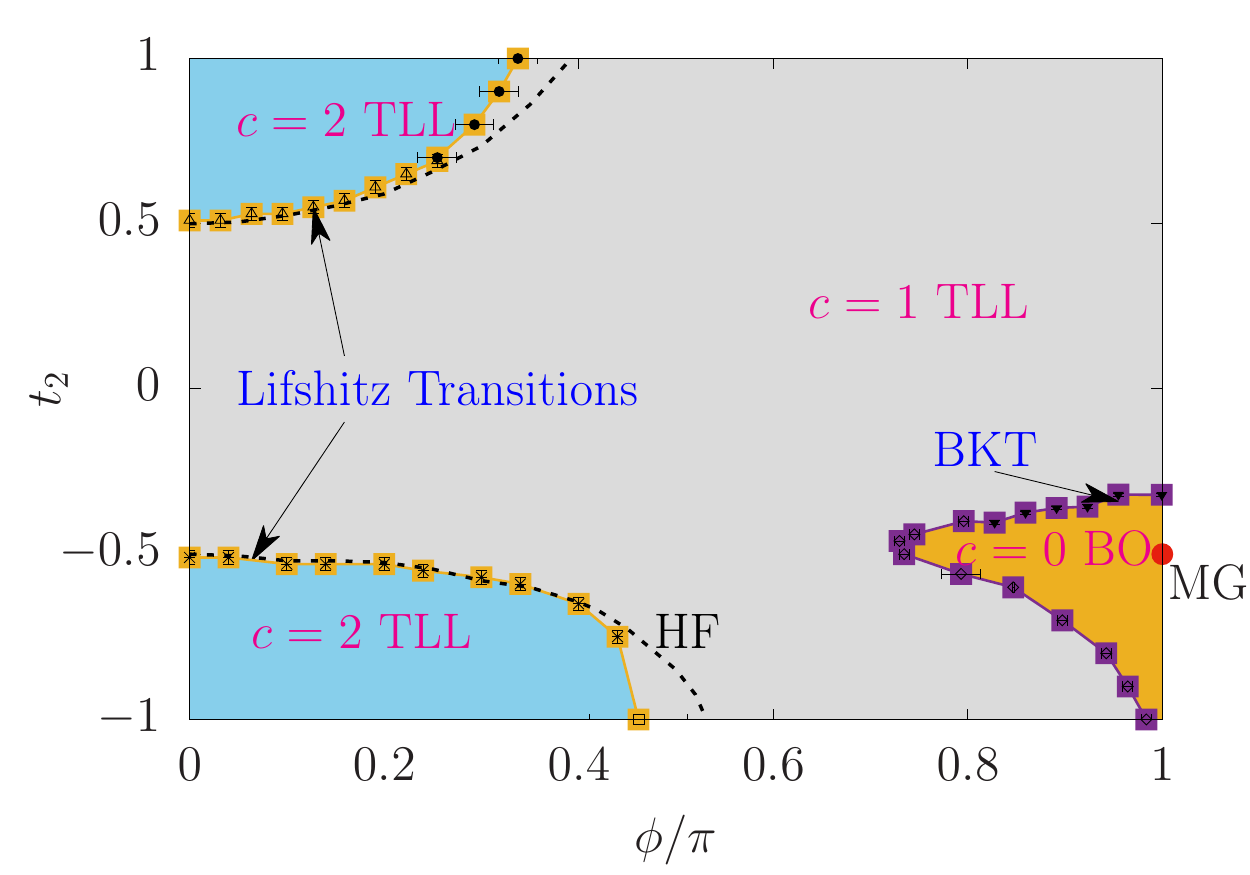}
	\caption{{\bf Phase diagram}: Phases of the model in the $\phi-t_2$ plane ($t_1=1$). At $\phi=0$ the free Fermi system encounters Lifshitz transitions where the number of Fermi points changes. At $\phi \neq 0$ these evolve into central charge($c$) $=1$ and $=2$ Tomonaga Luttinger liquids(TLL). Near $\phi=\pi$ system undergoes a BKT transition from $c=1$ TLL to a gapped bond-ordered (BO) phase which spontaneously breaks lattice parity. At $\phi=\pi$ and $t_2=-0.5$ the system has an exactly solvable Majumdar-Ghosh (MG) point. The phase boundaries are determined studying the excitation gap. Dashed line represents the Lifshtiz transitions under the Hartree-Fock(HF) approximation (see text).}
	\label{SchemPhase}
\end{figure}

In this paper, we demonstrate that even a deceptively simple quadratic system of pseudofermions is host to much interesting physics. Indeed, such a system, described by the Hamiltonian
\begin{align}
H=-\sum_i\left[t_1 a^\dagger_ia_{i+1}+t_2~a^\dagger_ia_{i+2}\right]+{\rm h.c.}
\label{eq_anyham}
\end{align}
where $t_1$ ($t_2$) denotes the nearest (next nearest) neighbor hopping on the 1D chain, has a rich phase diagram  (see fig.~\ref{SchemPhase}) with interesting gapless and gapped phases.  Phases realized include an inversion symmetry broken gapped bond-ordered (BO) phase in addition to the two Tomonaga-Luttinger liquids (TLL) with central charges $c=1$ and $c=2$ respectively. Most interestingly, in a regime of $t_2/t_1$, unconventional quantum phase transitions can be engendered by {\em tuning the  statistical parameter} $\phi$. The continuous quantum phase transition between the $c=1$ TLL and the BO phase is of Berezinskii-Kosterlitz-Thouless~(BKT) type %for generic values of $\phi$ 
with subtle {\it Berry phase} effects leading to inversion symmetry broken BO phase. Crucially as a function of the hopping amplitudes the {\it pseudofermions} show direct {\it Lifshitz} phase transition \cite{Lifshitz_JETP_1960, Blanter_PR_1994, Yamaji_JPSJ_2006, Rodney_PRB_2013} between $c=1$ and $c=2$ TLLs. This Lifshitz transition thus provides an example of a phase transition between two non-Fermi liquids described by conformal field theories (CFTs). We provide a comprehensive understanding of the phase diagram using a combination of approaches such as density matrix renormalisation group (DMRG) (numerically corroborated with exact diagonalization for smaller system sizes), Hartree-Fock (HF) theory, bosonization approaches and $(1+1)$ dimensional $XY$ duality. Our results can motivate experimental work in ultracold fermions using synthetic dimensions\cite{Celi_PRL_2014}.

To uncover the physics of \eqn{eq_anyham} we exploit the well known idea of interchanging statistics and interactions in 1D by  introducing  fractional Jordan-Wigner strings $K_i$, and defining operators
\begin{align}
c_i=K_i a_i,\quad c_i^\dagger=a_i^\dagger K_i^\dagger~~~{\rm with}~~~K_i=e^{-\ci \phi \sum_{j < i} n_{j}}.
\label{eq_fjw}
\end{align}
Eq. \ref{eq_anyham} is thus mapped into a {\em fermionic} Hamiltonian
\bea
{\cal H}  & = & -\sum_{i} \left[t_1 c_i^\dagger c_{i+1}  + t_2 e^{\ci \phi n_{i+1}} c^\dagger_i c_{i+2} \right] +{\rm h.c.}\,.
\label{eq:Ham}	       
\eea
where $c_i^\dagger, c_i$ are fermionic creation/annihilation operators at site $i$ obeying usual fermion anti-commutation algebra with the number density $n_i=c_i^\dagger c_i$.   We note that while at $\phi=0,\pi$ both time-reversal (TRS) and parity symmetries are separately present, at any generic $\phi$ only a combination of both is a symmetry (see Supplemental Material (SM) \footnote{Supplemental Material, which includes citations to \cite{Dalmonte_PRB_2015, Carrasquilla_PRA_2013, Gu_IJMB_2010, Chung_PRB_2001, Cheong_PRB_2004, Peschel_JPA_2003, Vidal_PRL_2003, Sachdev_book}}, Sec.~S1).  While the first term in \eqn{eq:Ham} is the nearest neighbor hopping, the second term contains the physics of {\em correlated hopping} between next-nearest-neighboring sites -- fermions hop with a phase of $0$ ($\phi$) in absence (presence) of another fermion at the intermediate site. We note that a finite $t_2$ is crucial to realization of non-trivial phases\cite{Hao_PRA_2009,Hao_PRA_2012}. Interestingly, correlated hoppings are known to arise in strongly correlated systems with constrained kinetic energies leading to frustration \cite{Foglio_PRB_1979,Arrachea_PRL_1994,Boer_PRL_1995,Vidal_PRB_2001}.

 %{{Even though in absence of $t_2$ the ground state does not show nontrivial many-body phases, it can still be interestingly viewed in terms of {\it pseudofermions} \cite{Hao_PRA_2009,Hao_PRA_2012}. Crucially, a finite $t_2$ brings in correlated hopping}} which are also known to arise in strongly correlated electronic systems with constrained kinetic energies leading to frustration \cite{Foglio_PRB_1979,Arrachea_PRL_1994,Boer_PRL_1995,Vidal_PRB_2001}.
 
Under the above non-local transformation the {\it pseudofermion} number density operator $N_i=a_i^\dagger a_i$ is equal to the fermion density operator $n_i=c_i^\dagger c_i$ and hence the {\it filling fraction} remains unchanged. Here, we shall concentrate on $1/2$ filling. In the remainder, we set  $t_1=1$ and study the phase diagram as a function of $t_2$ and $\phi$. Eq.~\ref{eq:Ham} is studied by analytical and numerical techniques.% DMRG \ttd{and HF} (numerical details in SM). 
 %%%%%%%%%%%%%%%%%%%%
 \paragraph*{\underline{Phase diagram}:}  Along the $\phi=0$ line %and $\phi=\pi$ limits are 
 {{the system reduces to that of free fermions with nearest and next-nearest-neighbor hopping with a single particle dispersion given by $E(k) = -2t_1 \cos k -2 t_2 \cos 2k$ with $k\in [-\pi,\pi]$.  %Due to parity for $\phi=0$, the dispersion is always symmetric about $k \rightarrow -k$. 
 There is a change in the number of the Fermi points as the system undergoes a Lifshitz transition at $t_2/t_1=\pm 0.5$. For $|t_2/t_1|>0.5 (<0.5)$, there are four (two) Fermi points corresponding to the left most extremum of \Fig{SchemPhase}.}}
  \begin{figure}
 	\includegraphics[width=1.0\columnwidth]{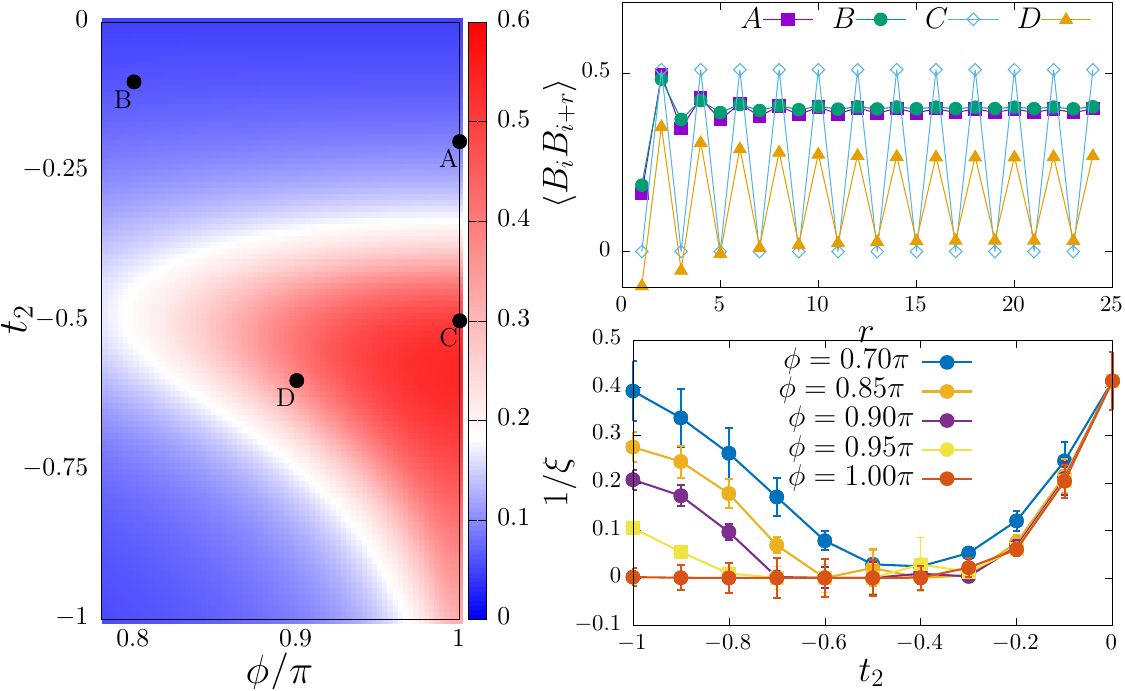}
 	\caption{{\bf{Gapped phase:}} (Left) $|O_{BO}|$ as a function of $\phi$ and $t_2$ in the gapped regime. (Top-Right) Dimer-dimer correlator $\langle B_i B_{i+r}\rangle$ and its behavior as a function of $r$ is shown at the marked points (A-D) on (Left). (Bottom-Right) Fitting $\langle B_i B_{i+r}\rangle \propto e^{-r/\xi}$; behavior of $1/\xi$ with $t_2$ for different values of $\phi$. }
 	\label{bondorfig}
 \end{figure}
  
  For $\phi = \pi$, we recover the familiar fermion to (hard-core) boson mapping evident from Eqs.~\ref{eq_comm} and \ref{eq_fjw}. Thus we have a $1/2$-filled system of hard-core bosons with nearest and next nearest neighbor hoppings corresponding to the right extremum of \Fig{SchemPhase}. This is the easy-plane limit of the $J_1-J_2$ spin-$1/2$ chain\cite{Mishra_PRB_2013,Dhar_PRA_2013}. It has two phases-- $c=1$ TLL %(or powerlaw SF) 
 for $t_2/t_1\gtrsim -0.3$ and a BO phase which spontaneously breaks inversion symmetry about a site which is characterised by a finite value of the order parameter 
 %\beq
 $O_{BO} = \frac{1}{L}\sum_i (-1)^i B_i, $
 %\eeq
 $B_i = \langle a^\dagger_i a_{i+1} + a^\dagger_{i+1} a_{i}   \rangle$. Since the phase has a finite excitation gap, we expect it to be stable for small deviation of $\phi$ from $\pi$. DMRG results (\cite{Note1}, Sec.~S3) are plotted in \Fig{bondorfig} where we show both the BO order parameter as well as the two-point correlation function for the bond order. The lobe of BO order is roughly centered about $t_2/t_1=-0.5$ which is the Majumdar-Ghosh point \cite{Majumdar_JMP_1969, Majumdar_JMP_1969_2, Mishra_PRB_2013} for which the the BO ground state is exact at $\phi=\pi$. The structure of the lobe shows that at a fixed $0.79\pi\lesssim\phi<\pi$ there is a reentrant transition into a $c=1$ TLL as we tune $t_2/t_1$ from positive to negative. At $t_2/t_1 = -\infty$ we have two decoupled chains which are in separate TLL phase. Turning on a positive $t_1$ destroys this state in favor of a bond order. However, we note that our calculations suggest that turning on a $\phi$ away from $\phi=\pi$ instead favors an instability to a $c=1$ TLL which competes with BO leading to a dome like structure.   
 
Within a self-consistent Hartree-Fock (HF) treatment (\cite{Note1}, Sec.~S2)) of the correlated hopping term (for $\phi\neq 0$) by decoupling it in the $k$-mode density $\langle n(k) \rangle=\langle c^\dagger_k c_k\rangle$, we find that for $|t_2/t_1|<0.5$, the centre of the Fermi surface shifts away from zero for $\phi\neq 0,\pi$ due to the absence of TRS. For higher values of $|t_2/t_1|$ the HF calculations show that the Lifshitz transition continues into the interacting regime. The Lifshitz transitions of the effective HF Hamiltonian at any $\{t_2, \phi\}$ traces a continuous quantum phase transition between the two gapless metallic phases (see the dashed curve in \Fig{SchemPhase}). However, not unexpectedly, this HF analysis breaks down in the gapped phase obtained in the vicinity of $\phi=\pi$. 
%%%%%%%%%%%%%
\begin{figure}
	\includegraphics[width=1.0\columnwidth]{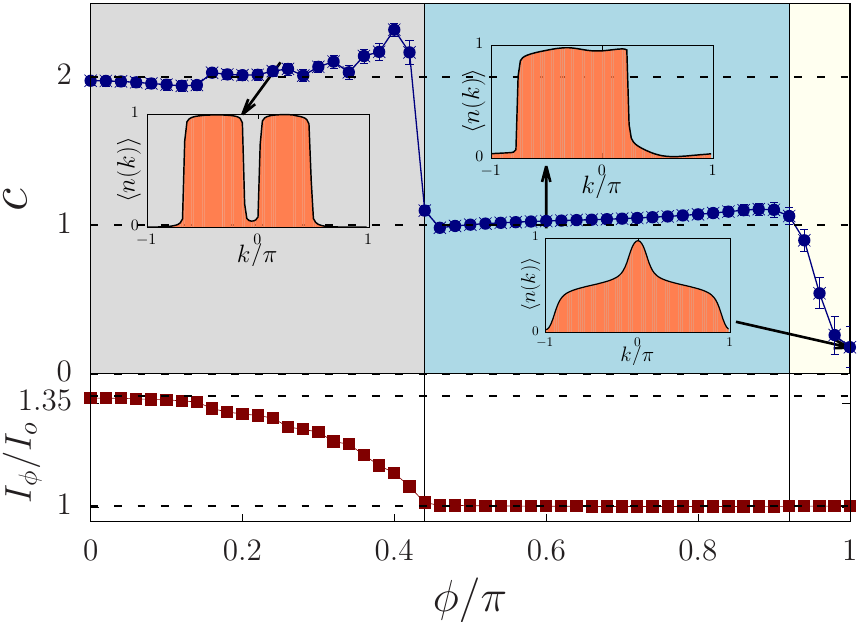}
	\caption{{\bf Lifshitz Transition}: (Top) Central charge as a function of $\phi$ at $t_2=-0.75$ shows a transition from $c=2$ to a $c=1$ plateau. Fermion occupancy $\langle n(k) \rangle$ is shown as a function of $k$ at three representative values of $\phi$ as pointed. (Bottom) Variation of $I_\phi \equiv \int_{-\pi}^{\pi} \sin^2 k \langle n(k) \rangle dk$ captures the Lifshitz transition. $I_o$ is the corresponding value for a half-filled Fermi sea at $t_2=0$. 
	}
	\label{Lifs}
\end{figure}

The phase diagram (see \fig{SchemPhase}) for $\phi$ away from $\pi$, is occupied by two gapless TLLs. Interestingly, the HF theory does reproduce DMRG fermion occupancy of these phases remarkably well. 
In order to characterize the phases further, we calculate their central charges (using Calabrese-Cardy-formula for the entanglement entropy \cite{Calabrese_JPA_2009}) and various correlation functions to determine the Luttinger parameters using DMRG results. \Fig{Lifs}  shows that even for finite $\phi$, the Lifshitz transition survives for the pseudofermions that separates the $c=1$ and the $c=2$ TLLs. 
The $c=1$ TLL can be understood within a low energy linearized (about the left and the right Fermi points ($k^L_F$ and $k^R_F$)) theory about the HF ground state given by $\mathcal{H}_0= -\ci\int dx\left[v_f^R\Psi_R^\dagger\partial_x\Psi_R-v_f^L\Psi_L^\dagger\partial_x\Psi_L\right]$ where, ($\Psi_R, \Psi_L$) are left and right moving fermions and  in the present case for $\phi,t_2\neq 0$ the corresponding Fermi velocities ($v^R_f$ and $v^L_f$) are different.  Further Luttinger theorem restricts $k^R_F-k^L_F=\pi$ at half filling. While a similar construction can be obtained for the $c=2$ TLL by linearising about the four Fermi points and introducing two pairs of left and right moving fermions, characterizing this low energy theory  requires characterizing the $2\times 2$ {\it matrix Luttinger parameter} \cite{sule2015determination}, which we do not pursue here.

\begin{figure}
	\includegraphics[width=0.99\columnwidth]{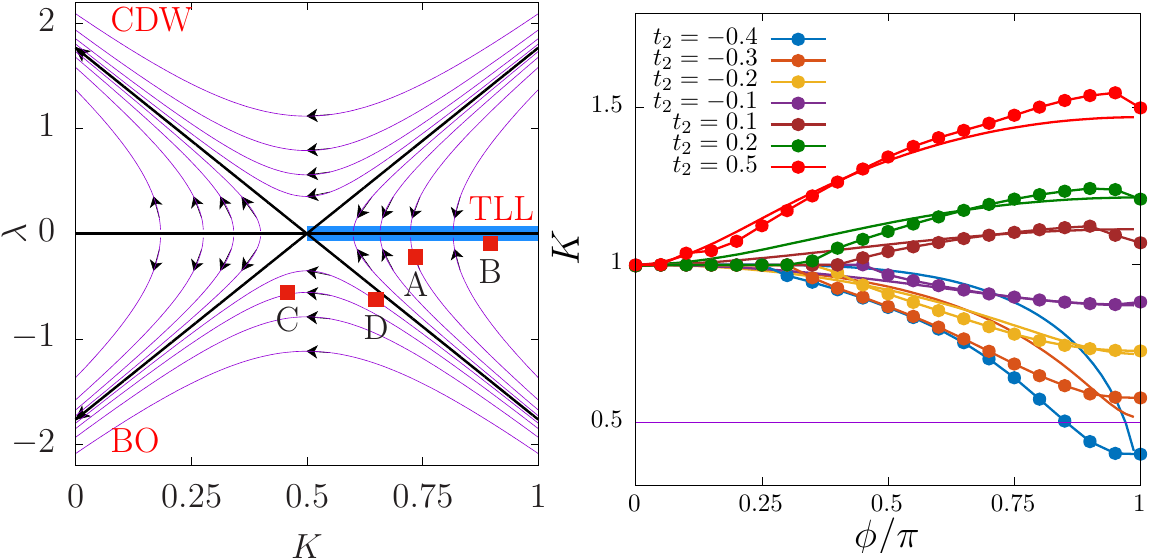}
     \caption{{\bf{Bosonization and BKT Transition}}(Left) Luttinger parameter ($K$) and $\lambda$ calculated from the bosonization  result for a set of parameters of $\{t_2, \phi/\pi\}$ as shown by (A-D) in \Fig{bondorfig} (see near \eqn{eq_bosonised_ham}) are shown by square points on the RG flow diagram of the BKT transition with TLL, BO and charge density wave (CDW) phases. $V$ and $\lambda$ are scaled by multiplicative prefactors of $\alpha/2\pi$ and $\alpha$ respectively where $\alpha \approx 0.3$. (Right) A plot of numerically extracted $K$ from the DMRG using $\langle c^\dagger_i c_{i+r} \rangle$ correlators and compared with the bosonization result.}
     \label{fig_lut_k}
\end{figure}

At finite $\phi>0$ and $t_2\neq 0$, none of TLLs have well defined quasi-particles--both of them being non-Fermi liquid metals. This is best seen by studying the effect of fluctuations over the HF theory \cite{Note1} within the framework of Abelian bosonization obtained by introducing the bosonized field $\Phi(x)$ and $\Theta(x)$ which obey the algebra $[\nabla \Phi(x), \Theta(y)] = [\nabla \Theta(x), \Phi(y)] = i \pi \delta (x-y)$. The effective low energy bosonized Hamiltonian for the $c=1$ TLL is given by 
\begin{align}
	H=&\frac{v_f}{2\pi}\int dx\left[\frac{1}{K}(\partial_x\Phi)^2+K(\partial_x\Theta)^2\right]\nonumber\\
	&-\frac{W}{2\pi}\int dx\left[\partial_x\Phi\partial_x\Theta+\partial_x\Theta\partial_x\Phi\right]-\lambda\int dx\cos4\Phi
	\label{eq_bosonised_ham}
\end{align}
where $v_f=\sqrt{(V_F)^2-4\pi^2V^2}$ is the renormalised average Fermi velocity with $V_F=\frac{v_f^R+v_f^L}{2}$ and $W=\frac{v_f^R-v_f^L}{2}$. Here, $V$ and $\lambda$ arises due to interactions with $V$ and $\lambda$ being the forward and Umklapp scattering amplitude respectively given in terms of the microscopic correlated hopping parameters \cite{Note1}. While $V$ kills the quasiparticles by renormalizing the Luttinger parameter $K$, $\lambda$ destabilizes the TLL leading to bond order. We use the relation between Luttinger parameter $K$ and the fermion two-point correlator
\beq
C(r) := \frac{1}{L}\sum_i \langle c^\dagger_i c_{i+r} \rangle \sim \frac{1}{r^{(K+1/K)/2}}
\label{fermCorr}
\eeq
to extract $K$ from DMRG results. Parameters $K$ and $\lambda$ can also be calculated from the bosonized theory $K\left(=\sqrt{\frac{V_F-2\pi V}{V_F+2\pi V}}\right)$ .  Away from the gapped region, where the Umklapp processes are small, the bosonization result compares well with DMRG results as seen in \Fig{fig_lut_k}.  For a fixed $t_2$, the Umklapp amplitude, $\lambda$,  increases monotonically with $\phi$ such that the Umklapp scattering becomes important ultimately making the TLL unstable to a gapped BO phase. In the latter case, the bosonization values of $K,\lambda$ should be viewed as initial points of the RG flow as shown in \Fig{fig_lut_k}. The instability to a gapped phase is also manifested through the Luttinger parameter  reaching a critical value of $K=1/2$ as anticipated from a perturbative RG calculation. Away from the bond-ordered phase, for generic values of $2K>1$  our numerical calculation suggest a direct transition between the two TLL.
%%%%%%%%%%%%%%%%%%%%%
\paragraph*{\underline{Phase boundaries and the phase transitions} :} In addition to the central charge, we use the excitation gap \cite{Mishra_PRB_2013}: $\Delta_L =  E(L, N+1) + E(L,N-1) -2E(L,N)$ (where $E(L, N)$ is the ground state energy for a system with $N$ particles on $L$ sites) to obtain the phase boundaries in Fig. \ref{SchemPhase} numerically. These calculations show that there are two types of phase boundaries denoting quantum phase transition between -- (1) $c=1$ TLL (equivalently a powerlaw superfluid (SF)) to the symmetry broken bond ordered phase, and (2) $c=1$ and $c=2$ TLL phases. We now explore the nature these phase transitions.
%%%%%%%%
\paragraph{\ul{Transition between the powerlaw SF to BO phase :} } Our numerical calculations show that the entire phase boundary is captured by a BKT type phase transition (data collapse of gap scaling to BKT form shown in \cite{Note1}). This is expected in the $\phi=\pi$ limit, where we have hard-core bosons and an application of XY duality in $(1+1)D$\cite{baeriswyl2005strong} {{captures both the SF and the BO phases where extra Berry phases induce breaking of inversion symmetry in the}} BO phase \cite{Note1,sachdev2002quantum}. To understand this transition to general $\phi$, it is convenient to use the bosonized field theory by writing down the Euclidean action corresponding to \eqn{eq_bosonised_ham}. The effect of finite $\phi$ is to provide a ``boosted" field theory where the effect of the ``boost" can then be gauged away \cite{Ray_AP_2017}. The Umklapp scattering then drives the BO instability gapping out the TLL as can be seen in the behavior of the Luttinger parameter (see \Fig{fig_lut_k}). The corresponding RG flows based on the Sine-Gordon theory (\Fig{fig_lut_k}) effectively capture the phase transition along with the phases.

%%%%%%%%%
\paragraph{\ul{Transition between $c=1$ and $c=2$ TLL} :} Our central charge calculations suggest that the Lifshitz transition for finite $\phi$  from $c=1$ CFT (TLL with two Fermi points) to $c=2$ CFT (TLL with four Fermi points) is a rather sudden one when compared to the BKT transition.  However, crucially there is no local real space order parameter based description for this quantum phase transition which then requires careful examination regarding its nature.  The HF band structure suggests that the low energy modes near this transition contains the linearly dispersing left, $\psi_L$, and right, $\psi_R$, fermions along with quadratically dispersing holes of the central lobe, $\psi_c$. At $\phi=0$, these modes are non interacting and the transition at $t_2/t_1=-0.5$ is given by a dynamical exponent $z=2$ theory where changing $t_2$  has the primary effect of changing the hole chemical potential leading to finite density of holes. Strictly speaking because of the overall Luttinger theory this leads to renormalisation of the Fermi-velocities of both $\psi_L$ and $\psi_R$, but, they lead to innocuous renormalisation of various quantities as these modes do not couple to $\psi_c$. Assuming that the above picture holds at least for finite $\phi$, we bosonize the left and right fermions to get the following effective low energy Hamiltonian $\mathcal{H}=H+H'$, where where $H$ is given by \eqn{eq_bosonised_ham} and
{\small \begin{align}
H'=\int dx~\psi_c^\dagger\left[\frac{\partial_x^2}{2m^*}-\mu\right]\psi_c  +\int dx\left( g_1\partial_x \Phi + g_2 \partial_x \Theta\right)\psi^\dagger_c  \psi_c.
\end{align}}
The first term in $H'$ is the free action of the quadratically dispersing fermionic holes with effective mass $m^*$ and  chemical potential $\mu$ which is zero at $t_2/t_1=-0.5$ and increases (decreases) for $t_2/t_1<-0.5 (>-0.5)$ and thereby capturing the HF phase transition.  $g_1$ and $g_2$ are two symmetry allowed coupling constants (details in \cite{Note1}, Sec.~S5) which go to zero as $\phi\rightarrow0$. Perturbative one-loop RG calculations around the $\mu=g_1=g_2=0$ shows that this critical point is stable\cite{Note1}. Similar field theories are suggested in context of multimode wires\cite{Sitte_PRL_2009,Meng_PRB_2011,Rodney_PRB_2013}; However, a full characterization requires further studies. In absence of any local order-parameter, we define ``moment-of-inertia" of the Fermi sea as $\int n(k) (\sin(k))^2 dk$ and find that it shows a smooth variation across the Lifshitz transition (see \Fig{Lifs}).
Remarkably just tuning the statistical phase of {\it pseduofermions} can mediate this transition between two non-Fermi liquids neither of which has low energy quasi-particles. 

%%%%%%%%%%%%%%%%%%%%
\paragraph*{\ul{Summary and outlook} :} We have shown that a simple quadratic system of {\it pseudofermions} hopping on $1$D lattice with nearest and next nearest neighbor hopping following fractionalized algebra (eq. \ref{eq_comm}), has a rich phase diagram with gapless TLL phases and gapped BO phases which can be accessed by tuning the {\it statistical parameter} and ratio of lattice hopping. The phase transitions include a BKT transition between a $c=1$ TLL and a BO phase as well as a possible continuous Lifshitz phase transition between two TLLs with central charges $c=1$ and $2$. 
The fractionalized algebra is naturally obtained in a system of fermions with correlated hopping. Recent proposals \cite{Liberto_PRA_2014, Cardarelli_PRA_2016, Ghosh_PRA_2017} of generating such off-site correlated hopping in fermionic ultra-cold atoms can potentially realise the above quadratic system of such particles enabling us to provide more comprehensive understanding of unconventional phases and phase transitions in lower dimensional systems. 
%%%%%%%%%%%%%%%%%%%%%%

\paragraph*{\ul{Acknowledgements} :} We acknowledge fruitful discussions with Diptiman Sen, H.R. Krishnamurthy, Krishnendu Sengupta, Subroto Mukerjee, Smitha Vishveshwara, R. Loganayagam, Avinash Dhar,  R. Moessner, Tapan Mishra, Arun Parmakanti and Sumilan Banerjee. A. A. and S. B. acknowledges MPG for funding through the Max Planck Partner group on strongly correlated systems at ICTS.  S. B. acknowledges SERB-DST (India) for funding through project grant No. ECR/2017/000504. V.B.S. acknowledges support from DST(India). DMRG calculations are performed using ITensor \cite{Itensor}. The numerical calculations were done on the cluster {\it Mowgli, boson} and {\it zero} at the ICTS and {\it Sahastra} at IISc. We acknowledge ICTS and the discussion meeting {\it ``New questions in quantum field theory from condensed matter theory"} during which some of these ideas originated. 

\let\oldaddcontentsline\addcontentsline% Store \addcontentsline
\renewcommand{\addcontentsline}[3]{}% Make \addcontentsline a no-op
\bibliography{ref1DCorr}

\begin{thebibliography}{66}%
\makeatletter
\providecommand \@ifxundefined [1]{%
 \@ifx{#1\undefined}
}%
\providecommand \@ifnum [1]{%
 \ifnum #1\expandafter \@firstoftwo
 \else \expandafter \@secondoftwo
 \fi
}%
\providecommand \@ifx [1]{%
 \ifx #1\expandafter \@firstoftwo
 \else \expandafter \@secondoftwo
 \fi
}%
\providecommand \natexlab [1]{#1}%
\providecommand \enquote  [1]{``#1''}%
\providecommand \bibnamefont  [1]{#1}%
\providecommand \bibfnamefont [1]{#1}%
\providecommand \citenamefont [1]{#1}%
\providecommand \href@noop [0]{\@secondoftwo}%
\providecommand \href [0]{\begingroup \@sanitize@url \@href}%
\providecommand \@href[1]{\@@startlink{#1}\@@href}%
\providecommand \@@href[1]{\endgroup#1\@@endlink}%
\providecommand \@sanitize@url [0]{\catcode `\\12\catcode `\$12\catcode
  `\&12\catcode `\#12\catcode `\^12\catcode `\_12\catcode `\%12\relax}%
\providecommand \@@startlink[1]{}%
\providecommand \@@endlink[0]{}%
\providecommand \url  [0]{\begingroup\@sanitize@url \@url }%
\providecommand \@url [1]{\endgroup\@href {#1}{\urlprefix }}%
\providecommand \urlprefix  [0]{URL }%
\providecommand \Eprint [0]{\href }%
\providecommand \doibase [0]{http://dx.doi.org/}%
\providecommand \selectlanguage [0]{\@gobble}%
\providecommand \bibinfo  [0]{\@secondoftwo}%
\providecommand \bibfield  [0]{\@secondoftwo}%
\providecommand \translation [1]{[#1]}%
\providecommand \BibitemOpen [0]{}%
\providecommand \bibitemStop [0]{}%
\providecommand \bibitemNoStop [0]{.\EOS\space}%
\providecommand \EOS [0]{\spacefactor3000\relax}%
\providecommand \BibitemShut  [1]{\csname bibitem#1\endcsname}%
\let\auto@bib@innerbib\@empty
%</preamble>
\bibitem [{\citenamefont {Kitaev}(2001)}]{Kitaev_PU_2001}%
  \BibitemOpen
  \bibfield  {author} {\bibinfo {author} {\bibfnamefont {A.~Y.}\ \bibnamefont
  {Kitaev}},\ }\href {http://stacks.iop.org/1063-7869/44/i=10S/a=S29}
  {\bibfield  {journal} {\bibinfo  {journal} {Physics-Uspekhi}\ }\textbf
  {\bibinfo {volume} {44}},\ \bibinfo {pages} {131} (\bibinfo {year}
  {2001})}\BibitemShut {NoStop}%
\bibitem [{\citenamefont {Alicea}(2012)}]{Alicea_RPP_2012}%
  \BibitemOpen
  \bibfield  {author} {\bibinfo {author} {\bibfnamefont {J.}~\bibnamefont
  {Alicea}},\ }\href {http://stacks.iop.org/0034-4885/75/i=7/a=076501}
  {\bibfield  {journal} {\bibinfo  {journal} {Reports on Progress in Physics}\
  }\textbf {\bibinfo {volume} {75}},\ \bibinfo {pages} {076501} (\bibinfo
  {year} {2012})}\BibitemShut {NoStop}%
\bibitem [{\citenamefont {Mourik}\ \emph {et~al.}(2012)\citenamefont {Mourik},
  \citenamefont {Zuo}, \citenamefont {Frolov}, \citenamefont {Plissard},
  \citenamefont {Bakkers},\ and\ \citenamefont
  {Kouwenhoven}}]{mourik2012signatures}%
  \BibitemOpen
  \bibfield  {author} {\bibinfo {author} {\bibfnamefont {V.}~\bibnamefont
  {Mourik}}, \bibinfo {author} {\bibfnamefont {K.}~\bibnamefont {Zuo}},
  \bibinfo {author} {\bibfnamefont {S.~M.}\ \bibnamefont {Frolov}}, \bibinfo
  {author} {\bibfnamefont {S.}~\bibnamefont {Plissard}}, \bibinfo {author}
  {\bibfnamefont {E.~P.}\ \bibnamefont {Bakkers}}, \ and\ \bibinfo {author}
  {\bibfnamefont {L.~P.}\ \bibnamefont {Kouwenhoven}},\ }\href
  {http://science.sciencemag.org/content/336/6084/1003.full} {\bibfield
  {journal} {\bibinfo  {journal} {Science}\ }\textbf {\bibinfo {volume}
  {336}},\ \bibinfo {pages} {1003} (\bibinfo {year} {2012})}\BibitemShut
  {NoStop}%
\bibitem [{\citenamefont {Leinaas}\ and\ \citenamefont
  {Myrheim}(1977)}]{Leinaas_NC_1977}%
  \BibitemOpen
  \bibfield  {author} {\bibinfo {author} {\bibfnamefont {J.~M.}\ \bibnamefont
  {Leinaas}}\ and\ \bibinfo {author} {\bibfnamefont {J.}~\bibnamefont
  {Myrheim}},\ }\href {\doibase 10.1007/BF02727953} {\bibfield  {journal}
  {\bibinfo  {journal} {Il Nuovo Cimento B (1971-1996)}\ }\textbf {\bibinfo
  {volume} {37}},\ \bibinfo {pages} {1} (\bibinfo {year} {1977})}\BibitemShut
  {NoStop}%
\bibitem [{\citenamefont {Lieb}\ and\ \citenamefont
  {Liniger}(1963)}]{Lieb_PR_1963}%
  \BibitemOpen
  \bibfield  {author} {\bibinfo {author} {\bibfnamefont {E.~H.}\ \bibnamefont
  {Lieb}}\ and\ \bibinfo {author} {\bibfnamefont {W.}~\bibnamefont {Liniger}},\
  }\href {\doibase 10.1103/PhysRev.130.1605} {\bibfield  {journal} {\bibinfo
  {journal} {Phys. Rev.}\ }\textbf {\bibinfo {volume} {130}},\ \bibinfo {pages}
  {1605} (\bibinfo {year} {1963})}\BibitemShut {NoStop}%
\bibitem [{\citenamefont {Kundu}(1999)}]{Kundu_PRL_1999}%
  \BibitemOpen
  \bibfield  {author} {\bibinfo {author} {\bibfnamefont {A.}~\bibnamefont
  {Kundu}},\ }\href {\doibase 10.1103/PhysRevLett.83.1275} {\bibfield
  {journal} {\bibinfo  {journal} {Phys. Rev. Lett.}\ }\textbf {\bibinfo
  {volume} {83}},\ \bibinfo {pages} {1275} (\bibinfo {year}
  {1999})}\BibitemShut {NoStop}%
\bibitem [{\citenamefont {Pasquier}(1994)}]{Pasquier_IMS_1994}%
  \BibitemOpen
  \bibfield  {author} {\bibinfo {author} {\bibfnamefont {V.}~\bibnamefont
  {Pasquier}},\ }in\ \href@noop {} {\emph {\bibinfo {booktitle} {Integrable
  Models and Strings}}}\ (\bibinfo  {publisher} {Springer},\ \bibinfo {year}
  {1994})\ pp.\ \bibinfo {pages} {36--48}\BibitemShut {NoStop}%
\bibitem [{\citenamefont {Ha}(1994)}]{Ha_PRL_1994}%
  \BibitemOpen
  \bibfield  {author} {\bibinfo {author} {\bibfnamefont {Z.~N.~C.}\
  \bibnamefont {Ha}},\ }\href {\doibase 10.1103/PhysRevLett.73.1574} {\bibfield
   {journal} {\bibinfo  {journal} {Phys. Rev. Lett.}\ }\textbf {\bibinfo
  {volume} {73}},\ \bibinfo {pages} {1574} (\bibinfo {year}
  {1994})}\BibitemShut {NoStop}%
\bibitem [{\citenamefont {Ha}(1995)}]{Ha_NPB_1995}%
  \BibitemOpen
  \bibfield  {author} {\bibinfo {author} {\bibfnamefont {Z.}~\bibnamefont
  {Ha}},\ }\href {\doibase https://doi.org/10.1016/0550-3213(94)00537-O}
  {\bibfield  {journal} {\bibinfo  {journal} {Nuclear Physics B}\ }\textbf
  {\bibinfo {volume} {435}},\ \bibinfo {pages} {604 } (\bibinfo {year}
  {1995})}\BibitemShut {NoStop}%
\bibitem [{\citenamefont {Keilmann}\ \emph {et~al.}(2011)\citenamefont
  {Keilmann}, \citenamefont {Lanzmich}, \citenamefont {McCulloch},\ and\
  \citenamefont {Roncaglia}}]{Keilmann_NatCom_2011}%
  \BibitemOpen
  \bibfield  {author} {\bibinfo {author} {\bibfnamefont {T.}~\bibnamefont
  {Keilmann}}, \bibinfo {author} {\bibfnamefont {S.}~\bibnamefont {Lanzmich}},
  \bibinfo {author} {\bibfnamefont {I.}~\bibnamefont {McCulloch}}, \ and\
  \bibinfo {author} {\bibfnamefont {M.}~\bibnamefont {Roncaglia}},\ }\href
  {https://www.nature.com/articles/ncomms1353} {\bibfield  {journal} {\bibinfo
  {journal} {Nature communications}\ }\textbf {\bibinfo {volume} {2}},\
  \bibinfo {pages} {361} (\bibinfo {year} {2011})}\BibitemShut {NoStop}%
\bibitem [{\citenamefont {Str\"ater}\ \emph {et~al.}(2016)\citenamefont
  {Str\"ater}, \citenamefont {Srivastava},\ and\ \citenamefont
  {Eckardt}}]{Strater_PRL_2016}%
  \BibitemOpen
  \bibfield  {author} {\bibinfo {author} {\bibfnamefont {C.}~\bibnamefont
  {Str\"ater}}, \bibinfo {author} {\bibfnamefont {S.~C.~L.}\ \bibnamefont
  {Srivastava}}, \ and\ \bibinfo {author} {\bibfnamefont {A.}~\bibnamefont
  {Eckardt}},\ }\href {\doibase 10.1103/PhysRevLett.117.205303} {\bibfield
  {journal} {\bibinfo  {journal} {Phys. Rev. Lett.}\ }\textbf {\bibinfo
  {volume} {117}},\ \bibinfo {pages} {205303} (\bibinfo {year}
  {2016})}\BibitemShut {NoStop}%
\bibitem [{\citenamefont {Greschner}\ and\ \citenamefont
  {Santos}(2015)}]{Greschner_PRL_2015}%
  \BibitemOpen
  \bibfield  {author} {\bibinfo {author} {\bibfnamefont {S.}~\bibnamefont
  {Greschner}}\ and\ \bibinfo {author} {\bibfnamefont {L.}~\bibnamefont
  {Santos}},\ }\href {\doibase 10.1103/PhysRevLett.115.053002} {\bibfield
  {journal} {\bibinfo  {journal} {Phys. Rev. Lett.}\ }\textbf {\bibinfo
  {volume} {115}},\ \bibinfo {pages} {053002} (\bibinfo {year}
  {2015})}\BibitemShut {NoStop}%
\bibitem [{\citenamefont {Cardarelli}\ \emph {et~al.}(2016)\citenamefont
  {Cardarelli}, \citenamefont {Greschner},\ and\ \citenamefont
  {Santos}}]{Cardarelli_PRA_2016}%
  \BibitemOpen
  \bibfield  {author} {\bibinfo {author} {\bibfnamefont {L.}~\bibnamefont
  {Cardarelli}}, \bibinfo {author} {\bibfnamefont {S.}~\bibnamefont
  {Greschner}}, \ and\ \bibinfo {author} {\bibfnamefont {L.}~\bibnamefont
  {Santos}},\ }\href {\doibase 10.1103/PhysRevA.94.023615} {\bibfield
  {journal} {\bibinfo  {journal} {Phys. Rev. A}\ }\textbf {\bibinfo {volume}
  {94}},\ \bibinfo {pages} {023615} (\bibinfo {year} {2016})}\BibitemShut
  {NoStop}%
\bibitem [{\citenamefont {Greschner}\ \emph {et~al.}(2018)\citenamefont
  {Greschner}, \citenamefont {Cardarelli},\ and\ \citenamefont
  {Santos}}]{Greschner_PRA_2018}%
  \BibitemOpen
  \bibfield  {author} {\bibinfo {author} {\bibfnamefont {S.}~\bibnamefont
  {Greschner}}, \bibinfo {author} {\bibfnamefont {L.}~\bibnamefont
  {Cardarelli}}, \ and\ \bibinfo {author} {\bibfnamefont {L.}~\bibnamefont
  {Santos}},\ }\href {\doibase 10.1103/PhysRevA.97.053605} {\bibfield
  {journal} {\bibinfo  {journal} {Phys. Rev. A}\ }\textbf {\bibinfo {volume}
  {97}},\ \bibinfo {pages} {053605} (\bibinfo {year} {2018})}\BibitemShut
  {NoStop}%
\bibitem [{\citenamefont {{Aneziris}}\ \emph {et~al.}(1991)\citenamefont
  {{Aneziris}}, \citenamefont {{Balachandran}},\ and\ \citenamefont
  {{Sen}}}]{Aneziris_IJMP_1991}%
  \BibitemOpen
  \bibfield  {author} {\bibinfo {author} {\bibfnamefont {C.}~\bibnamefont
  {{Aneziris}}}, \bibinfo {author} {\bibfnamefont {A.~P.}\ \bibnamefont
  {{Balachandran}}}, \ and\ \bibinfo {author} {\bibfnamefont {D.}~\bibnamefont
  {{Sen}}},\ }\href {\doibase 10.1142/S0217751X91002240} {\bibfield  {journal}
  {\bibinfo  {journal} {International Journal of Modern Physics A}\ }\textbf
  {\bibinfo {volume} {6}},\ \bibinfo {pages} {4721} (\bibinfo {year}
  {1991})}\BibitemShut {NoStop}%
\bibitem [{\citenamefont {Posske}\ \emph {et~al.}(2017)\citenamefont {Posske},
  \citenamefont {Trauzettel},\ and\ \citenamefont
  {Thorwart}}]{Posske_PRB_2017}%
  \BibitemOpen
  \bibfield  {author} {\bibinfo {author} {\bibfnamefont {T.}~\bibnamefont
  {Posske}}, \bibinfo {author} {\bibfnamefont {B.}~\bibnamefont {Trauzettel}},
  \ and\ \bibinfo {author} {\bibfnamefont {M.}~\bibnamefont {Thorwart}},\
  }\href {\doibase 10.1103/PhysRevB.96.195422} {\bibfield  {journal} {\bibinfo
  {journal} {Phys. Rev. B}\ }\textbf {\bibinfo {volume} {96}},\ \bibinfo
  {pages} {195422} (\bibinfo {year} {2017})}\BibitemShut {NoStop}%
\bibitem [{\citenamefont {Frau}\ \emph {et~al.}(1994)\citenamefont {Frau},
  \citenamefont {Lerda},\ and\ \citenamefont {Sciuto}}]{Frau_arXiv_1994}%
  \BibitemOpen
  \bibfield  {author} {\bibinfo {author} {\bibfnamefont {M.}~\bibnamefont
  {Frau}}, \bibinfo {author} {\bibfnamefont {A.}~\bibnamefont {Lerda}}, \ and\
  \bibinfo {author} {\bibfnamefont {S.}~\bibnamefont {Sciuto}},\ }\href
  {https://arxiv.org/abs/hep-th/9407161} {\bibfield  {journal} {\bibinfo
  {journal} {arXiv preprint hep-th/9407161}\ } (\bibinfo {year}
  {1994})}\BibitemShut {NoStop}%
\bibitem [{\citenamefont {El~Baz}\ and\ \citenamefont
  {Hassouni}(2003)}]{Baz_IJMP_2003}%
  \BibitemOpen
  \bibfield  {author} {\bibinfo {author} {\bibfnamefont {M.}~\bibnamefont
  {El~Baz}}\ and\ \bibinfo {author} {\bibfnamefont {Y.}~\bibnamefont
  {Hassouni}},\ }\href
  {https://www.worldscientific.com/doi/abs/10.1142/S0217751X03015386}
  {\bibfield  {journal} {\bibinfo  {journal} {International Journal of Modern
  Physics A}\ }\textbf {\bibinfo {volume} {18}},\ \bibinfo {pages} {3015}
  (\bibinfo {year} {2003})}\BibitemShut {NoStop}%
\bibitem [{\citenamefont {Haldane}(1991)}]{Haldane_PRL_1991}%
  \BibitemOpen
  \bibfield  {author} {\bibinfo {author} {\bibfnamefont {F.~D.~M.}\
  \bibnamefont {Haldane}},\ }\href {\doibase 10.1103/PhysRevLett.67.937}
  {\bibfield  {journal} {\bibinfo  {journal} {Phys. Rev. Lett.}\ }\textbf
  {\bibinfo {volume} {67}},\ \bibinfo {pages} {937} (\bibinfo {year}
  {1991})}\BibitemShut {NoStop}%
\bibitem [{\citenamefont {Wu}(1994)}]{Wu_PRL_1994}%
  \BibitemOpen
  \bibfield  {author} {\bibinfo {author} {\bibfnamefont {Y.-S.}\ \bibnamefont
  {Wu}},\ }\href {\doibase 10.1103/PhysRevLett.73.922} {\bibfield  {journal}
  {\bibinfo  {journal} {Phys. Rev. Lett.}\ }\textbf {\bibinfo {volume} {73}},\
  \bibinfo {pages} {922} (\bibinfo {year} {1994})}\BibitemShut {NoStop}%
\bibitem [{\citenamefont {Murthy}\ and\ \citenamefont
  {Shankar}(1994)}]{Murthy_PRL_1994}%
  \BibitemOpen
  \bibfield  {author} {\bibinfo {author} {\bibfnamefont {M.~V.~N.}\
  \bibnamefont {Murthy}}\ and\ \bibinfo {author} {\bibfnamefont
  {R.}~\bibnamefont {Shankar}},\ }\href {\doibase 10.1103/PhysRevLett.73.3331}
  {\bibfield  {journal} {\bibinfo  {journal} {Phys. Rev. Lett.}\ }\textbf
  {\bibinfo {volume} {73}},\ \bibinfo {pages} {3331} (\bibinfo {year}
  {1994})}\BibitemShut {NoStop}%
\bibitem [{\citenamefont {Batchelor}\ \emph {et~al.}(2006)\citenamefont
  {Batchelor}, \citenamefont {Guan},\ and\ \citenamefont
  {Oelkers}}]{Batchelor_PRL_2006}%
  \BibitemOpen
  \bibfield  {author} {\bibinfo {author} {\bibfnamefont {M.~T.}\ \bibnamefont
  {Batchelor}}, \bibinfo {author} {\bibfnamefont {X.-W.}\ \bibnamefont {Guan}},
  \ and\ \bibinfo {author} {\bibfnamefont {N.}~\bibnamefont {Oelkers}},\ }\href
  {\doibase 10.1103/PhysRevLett.96.210402} {\bibfield  {journal} {\bibinfo
  {journal} {Phys. Rev. Lett.}\ }\textbf {\bibinfo {volume} {96}},\ \bibinfo
  {pages} {210402} (\bibinfo {year} {2006})}\BibitemShut {NoStop}%
\bibitem [{\citenamefont {Guo}\ \emph {et~al.}(2009)\citenamefont {Guo},
  \citenamefont {Hao},\ and\ \citenamefont {Chen}}]{Guo_PRA_2009}%
  \BibitemOpen
  \bibfield  {author} {\bibinfo {author} {\bibfnamefont {H.}~\bibnamefont
  {Guo}}, \bibinfo {author} {\bibfnamefont {Y.}~\bibnamefont {Hao}}, \ and\
  \bibinfo {author} {\bibfnamefont {S.}~\bibnamefont {Chen}},\ }\href {\doibase
  10.1103/PhysRevA.80.052332} {\bibfield  {journal} {\bibinfo  {journal} {Phys.
  Rev. A}\ }\textbf {\bibinfo {volume} {80}},\ \bibinfo {pages} {052332}
  (\bibinfo {year} {2009})}\BibitemShut {NoStop}%
\bibitem [{\citenamefont {Eckholt}\ and\ \citenamefont
  {Garc{\'\i}a-Ripoll}(2009)}]{Eckholt_NJP_2009}%
  \BibitemOpen
  \bibfield  {author} {\bibinfo {author} {\bibfnamefont {M.}~\bibnamefont
  {Eckholt}}\ and\ \bibinfo {author} {\bibfnamefont {J.~J.}\ \bibnamefont
  {Garc{\'\i}a-Ripoll}},\ }\href
  {http://iopscience.iop.org/article/10.1088/1367-2630/11/9/093028/meta}
  {\bibfield  {journal} {\bibinfo  {journal} {New Journal of Physics}\ }\textbf
  {\bibinfo {volume} {11}},\ \bibinfo {pages} {093028} (\bibinfo {year}
  {2009})}\BibitemShut {NoStop}%
\bibitem [{\citenamefont {Eckholt}\ and\ \citenamefont
  {Garc\'{\i}a-Ripoll}(2008)}]{Eckholt_PRA_2008}%
  \BibitemOpen
  \bibfield  {author} {\bibinfo {author} {\bibfnamefont {M.}~\bibnamefont
  {Eckholt}}\ and\ \bibinfo {author} {\bibfnamefont {J.~J.}\ \bibnamefont
  {Garc\'{\i}a-Ripoll}},\ }\href {\doibase 10.1103/PhysRevA.77.063603}
  {\bibfield  {journal} {\bibinfo  {journal} {Phys. Rev. A}\ }\textbf {\bibinfo
  {volume} {77}},\ \bibinfo {pages} {063603} (\bibinfo {year}
  {2008})}\BibitemShut {NoStop}%
\bibitem [{\citenamefont {Zhang}\ \emph {et~al.}(2017)\citenamefont {Zhang},
  \citenamefont {Greschner}, \citenamefont {Fan}, \citenamefont {Scott},\ and\
  \citenamefont {Zhang}}]{Zhang_PRA_2017}%
  \BibitemOpen
  \bibfield  {author} {\bibinfo {author} {\bibfnamefont {W.}~\bibnamefont
  {Zhang}}, \bibinfo {author} {\bibfnamefont {S.}~\bibnamefont {Greschner}},
  \bibinfo {author} {\bibfnamefont {E.}~\bibnamefont {Fan}}, \bibinfo {author}
  {\bibfnamefont {T.~C.}\ \bibnamefont {Scott}}, \ and\ \bibinfo {author}
  {\bibfnamefont {Y.}~\bibnamefont {Zhang}},\ }\href {\doibase
  10.1103/PhysRevA.95.053614} {\bibfield  {journal} {\bibinfo  {journal} {Phys.
  Rev. A}\ }\textbf {\bibinfo {volume} {95}},\ \bibinfo {pages} {053614}
  (\bibinfo {year} {2017})}\BibitemShut {NoStop}%
\bibitem [{\citenamefont {Lange}\ \emph
  {et~al.}(2017{\natexlab{a}})\citenamefont {Lange}, \citenamefont {Ejima},\
  and\ \citenamefont {Fehske}}]{Lange_PRL_2017}%
  \BibitemOpen
  \bibfield  {author} {\bibinfo {author} {\bibfnamefont {F.}~\bibnamefont
  {Lange}}, \bibinfo {author} {\bibfnamefont {S.}~\bibnamefont {Ejima}}, \ and\
  \bibinfo {author} {\bibfnamefont {H.}~\bibnamefont {Fehske}},\ }\href
  {\doibase 10.1103/PhysRevLett.118.120401} {\bibfield  {journal} {\bibinfo
  {journal} {Phys. Rev. Lett.}\ }\textbf {\bibinfo {volume} {118}},\ \bibinfo
  {pages} {120401} (\bibinfo {year} {2017}{\natexlab{a}})}\BibitemShut
  {NoStop}%
\bibitem [{\citenamefont {Lange}\ \emph
  {et~al.}(2017{\natexlab{b}})\citenamefont {Lange}, \citenamefont {Ejima},\
  and\ \citenamefont {Fehske}}]{Lange_PRA_2017}%
  \BibitemOpen
  \bibfield  {author} {\bibinfo {author} {\bibfnamefont {F.}~\bibnamefont
  {Lange}}, \bibinfo {author} {\bibfnamefont {S.}~\bibnamefont {Ejima}}, \ and\
  \bibinfo {author} {\bibfnamefont {H.}~\bibnamefont {Fehske}},\ }\href
  {\doibase 10.1103/PhysRevA.95.063621} {\bibfield  {journal} {\bibinfo
  {journal} {Phys. Rev. A}\ }\textbf {\bibinfo {volume} {95}},\ \bibinfo
  {pages} {063621} (\bibinfo {year} {2017}{\natexlab{b}})}\BibitemShut
  {NoStop}%
\bibitem [{\citenamefont {Arcila-Forero}\ \emph {et~al.}(2018)\citenamefont
  {Arcila-Forero}, \citenamefont {Franco},\ and\ \citenamefont
  {Silva-Valencia}}]{Forero_PRA_2018}%
  \BibitemOpen
  \bibfield  {author} {\bibinfo {author} {\bibfnamefont {J.}~\bibnamefont
  {Arcila-Forero}}, \bibinfo {author} {\bibfnamefont {R.}~\bibnamefont
  {Franco}}, \ and\ \bibinfo {author} {\bibfnamefont {J.}~\bibnamefont
  {Silva-Valencia}},\ }\href {\doibase 10.1103/PhysRevA.97.023631} {\bibfield
  {journal} {\bibinfo  {journal} {Phys. Rev. A}\ }\textbf {\bibinfo {volume}
  {97}},\ \bibinfo {pages} {023631} (\bibinfo {year} {2018})}\BibitemShut
  {NoStop}%
\bibitem [{\citenamefont {Zuo}\ \emph {et~al.}(2018)\citenamefont {Zuo},
  \citenamefont {Li},\ and\ \citenamefont {Li}}]{Zuo_PRB_2018}%
  \BibitemOpen
  \bibfield  {author} {\bibinfo {author} {\bibfnamefont {Z.-W.}\ \bibnamefont
  {Zuo}}, \bibinfo {author} {\bibfnamefont {G.-L.}\ \bibnamefont {Li}}, \ and\
  \bibinfo {author} {\bibfnamefont {L.}~\bibnamefont {Li}},\ }\href {\doibase
  10.1103/PhysRevB.97.115126} {\bibfield  {journal} {\bibinfo  {journal} {Phys.
  Rev. B}\ }\textbf {\bibinfo {volume} {97}},\ \bibinfo {pages} {115126}
  (\bibinfo {year} {2018})}\BibitemShut {NoStop}%
\bibitem [{\citenamefont {Ketterle}\ and\ \citenamefont
  {Zwierlein}(2008)}]{Ketterle_arXiv_2008}%
  \BibitemOpen
  \bibfield  {author} {\bibinfo {author} {\bibfnamefont {W.}~\bibnamefont
  {Ketterle}}\ and\ \bibinfo {author} {\bibfnamefont {M.~W.}\ \bibnamefont
  {Zwierlein}},\ }\href {https://arxiv.org/abs/0801.2500} {\bibfield  {journal}
  {\bibinfo  {journal} {arXiv preprint arXiv:0801.2500}\ } (\bibinfo {year}
  {2008})}\BibitemShut {NoStop}%
\bibitem [{\citenamefont {Giorgini}\ \emph {et~al.}(2008)\citenamefont
  {Giorgini}, \citenamefont {Pitaevskii},\ and\ \citenamefont
  {Stringari}}]{Giorgini_RMP_2008}%
  \BibitemOpen
  \bibfield  {author} {\bibinfo {author} {\bibfnamefont {S.}~\bibnamefont
  {Giorgini}}, \bibinfo {author} {\bibfnamefont {L.~P.}\ \bibnamefont
  {Pitaevskii}}, \ and\ \bibinfo {author} {\bibfnamefont {S.}~\bibnamefont
  {Stringari}},\ }\href {\doibase 10.1103/RevModPhys.80.1215} {\bibfield
  {journal} {\bibinfo  {journal} {Rev. Mod. Phys.}\ }\textbf {\bibinfo {volume}
  {80}},\ \bibinfo {pages} {1215} (\bibinfo {year} {2008})}\BibitemShut
  {NoStop}%
\bibitem [{\citenamefont {Lifshitz}\ \emph {et~al.}(1960)\citenamefont
  {Lifshitz} \emph {et~al.}}]{Lifshitz_JETP_1960}%
  \BibitemOpen
  \bibfield  {author} {\bibinfo {author} {\bibfnamefont {I.}~\bibnamefont
  {Lifshitz}} \emph {et~al.},\ }\href
  {http://www.jetp.ac.ru/cgi-bin/e/index/e/11/5/p1130?a=list} {\bibfield
  {journal} {\bibinfo  {journal} {Sov. Phys. JETP}\ }\textbf {\bibinfo {volume}
  {11}},\ \bibinfo {pages} {1130} (\bibinfo {year} {1960})}\BibitemShut
  {NoStop}%
\bibitem [{\citenamefont {Blanter}\ \emph {et~al.}(1994)\citenamefont
  {Blanter}, \citenamefont {Kaganov}, \citenamefont {Pantsulaya},\ and\
  \citenamefont {Varlamov}}]{Blanter_PR_1994}%
  \BibitemOpen
  \bibfield  {author} {\bibinfo {author} {\bibfnamefont {Y.~M.}\ \bibnamefont
  {Blanter}}, \bibinfo {author} {\bibfnamefont {M.}~\bibnamefont {Kaganov}},
  \bibinfo {author} {\bibfnamefont {A.}~\bibnamefont {Pantsulaya}}, \ and\
  \bibinfo {author} {\bibfnamefont {A.}~\bibnamefont {Varlamov}},\ }\href
  {https://www.sciencedirect.com/science/article/pii/0370157394901031}
  {\bibfield  {journal} {\bibinfo  {journal} {Physics Reports}\ }\textbf
  {\bibinfo {volume} {245}},\ \bibinfo {pages} {159} (\bibinfo {year}
  {1994})}\BibitemShut {NoStop}%
\bibitem [{\citenamefont {Yamaji}\ \emph {et~al.}(2006)\citenamefont {Yamaji},
  \citenamefont {Misawa},\ and\ \citenamefont {Imada}}]{Yamaji_JPSJ_2006}%
  \BibitemOpen
  \bibfield  {author} {\bibinfo {author} {\bibfnamefont {Y.}~\bibnamefont
  {Yamaji}}, \bibinfo {author} {\bibfnamefont {T.}~\bibnamefont {Misawa}}, \
  and\ \bibinfo {author} {\bibfnamefont {M.}~\bibnamefont {Imada}},\ }\href
  {https://journals.jps.jp/doi/10.1143/JPSJ.75.094719} {\bibfield  {journal}
  {\bibinfo  {journal} {Journal of the Physical Society of Japan}\ }\textbf
  {\bibinfo {volume} {75}},\ \bibinfo {pages} {094719} (\bibinfo {year}
  {2006})}\BibitemShut {NoStop}%
\bibitem [{\citenamefont {Rodney}\ \emph {et~al.}(2013)\citenamefont {Rodney},
  \citenamefont {Song}, \citenamefont {Lee}, \citenamefont {Le~Hur},\ and\
  \citenamefont {S\o{}rensen}}]{Rodney_PRB_2013}%
  \BibitemOpen
  \bibfield  {author} {\bibinfo {author} {\bibfnamefont {M.}~\bibnamefont
  {Rodney}}, \bibinfo {author} {\bibfnamefont {H.~F.}\ \bibnamefont {Song}},
  \bibinfo {author} {\bibfnamefont {S.-S.}\ \bibnamefont {Lee}}, \bibinfo
  {author} {\bibfnamefont {K.}~\bibnamefont {Le~Hur}}, \ and\ \bibinfo {author}
  {\bibfnamefont {E.~S.}\ \bibnamefont {S\o{}rensen}},\ }\href {\doibase
  10.1103/PhysRevB.87.115132} {\bibfield  {journal} {\bibinfo  {journal} {Phys.
  Rev. B}\ }\textbf {\bibinfo {volume} {87}},\ \bibinfo {pages} {115132}
  (\bibinfo {year} {2013})}\BibitemShut {NoStop}%
\bibitem [{\citenamefont {Celi}\ \emph {et~al.}(2014)\citenamefont {Celi},
  \citenamefont {Massignan}, \citenamefont {Ruseckas}, \citenamefont {Goldman},
  \citenamefont {Spielman}, \citenamefont {Juzeli\ifmmode~\bar{u}\else
  \={u}\fi{}nas},\ and\ \citenamefont {Lewenstein}}]{Celi_PRL_2014}%
  \BibitemOpen
  \bibfield  {author} {\bibinfo {author} {\bibfnamefont {A.}~\bibnamefont
  {Celi}}, \bibinfo {author} {\bibfnamefont {P.}~\bibnamefont {Massignan}},
  \bibinfo {author} {\bibfnamefont {J.}~\bibnamefont {Ruseckas}}, \bibinfo
  {author} {\bibfnamefont {N.}~\bibnamefont {Goldman}}, \bibinfo {author}
  {\bibfnamefont {I.~B.}\ \bibnamefont {Spielman}}, \bibinfo {author}
  {\bibfnamefont {G.}~\bibnamefont {Juzeli\ifmmode~\bar{u}\else
  \={u}\fi{}nas}}, \ and\ \bibinfo {author} {\bibfnamefont {M.}~\bibnamefont
  {Lewenstein}},\ }\href {\doibase 10.1103/PhysRevLett.112.043001} {\bibfield
  {journal} {\bibinfo  {journal} {Phys. Rev. Lett.}\ }\textbf {\bibinfo
  {volume} {112}},\ \bibinfo {pages} {043001} (\bibinfo {year}
  {2014})}\BibitemShut {NoStop}%
\bibitem [{Note1()}]{Note1}%
  \BibitemOpen
  \bibinfo {note} {Supplemental Material, which includes citations to \cite
  {Dalmonte_PRB_2015, Carrasquilla_PRA_2013, Gu_IJMB_2010, Chung_PRB_2001,
  Cheong_PRB_2004, Peschel_JPA_2003, Vidal_PRL_2003, Sachdev_book}}\BibitemShut
  {NoStop}%
\bibitem [{\citenamefont {Hao}\ \emph {et~al.}(2009)\citenamefont {Hao},
  \citenamefont {Zhang},\ and\ \citenamefont {Chen}}]{Hao_PRA_2009}%
  \BibitemOpen
  \bibfield  {author} {\bibinfo {author} {\bibfnamefont {Y.}~\bibnamefont
  {Hao}}, \bibinfo {author} {\bibfnamefont {Y.}~\bibnamefont {Zhang}}, \ and\
  \bibinfo {author} {\bibfnamefont {S.}~\bibnamefont {Chen}},\ }\href {\doibase
  10.1103/PhysRevA.79.043633} {\bibfield  {journal} {\bibinfo  {journal} {Phys.
  Rev. A}\ }\textbf {\bibinfo {volume} {79}},\ \bibinfo {pages} {043633}
  (\bibinfo {year} {2009})}\BibitemShut {NoStop}%
\bibitem [{\citenamefont {Hao}\ and\ \citenamefont
  {Chen}(2012)}]{Hao_PRA_2012}%
  \BibitemOpen
  \bibfield  {author} {\bibinfo {author} {\bibfnamefont {Y.}~\bibnamefont
  {Hao}}\ and\ \bibinfo {author} {\bibfnamefont {S.}~\bibnamefont {Chen}},\
  }\href {\doibase 10.1103/PhysRevA.86.043631} {\bibfield  {journal} {\bibinfo
  {journal} {Phys. Rev. A}\ }\textbf {\bibinfo {volume} {86}},\ \bibinfo
  {pages} {043631} (\bibinfo {year} {2012})}\BibitemShut {NoStop}%
\bibitem [{\citenamefont {Foglio}\ and\ \citenamefont
  {Falicov}(1979)}]{Foglio_PRB_1979}%
  \BibitemOpen
  \bibfield  {author} {\bibinfo {author} {\bibfnamefont {M.~E.}\ \bibnamefont
  {Foglio}}\ and\ \bibinfo {author} {\bibfnamefont {L.~M.}\ \bibnamefont
  {Falicov}},\ }\href {\doibase 10.1103/PhysRevB.20.4554} {\bibfield  {journal}
  {\bibinfo  {journal} {Phys. Rev. B}\ }\textbf {\bibinfo {volume} {20}},\
  \bibinfo {pages} {4554} (\bibinfo {year} {1979})}\BibitemShut {NoStop}%
\bibitem [{\citenamefont {Arrachea}\ and\ \citenamefont
  {Aligia}(1994)}]{Arrachea_PRL_1994}%
  \BibitemOpen
  \bibfield  {author} {\bibinfo {author} {\bibfnamefont {L.}~\bibnamefont
  {Arrachea}}\ and\ \bibinfo {author} {\bibfnamefont {A.~A.}\ \bibnamefont
  {Aligia}},\ }\href {\doibase 10.1103/PhysRevLett.73.2240} {\bibfield
  {journal} {\bibinfo  {journal} {Phys. Rev. Lett.}\ }\textbf {\bibinfo
  {volume} {73}},\ \bibinfo {pages} {2240} (\bibinfo {year}
  {1994})}\BibitemShut {NoStop}%
\bibitem [{\citenamefont {de~Boer}\ \emph {et~al.}(1995)\citenamefont
  {de~Boer}, \citenamefont {Korepin},\ and\ \citenamefont
  {Schadschneider}}]{Boer_PRL_1995}%
  \BibitemOpen
  \bibfield  {author} {\bibinfo {author} {\bibfnamefont {J.}~\bibnamefont
  {de~Boer}}, \bibinfo {author} {\bibfnamefont {V.~E.}\ \bibnamefont
  {Korepin}}, \ and\ \bibinfo {author} {\bibfnamefont {A.}~\bibnamefont
  {Schadschneider}},\ }\href {\doibase 10.1103/PhysRevLett.74.789} {\bibfield
  {journal} {\bibinfo  {journal} {Phys. Rev. Lett.}\ }\textbf {\bibinfo
  {volume} {74}},\ \bibinfo {pages} {789} (\bibinfo {year} {1995})}\BibitemShut
  {NoStop}%
\bibitem [{\citenamefont {Vidal}\ and\ \citenamefont
  {Dou\ifmmode~\mbox{\c{c}}\else \c{c}\fi{}ot}(2001)}]{Vidal_PRB_2001}%
  \BibitemOpen
  \bibfield  {author} {\bibinfo {author} {\bibfnamefont {J.}~\bibnamefont
  {Vidal}}\ and\ \bibinfo {author} {\bibfnamefont {B.}~\bibnamefont
  {Dou\ifmmode~\mbox{\c{c}}\else \c{c}\fi{}ot}},\ }\href {\doibase
  10.1103/PhysRevB.65.045102} {\bibfield  {journal} {\bibinfo  {journal} {Phys.
  Rev. B}\ }\textbf {\bibinfo {volume} {65}},\ \bibinfo {pages} {045102}
  (\bibinfo {year} {2001})}\BibitemShut {NoStop}%
\bibitem [{\citenamefont {Mishra}\ \emph {et~al.}(2013)\citenamefont {Mishra},
  \citenamefont {Pai}, \citenamefont {Mukerjee},\ and\ \citenamefont
  {Paramekanti}}]{Mishra_PRB_2013}%
  \BibitemOpen
  \bibfield  {author} {\bibinfo {author} {\bibfnamefont {T.}~\bibnamefont
  {Mishra}}, \bibinfo {author} {\bibfnamefont {R.~V.}\ \bibnamefont {Pai}},
  \bibinfo {author} {\bibfnamefont {S.}~\bibnamefont {Mukerjee}}, \ and\
  \bibinfo {author} {\bibfnamefont {A.}~\bibnamefont {Paramekanti}},\ }\href
  {\doibase 10.1103/PhysRevB.87.174504} {\bibfield  {journal} {\bibinfo
  {journal} {Phys. Rev. B}\ }\textbf {\bibinfo {volume} {87}},\ \bibinfo
  {pages} {174504} (\bibinfo {year} {2013})}\BibitemShut {NoStop}%
\bibitem [{\citenamefont {Dhar}\ \emph {et~al.}(2013)\citenamefont {Dhar},
  \citenamefont {Mishra}, \citenamefont {Pai}, \citenamefont {Mukerjee},\ and\
  \citenamefont {Das}}]{Dhar_PRA_2013}%
  \BibitemOpen
  \bibfield  {author} {\bibinfo {author} {\bibfnamefont {A.}~\bibnamefont
  {Dhar}}, \bibinfo {author} {\bibfnamefont {T.}~\bibnamefont {Mishra}},
  \bibinfo {author} {\bibfnamefont {R.~V.}\ \bibnamefont {Pai}}, \bibinfo
  {author} {\bibfnamefont {S.}~\bibnamefont {Mukerjee}}, \ and\ \bibinfo
  {author} {\bibfnamefont {B.~P.}\ \bibnamefont {Das}},\ }\href {\doibase
  10.1103/PhysRevA.88.053625} {\bibfield  {journal} {\bibinfo  {journal} {Phys.
  Rev. A}\ }\textbf {\bibinfo {volume} {88}},\ \bibinfo {pages} {053625}
  (\bibinfo {year} {2013})}\BibitemShut {NoStop}%
\bibitem [{\citenamefont {Majumdar}\ and\ \citenamefont
  {Ghosh}(1969{\natexlab{a}})}]{Majumdar_JMP_1969}%
  \BibitemOpen
  \bibfield  {author} {\bibinfo {author} {\bibfnamefont {C.~K.}\ \bibnamefont
  {Majumdar}}\ and\ \bibinfo {author} {\bibfnamefont {D.~K.}\ \bibnamefont
  {Ghosh}},\ }\href {https://aip.scitation.org/doi/10.1063/1.1664979}
  {\bibfield  {journal} {\bibinfo  {journal} {Journal of Mathematical Physics}\
  }\textbf {\bibinfo {volume} {10}},\ \bibinfo {pages} {1399} (\bibinfo {year}
  {1969}{\natexlab{a}})}\BibitemShut {NoStop}%
\bibitem [{\citenamefont {Majumdar}\ and\ \citenamefont
  {Ghosh}(1969{\natexlab{b}})}]{Majumdar_JMP_1969_2}%
  \BibitemOpen
  \bibfield  {author} {\bibinfo {author} {\bibfnamefont {C.~K.}\ \bibnamefont
  {Majumdar}}\ and\ \bibinfo {author} {\bibfnamefont {D.~K.}\ \bibnamefont
  {Ghosh}},\ }\href {https://aip.scitation.org/doi/10.1063/1.1664978}
  {\bibfield  {journal} {\bibinfo  {journal} {Journal of Mathematical Physics}\
  }\textbf {\bibinfo {volume} {10}},\ \bibinfo {pages} {1388} (\bibinfo {year}
  {1969}{\natexlab{b}})}\BibitemShut {NoStop}%
\bibitem [{\citenamefont {Calabrese}\ and\ \citenamefont
  {Cardy}(2009)}]{Calabrese_JPA_2009}%
  \BibitemOpen
  \bibfield  {author} {\bibinfo {author} {\bibfnamefont {P.}~\bibnamefont
  {Calabrese}}\ and\ \bibinfo {author} {\bibfnamefont {J.}~\bibnamefont
  {Cardy}},\ }\href
  {http://iopscience.iop.org/article/10.1088/1751-8113/42/50/504005} {\bibfield
   {journal} {\bibinfo  {journal} {Journal of Physics A: Mathematical and
  Theoretical}\ }\textbf {\bibinfo {volume} {42}},\ \bibinfo {pages} {504005}
  (\bibinfo {year} {2009})}\BibitemShut {NoStop}%
\bibitem [{\citenamefont {Sule}\ \emph {et~al.}(2015)\citenamefont {Sule},
  \citenamefont {Changlani}, \citenamefont {Maruyama},\ and\ \citenamefont
  {Ryu}}]{sule2015determination}%
  \BibitemOpen
  \bibfield  {author} {\bibinfo {author} {\bibfnamefont {O.~M.}\ \bibnamefont
  {Sule}}, \bibinfo {author} {\bibfnamefont {H.~J.}\ \bibnamefont {Changlani}},
  \bibinfo {author} {\bibfnamefont {I.}~\bibnamefont {Maruyama}}, \ and\
  \bibinfo {author} {\bibfnamefont {S.}~\bibnamefont {Ryu}},\ }\href {\doibase
  10.1103/PhysRevB.92.075128} {\bibfield  {journal} {\bibinfo  {journal} {Phys.
  Rev. B}\ }\textbf {\bibinfo {volume} {92}},\ \bibinfo {pages} {075128}
  (\bibinfo {year} {2015})}\BibitemShut {NoStop}%
\bibitem [{\citenamefont {Baeriswyl}\ and\ \citenamefont
  {Degiorgi}(2005)}]{baeriswyl2005strong}%
  \BibitemOpen
  \bibfield  {author} {\bibinfo {author} {\bibfnamefont {D.}~\bibnamefont
  {Baeriswyl}}\ and\ \bibinfo {author} {\bibfnamefont {L.}~\bibnamefont
  {Degiorgi}},\ }\href@noop {} {\emph {\bibinfo {title} {Strong interactions in
  low dimensions}}},\ Vol.~\bibinfo {volume} {25}\ (\bibinfo  {publisher}
  {Springer Science \& Business Media},\ \bibinfo {year} {2005})\BibitemShut
  {NoStop}%
\bibitem [{\citenamefont {Sachdev}(2002)}]{sachdev2002quantum}%
  \BibitemOpen
  \bibfield  {author} {\bibinfo {author} {\bibfnamefont {S.}~\bibnamefont
  {Sachdev}},\ }\href
  {https://www.sciencedirect.com/science/article/pii/S0378437102010403}
  {\bibfield  {journal} {\bibinfo  {journal} {Physica A: Statistical Mechanics
  and its Applications}\ }\textbf {\bibinfo {volume} {313}},\ \bibinfo {pages}
  {252} (\bibinfo {year} {2002})}\BibitemShut {NoStop}%
\bibitem [{\citenamefont {Ray}\ \emph {et~al.}(2017)\citenamefont {Ray},
  \citenamefont {Mukerjee},\ and\ \citenamefont {Shenoy}}]{Ray_AP_2017}%
  \BibitemOpen
  \bibfield  {author} {\bibinfo {author} {\bibfnamefont {S.}~\bibnamefont
  {Ray}}, \bibinfo {author} {\bibfnamefont {S.}~\bibnamefont {Mukerjee}}, \
  and\ \bibinfo {author} {\bibfnamefont {V.~B.}\ \bibnamefont {Shenoy}},\
  }\href {\doibase https://doi.org/10.1016/j.aop.2017.07.002} {\bibfield
  {journal} {\bibinfo  {journal} {Annals of Physics}\ }\textbf {\bibinfo
  {volume} {384}},\ \bibinfo {pages} {71 } (\bibinfo {year}
  {2017})}\BibitemShut {NoStop}%
\bibitem [{\citenamefont {Sitte}\ \emph {et~al.}(2009)\citenamefont {Sitte},
  \citenamefont {Rosch}, \citenamefont {Meyer}, \citenamefont {Matveev},\ and\
  \citenamefont {Garst}}]{Sitte_PRL_2009}%
  \BibitemOpen
  \bibfield  {author} {\bibinfo {author} {\bibfnamefont {M.}~\bibnamefont
  {Sitte}}, \bibinfo {author} {\bibfnamefont {A.}~\bibnamefont {Rosch}},
  \bibinfo {author} {\bibfnamefont {J.~S.}\ \bibnamefont {Meyer}}, \bibinfo
  {author} {\bibfnamefont {K.~A.}\ \bibnamefont {Matveev}}, \ and\ \bibinfo
  {author} {\bibfnamefont {M.}~\bibnamefont {Garst}},\ }\href {\doibase
  10.1103/PhysRevLett.102.176404} {\bibfield  {journal} {\bibinfo  {journal}
  {Phys. Rev. Lett.}\ }\textbf {\bibinfo {volume} {102}},\ \bibinfo {pages}
  {176404} (\bibinfo {year} {2009})}\BibitemShut {NoStop}%
\bibitem [{\citenamefont {Meng}\ \emph {et~al.}(2011)\citenamefont {Meng},
  \citenamefont {Dixit}, \citenamefont {Garst},\ and\ \citenamefont
  {Meyer}}]{Meng_PRB_2011}%
  \BibitemOpen
  \bibfield  {author} {\bibinfo {author} {\bibfnamefont {T.}~\bibnamefont
  {Meng}}, \bibinfo {author} {\bibfnamefont {M.}~\bibnamefont {Dixit}},
  \bibinfo {author} {\bibfnamefont {M.}~\bibnamefont {Garst}}, \ and\ \bibinfo
  {author} {\bibfnamefont {J.~S.}\ \bibnamefont {Meyer}},\ }\href {\doibase
  10.1103/PhysRevB.83.125323} {\bibfield  {journal} {\bibinfo  {journal} {Phys.
  Rev. B}\ }\textbf {\bibinfo {volume} {83}},\ \bibinfo {pages} {125323}
  (\bibinfo {year} {2011})}\BibitemShut {NoStop}%
\bibitem [{\citenamefont {Liberto}\ \emph {et~al.}(2014)\citenamefont
  {Liberto}, \citenamefont {Creffield}, \citenamefont {Japaridze},\ and\
  \citenamefont {Smith}}]{Liberto_PRA_2014}%
  \BibitemOpen
  \bibfield  {author} {\bibinfo {author} {\bibfnamefont {M.~D.}\ \bibnamefont
  {Liberto}}, \bibinfo {author} {\bibfnamefont {C.~E.}\ \bibnamefont
  {Creffield}}, \bibinfo {author} {\bibfnamefont {G.~I.}\ \bibnamefont
  {Japaridze}}, \ and\ \bibinfo {author} {\bibfnamefont {C.~M.}\ \bibnamefont
  {Smith}},\ }\href {\doibase 10.1103/PhysRevA.89.013624} {\bibfield  {journal}
  {\bibinfo  {journal} {Phys. Rev. A}\ }\textbf {\bibinfo {volume} {89}},\
  \bibinfo {pages} {013624} (\bibinfo {year} {2014})}\BibitemShut {NoStop}%
\bibitem [{\citenamefont {Ghosh}\ \emph {et~al.}(2017)\citenamefont {Ghosh},
  \citenamefont {Greschner}, \citenamefont {Yadav}, \citenamefont {Mishra},
  \citenamefont {Rizzi},\ and\ \citenamefont {Shenoy}}]{Ghosh_PRA_2017}%
  \BibitemOpen
  \bibfield  {author} {\bibinfo {author} {\bibfnamefont {S.~K.}\ \bibnamefont
  {Ghosh}}, \bibinfo {author} {\bibfnamefont {S.}~\bibnamefont {Greschner}},
  \bibinfo {author} {\bibfnamefont {U.~K.}\ \bibnamefont {Yadav}}, \bibinfo
  {author} {\bibfnamefont {T.}~\bibnamefont {Mishra}}, \bibinfo {author}
  {\bibfnamefont {M.}~\bibnamefont {Rizzi}}, \ and\ \bibinfo {author}
  {\bibfnamefont {V.~B.}\ \bibnamefont {Shenoy}},\ }\href {\doibase
  10.1103/PhysRevA.95.063612} {\bibfield  {journal} {\bibinfo  {journal} {Phys.
  Rev. A}\ }\textbf {\bibinfo {volume} {95}},\ \bibinfo {pages} {063612}
  (\bibinfo {year} {2017})}\BibitemShut {NoStop}%
\bibitem [{\citenamefont {ITensor}()}]{Itensor}%
  \BibitemOpen
  \bibfield  {author} {\bibinfo {author} {\bibnamefont {ITensor}},\ }\href
  {http://itensor.org/} {\bibinfo  {journal} {http://itensor.org/}\
  }\BibitemShut {NoStop}%
\bibitem [{\citenamefont {Dalmonte}\ \emph {et~al.}(2015)\citenamefont
  {Dalmonte}, \citenamefont {Carrasquilla}, \citenamefont {Taddia},
  \citenamefont {Ercolessi},\ and\ \citenamefont {Rigol}}]{Dalmonte_PRB_2015}%
  \BibitemOpen
\bibfield  {journal} {  }\bibfield  {author} {\bibinfo {author} {\bibfnamefont
  {M.}~\bibnamefont {Dalmonte}}, \bibinfo {author} {\bibfnamefont
  {J.}~\bibnamefont {Carrasquilla}}, \bibinfo {author} {\bibfnamefont
  {L.}~\bibnamefont {Taddia}}, \bibinfo {author} {\bibfnamefont
  {E.}~\bibnamefont {Ercolessi}}, \ and\ \bibinfo {author} {\bibfnamefont
  {M.}~\bibnamefont {Rigol}},\ }\href {\doibase 10.1103/PhysRevB.91.165136}
  {\bibfield  {journal} {\bibinfo  {journal} {Phys. Rev. B}\ }\textbf {\bibinfo
  {volume} {91}},\ \bibinfo {pages} {165136} (\bibinfo {year}
  {2015})}\BibitemShut {NoStop}%
\bibitem [{\citenamefont {Carrasquilla}\ \emph {et~al.}(2013)\citenamefont
  {Carrasquilla}, \citenamefont {Manmana},\ and\ \citenamefont
  {Rigol}}]{Carrasquilla_PRA_2013}%
  \BibitemOpen
  \bibfield  {author} {\bibinfo {author} {\bibfnamefont {J.}~\bibnamefont
  {Carrasquilla}}, \bibinfo {author} {\bibfnamefont {S.~R.}\ \bibnamefont
  {Manmana}}, \ and\ \bibinfo {author} {\bibfnamefont {M.}~\bibnamefont
  {Rigol}},\ }\href {\doibase 10.1103/PhysRevA.87.043606} {\bibfield  {journal}
  {\bibinfo  {journal} {Phys. Rev. A}\ }\textbf {\bibinfo {volume} {87}},\
  \bibinfo {pages} {043606} (\bibinfo {year} {2013})}\BibitemShut {NoStop}%
\bibitem [{\citenamefont {Gu}(2010)}]{Gu_IJMB_2010}%
  \BibitemOpen
  \bibfield  {author} {\bibinfo {author} {\bibfnamefont {S.-J.}\ \bibnamefont
  {Gu}},\ }\href
  {https://www.worldscientific.com/doi/abs/10.1142/S0217979210056335}
  {\bibfield  {journal} {\bibinfo  {journal} {International Journal of Modern
  Physics B}\ }\textbf {\bibinfo {volume} {24}},\ \bibinfo {pages} {4371}
  (\bibinfo {year} {2010})}\BibitemShut {NoStop}%
\bibitem [{\citenamefont {Chung}\ and\ \citenamefont
  {Peschel}(2001)}]{Chung_PRB_2001}%
  \BibitemOpen
  \bibfield  {author} {\bibinfo {author} {\bibfnamefont {M.-C.}\ \bibnamefont
  {Chung}}\ and\ \bibinfo {author} {\bibfnamefont {I.}~\bibnamefont
  {Peschel}},\ }\href {\doibase 10.1103/PhysRevB.64.064412} {\bibfield
  {journal} {\bibinfo  {journal} {Phys. Rev. B}\ }\textbf {\bibinfo {volume}
  {64}},\ \bibinfo {pages} {064412} (\bibinfo {year} {2001})}\BibitemShut
  {NoStop}%
\bibitem [{\citenamefont {Cheong}\ and\ \citenamefont
  {Henley}(2004)}]{Cheong_PRB_2004}%
  \BibitemOpen
  \bibfield  {author} {\bibinfo {author} {\bibfnamefont {S.-A.}\ \bibnamefont
  {Cheong}}\ and\ \bibinfo {author} {\bibfnamefont {C.~L.}\ \bibnamefont
  {Henley}},\ }\href {\doibase 10.1103/PhysRevB.69.075111} {\bibfield
  {journal} {\bibinfo  {journal} {Phys. Rev. B}\ }\textbf {\bibinfo {volume}
  {69}},\ \bibinfo {pages} {075111} (\bibinfo {year} {2004})}\BibitemShut
  {NoStop}%
\bibitem [{\citenamefont {Peschel}(2003)}]{Peschel_JPA_2003}%
  \BibitemOpen
  \bibfield  {author} {\bibinfo {author} {\bibfnamefont {I.}~\bibnamefont
  {Peschel}},\ }\href {http://stacks.iop.org/0305-4470/36/i=14/a=101}
  {\bibfield  {journal} {\bibinfo  {journal} {Journal of Physics A:
  Mathematical and General}\ }\textbf {\bibinfo {volume} {36}},\ \bibinfo
  {pages} {L205} (\bibinfo {year} {2003})}\BibitemShut {NoStop}%
\bibitem [{\citenamefont {Vidal}\ \emph {et~al.}(2003)\citenamefont {Vidal},
  \citenamefont {Latorre}, \citenamefont {Rico},\ and\ \citenamefont
  {Kitaev}}]{Vidal_PRL_2003}%
  \BibitemOpen
  \bibfield  {author} {\bibinfo {author} {\bibfnamefont {G.}~\bibnamefont
  {Vidal}}, \bibinfo {author} {\bibfnamefont {J.~I.}\ \bibnamefont {Latorre}},
  \bibinfo {author} {\bibfnamefont {E.}~\bibnamefont {Rico}}, \ and\ \bibinfo
  {author} {\bibfnamefont {A.}~\bibnamefont {Kitaev}},\ }\href {\doibase
  10.1103/PhysRevLett.90.227902} {\bibfield  {journal} {\bibinfo  {journal}
  {Phys. Rev. Lett.}\ }\textbf {\bibinfo {volume} {90}},\ \bibinfo {pages}
  {227902} (\bibinfo {year} {2003})}\BibitemShut {NoStop}%
\bibitem [{\citenamefont {Sachdev}(2011)}]{Sachdev_book}%
  \BibitemOpen
  \bibfield  {author} {\bibinfo {author} {\bibfnamefont {S.}~\bibnamefont
  {Sachdev}},\ }\href@noop {} {\emph {\bibinfo {title} {Quantum phase
  transitions}}}\ (\bibinfo  {publisher} {Cambridge university press},\
  \bibinfo {year} {2011})\BibitemShut {NoStop}%
\end{thebibliography}%
\let\addcontentsline\oldaddcontentsline%

%\putbib[ref1DCorr]
%\end{bibunit}

\newpage

\onecolumngrid	
\relax
\clearpage
\newpage

\appendix

\setcounter{page}{1} 
\setcounter{figure}{0}
\setcounter{section}{0}

\renewcommand{\appendixname}{}
\renewcommand{\thesection}{S\arabic{section}}
\renewcommand{\thetable}{S\Roman{table}}
\renewcommand{\figurename}{Supplementary Figure}
\renewcommand{\thefigure}{S\arabic{figure}}
\renewcommand{\theequation}{\thesection.\arabic{equation}}

\centerline{\bf Supplemental Material}
\centerline{\bf for}
\centerline{\bf \titlename}
\centerline{Adhip Agarwala, Gaurav Kumar Gupta, Vijay B. Shenoy and Subhro Bhattacharjee}
%\affiliation{International Centre for Theoretical Sciences, Tata Institute of Fundamental Research, Bengaluru 560089, India}
%\author{Gaurav Kumar Gupta}
%\affiliation{Centre for Condensed Matter Theory, Department of Physics, Indian Institute of Science, Bangalore 560 012, India}
%\author{Vijay B. Shenoy}
%\affiliation{Centre for Condensed Matter Theory, Department of Physics, Indian Institute of Science, Bangalore 560 012, India}
%\author{Subhro Bhattacharjee}
%\affiliation{International Centre for Theoretical Sciences, Tata Institute of Fundamental Research, Bengaluru 560089, India}
%\title{\titlename}
%\author{Adhip Agarwala}
%\affiliation{International Centre for Theoretical Sciences, Tata Institute of Fundamental Research, Bengaluru 560089, India}
%\author{Gaurav Kumar Gupta}
%\affiliation{Centre for Condensed Matter Theory, Department of Physics, Indian Institute of Science, Bangalore 560 012, India}
%\author{Vijay B. Shenoy}
%\affiliation{Centre for Condensed Matter Theory, Department of Physics, Indian Institute of Science, Bangalore 560 012, India}
%\author{Subhro Bhattacharjee}
%\affiliation{International Centre for Theoretical Sciences, Tata Institute of Fundamental Research, Bengaluru 560089, India}

\tableofcontents
%%%%%%%%%%%%%%%%%%%%%%%%%
\section{Microscopic Hamiltonian :  symmetries and the free fermion limit ($\phi=0$)}
\label{SMsymm}

\begin{figure}
\centering
\begin{minipage}{0.1\textwidth}
\centering
\subfigure[]{
\includegraphics[scale=0.6,angle=90]{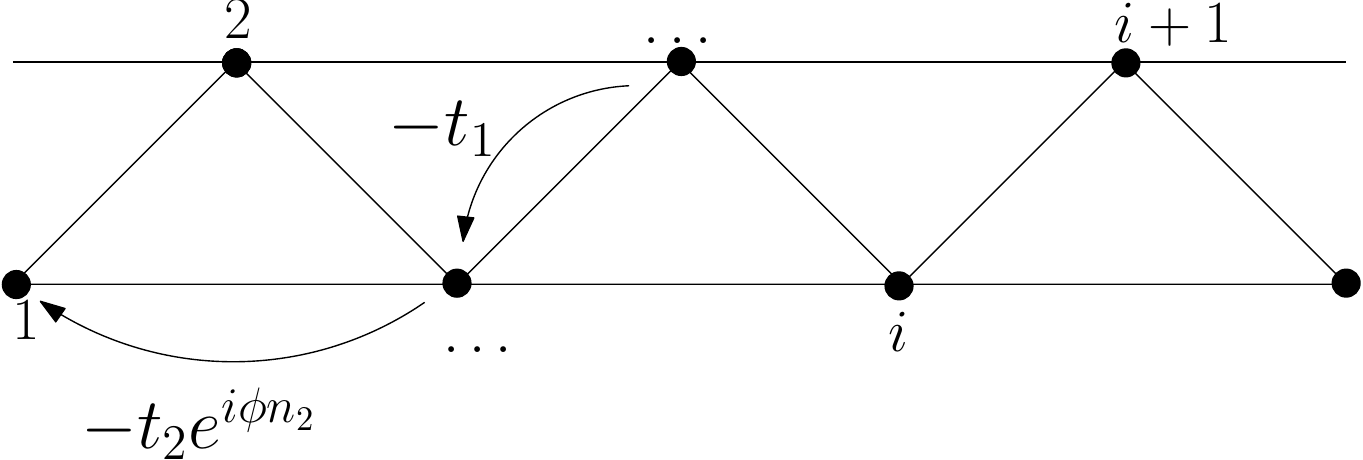}
\label{schemFig}
}
\end{minipage}
\begin{minipage}{0.7\textwidth}
\centering
\begin{subfigure}[Dispersion when $\phi=0$ and $t_2=2$]{
	\includegraphics[scale=0.4]{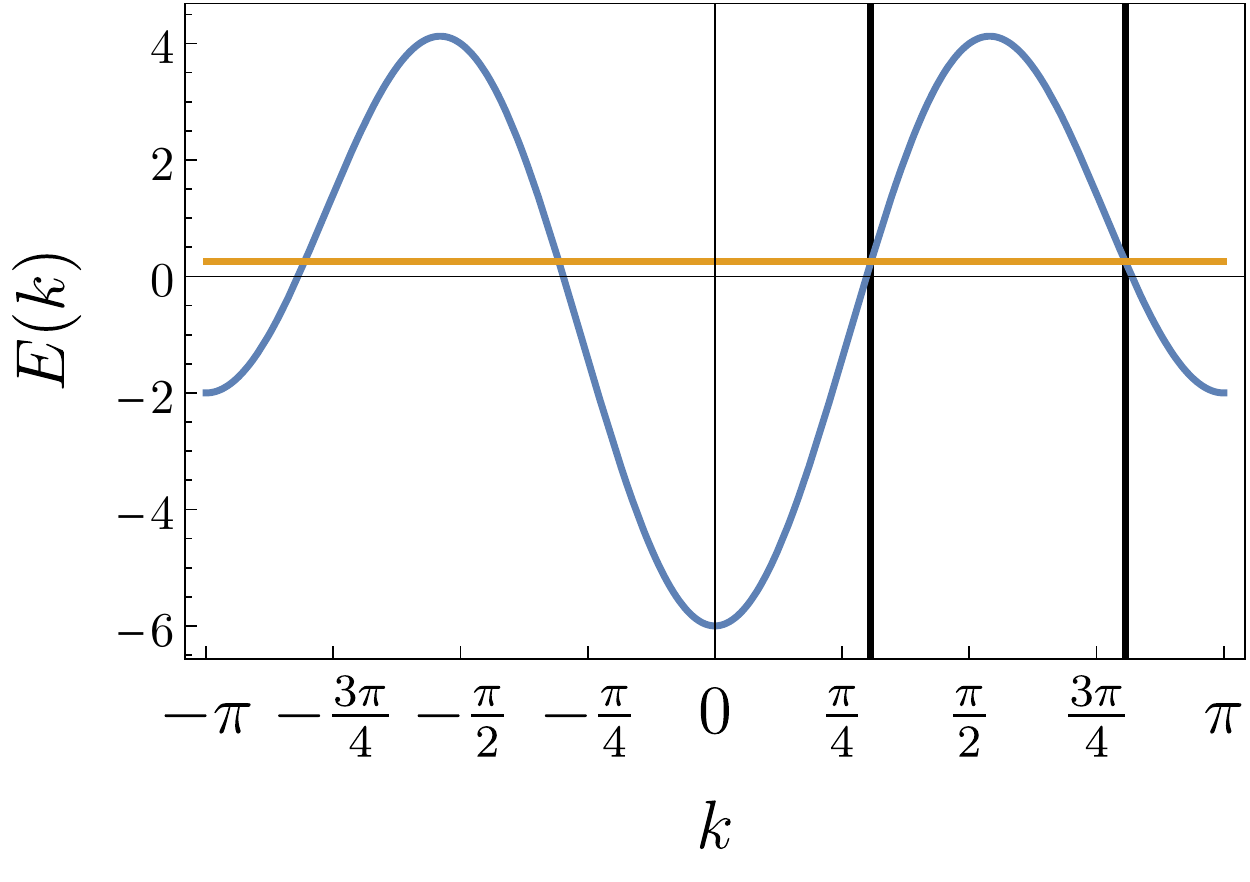}
			\mylabel{fig3}
		}
	\end{subfigure}
	\begin{subfigure}[Dispersion when $\phi=0$ and $t_2=0.5$]{
			\includegraphics[scale=0.4]{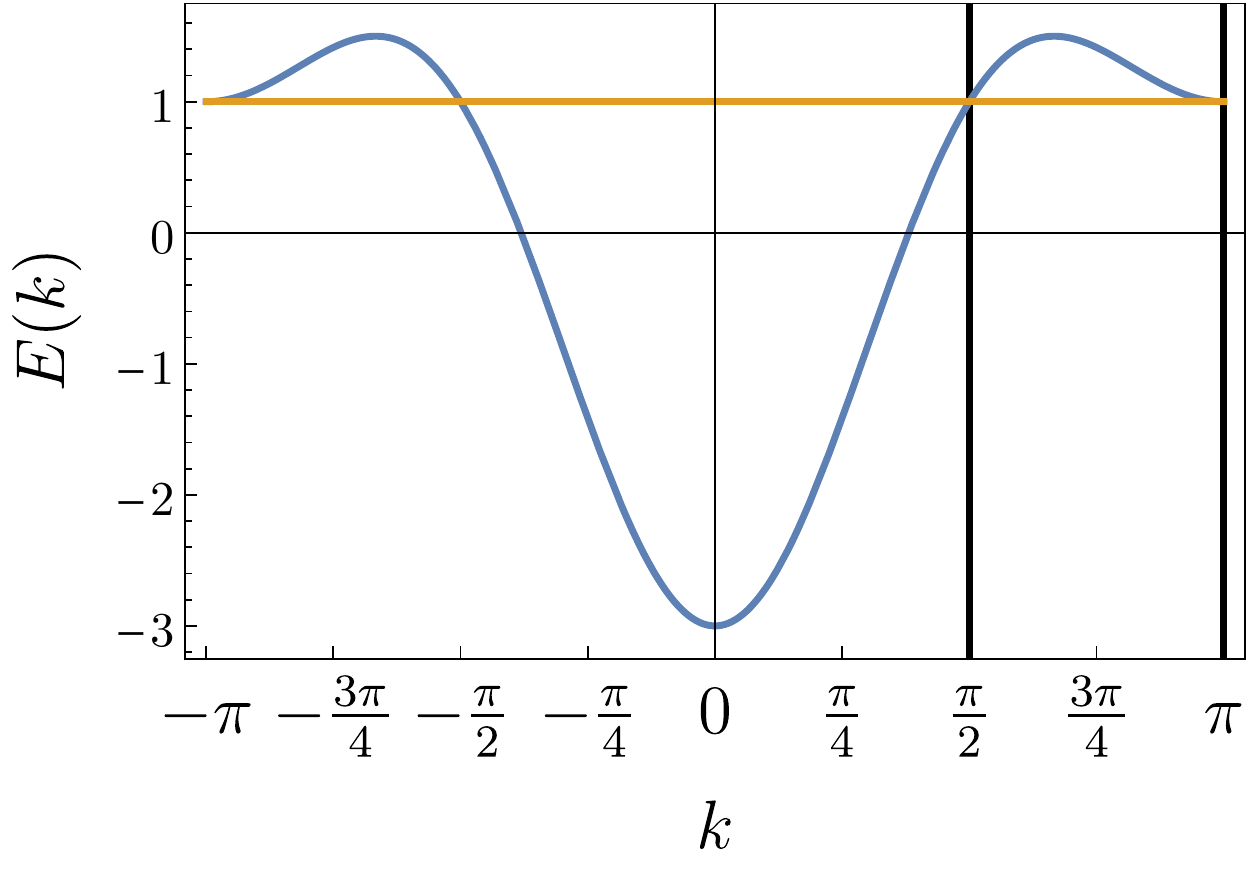}
			\label{fig1}
		}
	\end{subfigure}
	\begin{subfigure}[Dispersion when $\phi=0$ and $t_2=-0.5$]{
			\includegraphics[scale=0.4]{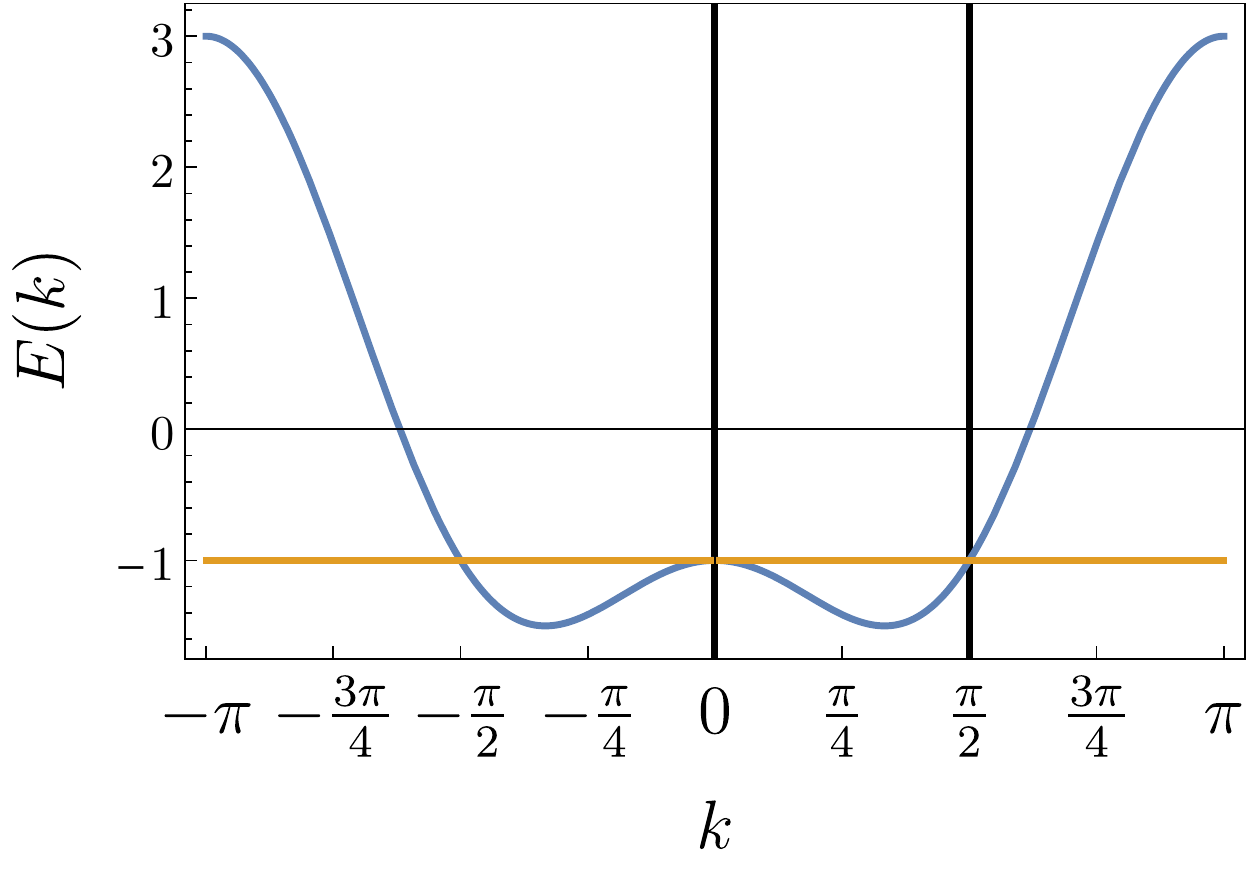}
			\mylabel{fig2}
		}
	\end{subfigure}
	\begin{subfigure}[Dispersion when $\phi=0$ and $t_2=-2$]{
			\includegraphics[scale=0.4]{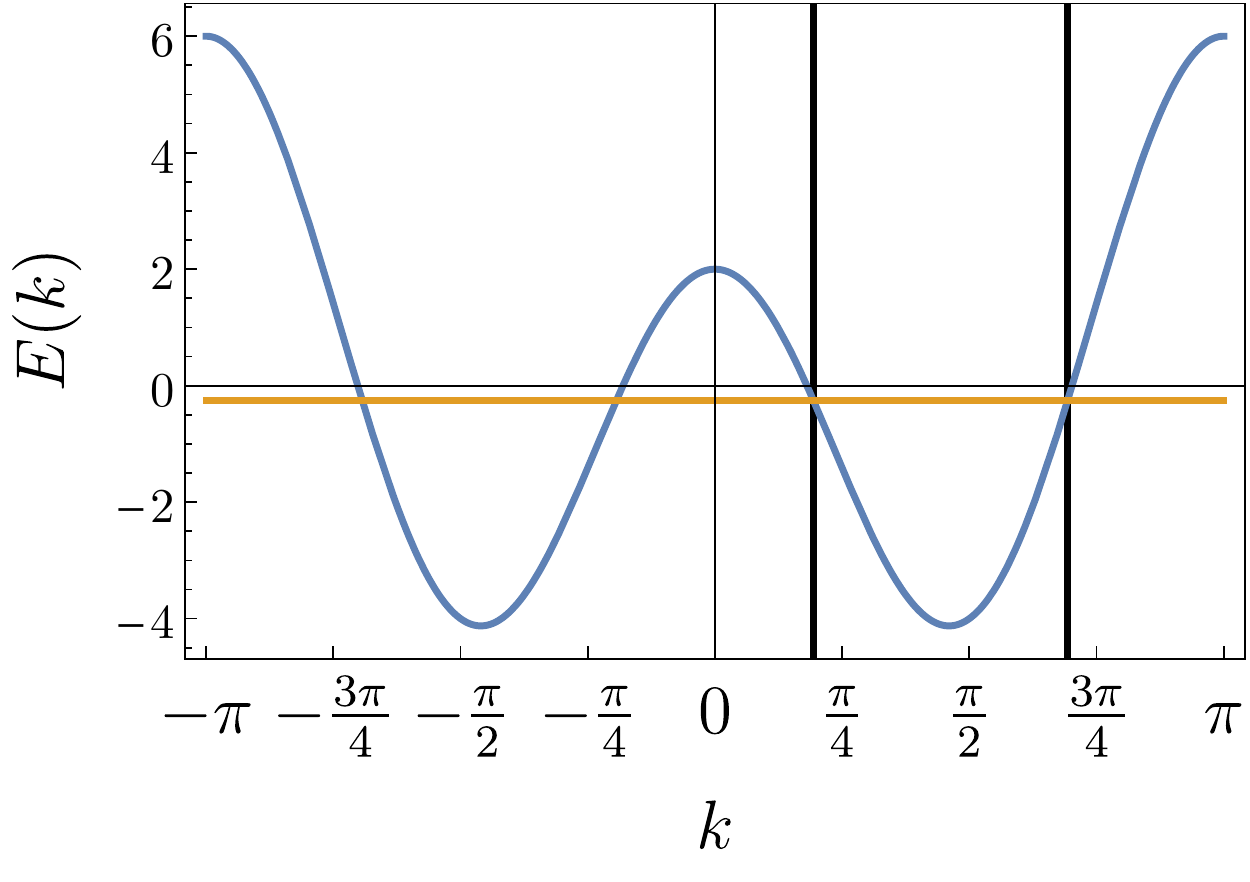}
			\mylabel{fig4}
		}
\end{subfigure}
\end{minipage}
\caption{(a) Schematic figure describing the Hamiltonian. (b)-(e) Single particle dispersion for the fermion problem in $\phi=0$ limit. For $t_2>0$  depending on the value of $t_2$ we can either have two (for $|t_2| < \frac{1}{2}$)  Fermi points (at $\pm \pi/2$) or four  (for $|t_2| > \frac{1}{2}$) Fermi points.}
\label{fig_disp}
\end{figure}

%%%%%%%%%%%%%%%%%%%%%%%%%
\subsection{Symmetries} 
The one dimensional chain with nearest and next nearest neighbor coupling can be alternatively thought of as a zig-zag ladder (fig. \ref{schemFig}) with the following symmetries : 
(1) Lattice translation by unit lattice spacing of the one dimensional chain, $T_x$, and (2) Combination of time-reversal (${\cal T}$) and parity ${(\cal P)}$ as  $\mathcal{Q}=\mathcal{P}\mathcal{T}$.

Under $T_x$, the fermions transform as, 
\bea
{T_1} c_i {T_1}^{-1} = c_{i+1}, \;\;\;\ {T_{-1}} c_i {T_{-1}}^{-1} = c_{i-1},\quad{T_1} c_L {T_1}^{-1} = c_{1}, \;\;\;\ {T_{-1}} c_1 {T_{-1}}^{-1} = c_{L}  .
\eea
Note that we have we have $L$ sites labeled $\{1 \ldots L\}$ where $L$ is taken to be an even number. Thus on the pseudofermions, $a_i$, translations act as
\bea
{T_1} a_i {T_1}^{-1} =  {T_1} K^\dagger_i c_i {T_1}^{-1} = e^{-i\phi n_1} K^\dagger_{i+1} c_{i+1} = e^{-i\phi N_1} a_{i+1} \\ 
{T_{-1}} a_i {T_{-1}}^{-1} =  {T_{-1}} K^\dagger_i c_i {T_{-1}}^{-1} = e^{i\phi n_L} K^\dagger_{i-1} c_{i-1} = e^{i\phi N_L} a_{i-1}. 
\eea
Under $\mathcal{T}$ and $\mathcal{P}$, the fermion operators transform in the following way
\bea
&&{\cal T} c_i {\cal T}^{-1} = c_i ,~~~~~~~~~~~~~~{\cal T} c^\dagger_i {\cal T}^{-1} = c^\dagger_i \\
&&{\cal P} c_i {\cal P}^{-1} = c_{L-i+1}~~~~~~~~~~~{\cal P} c^\dagger_i {\cal P}^{-1} = c^\dagger_{L-i+1} 
\eea
This results in the following transformation for the {\it pseudofermion} operators $a_i$,
\bea
{\cal T} a_i {\cal T}^{-1} &=& {\cal T} K^\dagger_i c_i {\cal T}^{-1} = K_i c_i \\
{\cal P} a_i {\cal P}^{-1} &=& {\cal P} K^\dagger_i c_i {\cal P}^{-1} = e^{i\phi {\cal N}} K_{L-i+2} c_{L-i+1} =  e^{i\phi {\cal N}} K_{L-i+1} c_{L-i+1} 
\eea
where ${\cal N} = \sum_i N_i = \sum_i n_i$. Thus, under the combination of these symmetries, $\mathcal{Q}$,  the $a_i$s transform as : 
\bea
{\cal Q}a_i{\cal Q}^{-1} = e^{-i \phi {\cal N}} a_{L-i+1}~~~~~~~~~~~~~~~~{\cal Q}a^\dagger_{i}{\cal Q}^{-1} = a^\dagger_{L-i+1} e^{i \phi {\cal N}} 
\eea
\subsection{Free fermions at $\phi=0$ }
\label{phi0free}
The free fermion dispersion at any value of $t_2$ is given by, 
\beq
E(k) = -2t_1 \cos k -2 t_2 \cos 2k
\eeq

The dispersions for few representative values of $t_2$ (keeping $t_1=1$) are shown in Fig. \ref{fig_disp}. For $-1/2<t_2<1/2$, there are two Fermi points at $k=\pm \pi/2$. For $|t_2|>1/2$, there are $4$ Fermi points ($k^{R\pm}_F, k^{L^\pm}_F$) given by
\begin{align}
&{\rm For}~t_2>1/2 : ~~~\quad k^{R+}_F=\arccos\left(\frac{-1}{2\sqrt{2}t_2}\right) - \frac{\pi}{4},~~k^{L+}_F = k^{R+}_F + \frac{\pi}{2},~~k^{R-}_F = -k^{L+}_F,~~k^{L-}_F = - k^{R+}_F\\
&{\rm For}~t_2<1/2 : ~~~\quad k^{R+}_F = \arccos\left(\frac{1}{2\sqrt{2}|t_2|}\right) + \frac{\pi}{4},~~k^{L+}_F = k^{R+}_F - \frac{\pi}{2},~~k^{R-}_F = -k^{L+}_F,~~k^{L-}_F = - k^{R+}_F
\end{align}

Therefore at $\phi=0$, as a function of $t_2$ we have Lifshitz transition akin to the tuning of the chemical potential.
%%%%%%%%%%%%%%%%%%%%%%%
\paragraph*{\ul{Fermi Velocity} :}

The Fermi velocity can be determined by the slope of the dispersing band at the Fermi points. For $|t_2|<\frac{1}{2}$ the Fermi points remain pinned at $k_F=\pm \frac{\pi}{2}$ and the corresponding Fermi velocities continue to be $v_F =  \frac{\partial E(k)}{\partial k}|_{k=k_F}  =  \pm 2t_1$. For $|t_2|>\frac{1}{2}$ the Fermi velocities at the four Fermi points depend on $t_2$ and are given by  $\left(2t_1 \sin k^{R/L \pm}_F + 4 t_2 \sin 2k^{R/L \pm}_F \right) $ respectively.

{

\section{Hartree-Fock theory}
\label{SMHFnk}

Rewriting the fermionic Hamiltonian as a sum of free quadratic part and an interacting quartic part, we get
\begin{align}
	H=H_0+H_{int}
\end{align}
where in the momentum space $k\in[-\pi,\pi]$, $H_0 = \sum_k \varepsilon_0(k)  c^\dagger_k c_k$ with $\varepsilon_0(k)=-2[t\cos(k) +t_2\cos(2k) ]$ being the bare dispersion and
\beq
H_{int} = -t_2 \sum_{k_1,k_2,k_3, k_4} \delta(k_1+k_2-k_3-k_4) [(e^{i\phi} -1)e^{i(-k_2+k_3+2k_4)}] c^\dagger_{k_1} c^\dagger_{k_2} c_{k_3}c_{k_4} + {\rm h.c.}
\label{intterm}
\eeq
Now concentrating on the Hartree-Fock decomposition of the interactions as
\beq
c^\dagger_{k_1} c^\dagger_{k_2} c_{k_3}c_{k_4} \rightarrow - \langle c^\dagger_{k_1}  c_{k_3}  \rangle c^\dagger_{k_2} c_{k_4}  - c^\dagger_{k_1}  c_{k_3} \langle c^\dagger_{k_2} c_{k_4} \rangle + \langle c^\dagger_{k_1} c_{k_4} \rangle c^\dagger_{k_2} c_{k_3} +  c^\dagger_{k_1} c_{k_4} \langle c^\dagger_{k_2} c_{k_3}  \rangle 
\eeq 
we get
\bea
H_{HF} &=& \sum_{k} \{\varepsilon_0(k) + A(k) + B + C(k) \} c^\dagger_k c_k
\eea
where,
\bea
A(k) &=&  {2}t_2 \sum_{k'} \{ (e^{i\phi}-1) e^{i(k+k')} + h.c. \} \langle n(k') \rangle \\
B &=& -2t_2 \sum_{k'}\{\cos(\phi+2k')-\cos(2k')\}\langle n(k') \rangle \\
C(k) &=& -2t_2 \{\sum_{k'} \langle n(k') \rangle \} \{\cos(\phi+2k) - \cos(2k) \}
\eea

\begin{figure}
\subfigure[]{
	\includegraphics[scale=0.5]{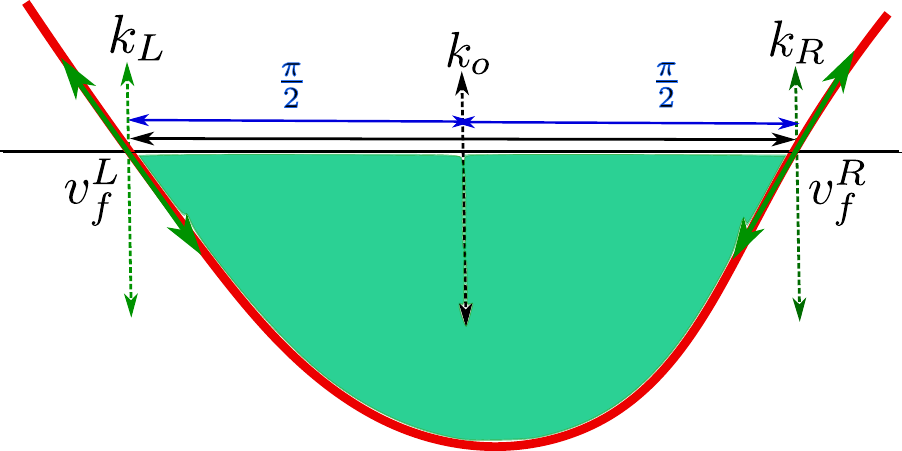}
	}
	\subfigure[]{
	\includegraphics[scale=0.5]{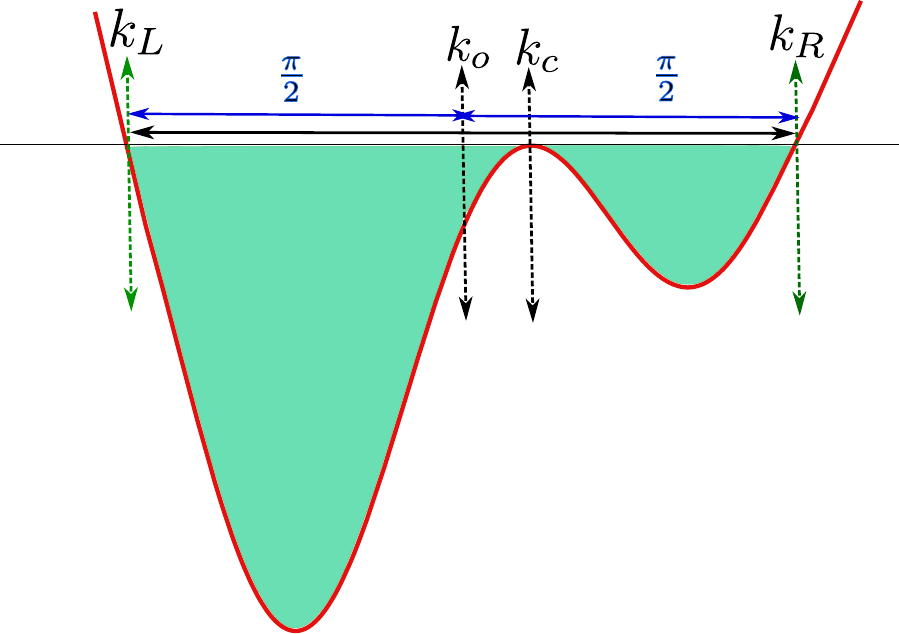}
	}
	\subfigure[]{
	\includegraphics[scale=0.5]{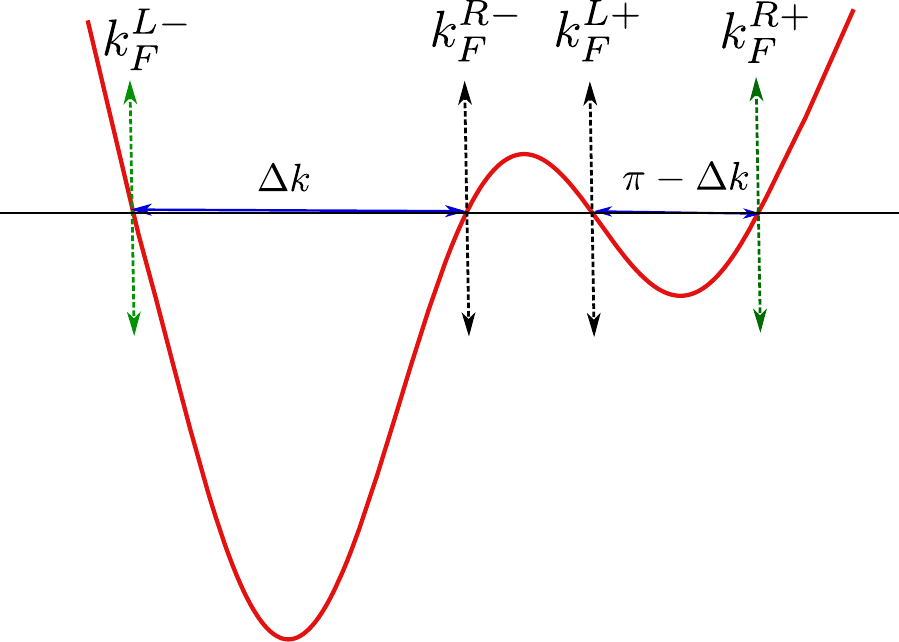}
	}
	\caption{(a) The Hartree-Fock (HF) dispersion is schematically shown in the TLL $c=1$ phase. (b) Schematic of the HF dispersion at a ``Lifshitz" transition when the system undergoes a  gapless-gapless transition between TLL ($c=1$) and TLL ($c=2$) phase.  (c) Schematic of the HF dispersion when the system is in $c=2$ TLL phase.}
	\label{HFSchemFerm}
\end{figure}

Solving this self-consistently for $\langle n(k)\rangle=\langle c_k^\dagger c_k\rangle$  produces the renormalised single particle dispersion for different values of the parameters. The general structure of the HF band is shown schematically in \Fig{HFSchemFerm} when the system is in either of the two kinds of TLL phases and at a Lifshitz transition.   \Fig{nofkMFT} shows the HF fermionic occupation for representative points in the parameter space. This should be compared with the DMRG results as shown in \Fig{nofk}. The HF theory clearly captures the gapless phases in $c=1$ and $c=2$ TLL regimes. It also captures the associated Fermi momenta and Lifshitz transitions as seen by a dashed line in Fig.~1 of the main text. 

\begin{figure}
	\includegraphics[width=0.23\columnwidth]{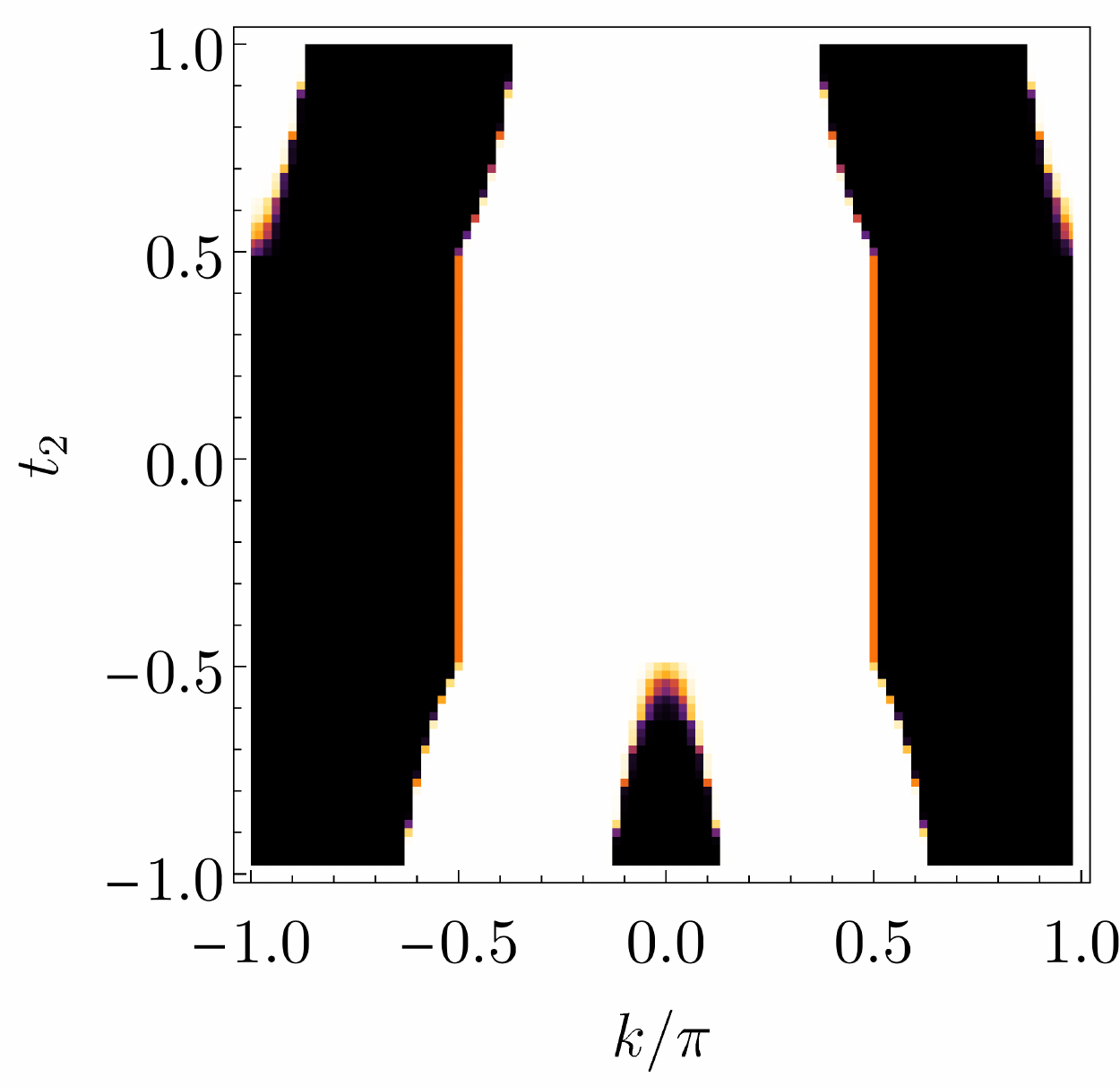}
	\includegraphics[width=0.23\columnwidth]{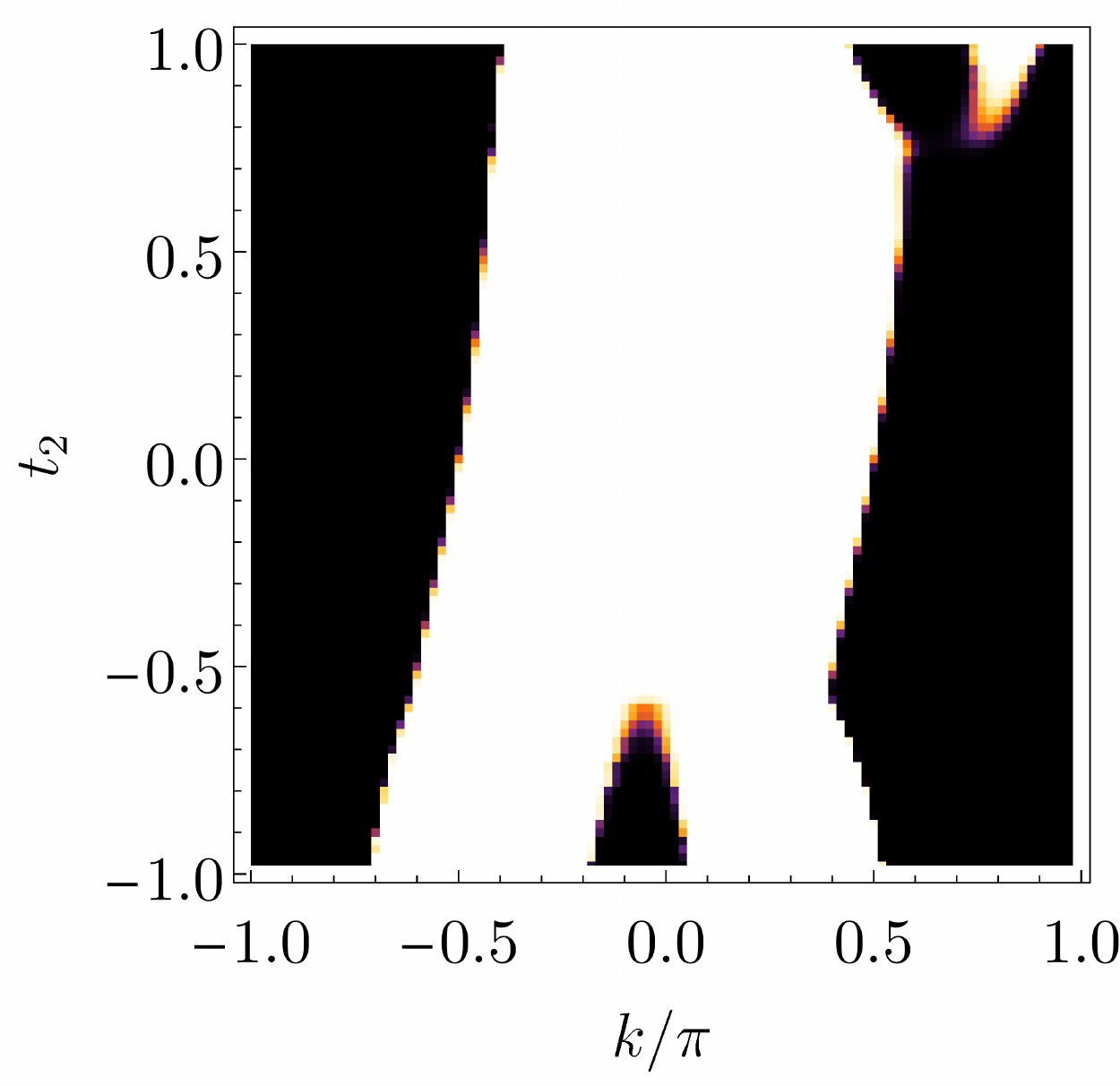}
	\includegraphics[width=0.23\columnwidth]{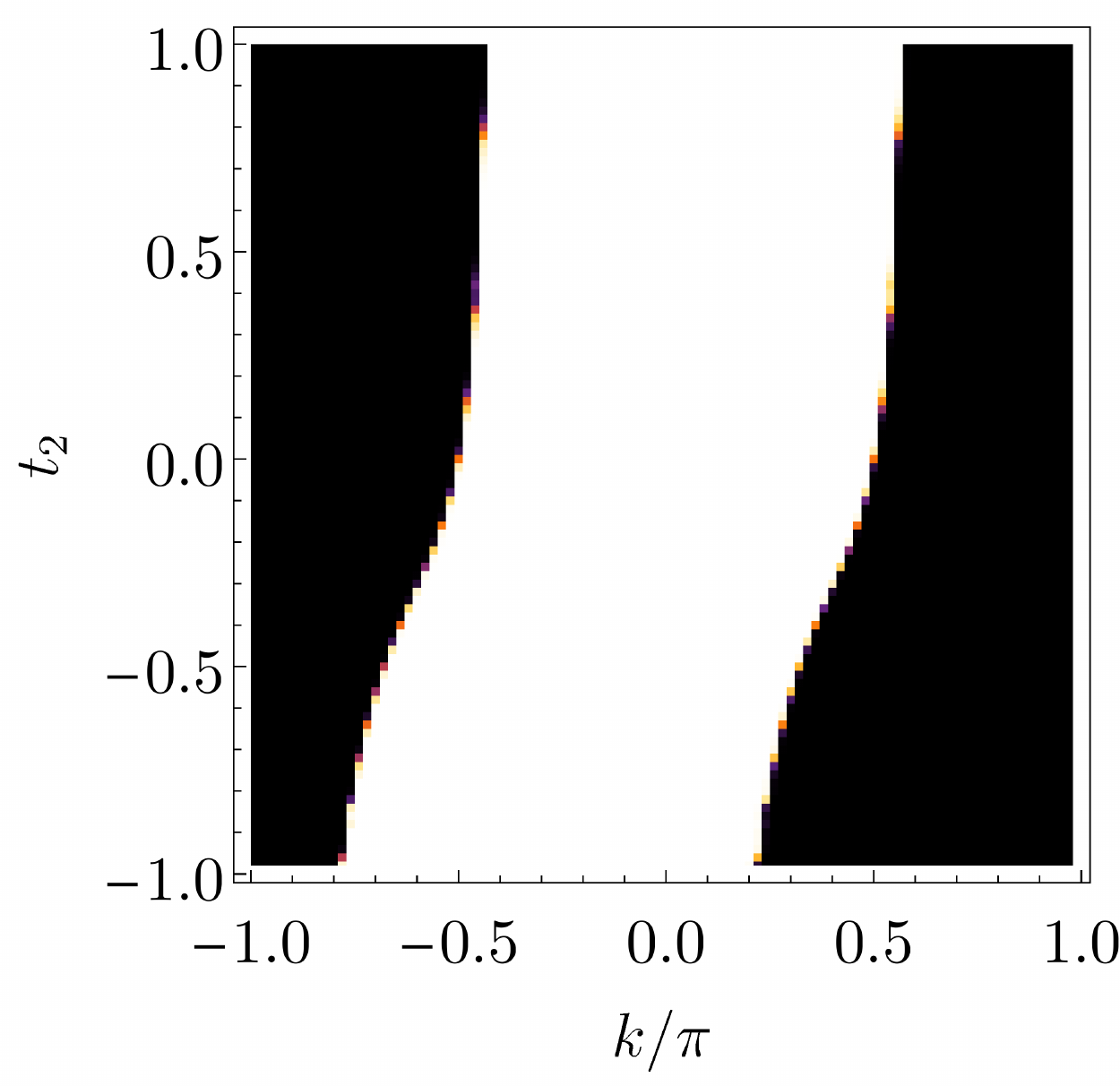}
	\includegraphics[width=0.23\columnwidth]{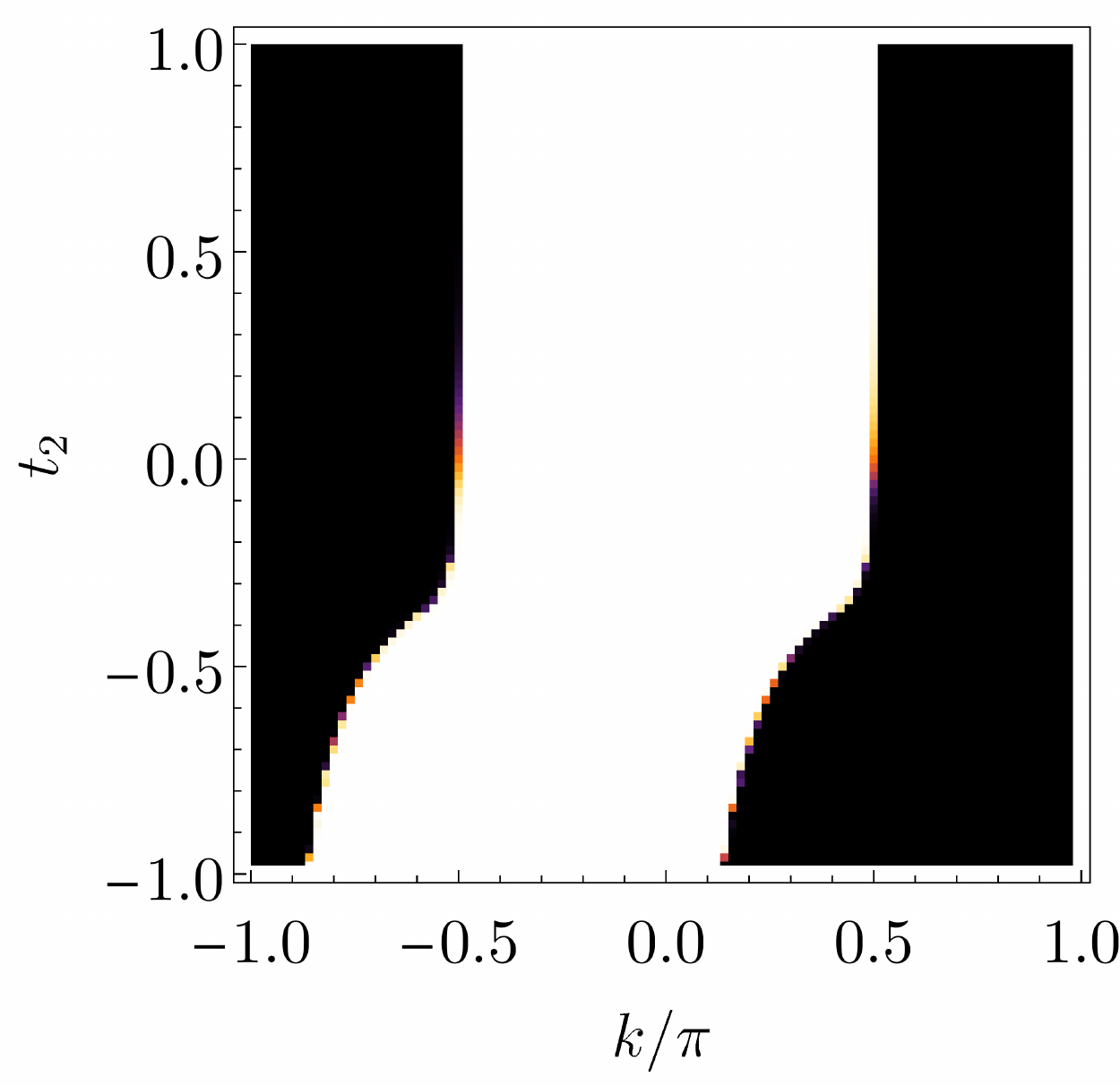}
	\caption{$\langle n(k) \rangle $ as a function of $k$ and $t_2$ for different values of $\phi$ using a self consistent mean field theory. The values of $\phi$ are $0.0, 1.0, 2.0$ and $3.0$.}
	\label{nofkMFT}
\end{figure}
In the $c=1$ phase the center of the Fermi sea is shifted and the occupied states have momenta $ k_o - \frac{\pi}{2} \le k \le k_o + \frac{\pi}{2}$ where the shift of the centre, $k_o$ is given by
\beq
\sin(k_o) = \frac{4t_2}{\pi} \sin\left(\frac{\phi}{2}\right) \cos \left(\frac{\phi}{2}+ 2k_o \right)
\label{fermshift}
\eeq 
In this phase, the HF Hamiltonian can be linearised about the two Fermi points by introducing the left and right moving fermion fields, $\Psi_L$ and $\Psi_R$ respectively to get linearised Hamiltonian in real space as
\begin{align}
	\mathcal{H}_0=i\int dx\left[v_f^R\Psi_R^\dagger\partial_x\Psi_R-v_f^L\Psi_L^\dagger\partial_x\Psi_L\right]
\end{align}
where $v_f^R(v_f^L)$ are the Fermi velocities for the right and left moving fermions respectively which are given by :
\begin{align}
	v^R_f=\frac{\partial\varepsilon_k}{\partial k_f^R}~~~~~v^L_f=-\frac{\partial\varepsilon_k}{\partial k^L_f}
\end{align}
Due to breaking of the time-reversal symmetry -- the Fermi sea center gets shifted from $k=0$ ($\equiv k_o$) and the two Fermi velocities at the Fermi points can in general be different. The variation of Fermi velocities and $k_o$ as a function of few parameters is shown in \Fig{diffvpar}.

\begin{figure}
	\includegraphics[width=0.23\columnwidth]{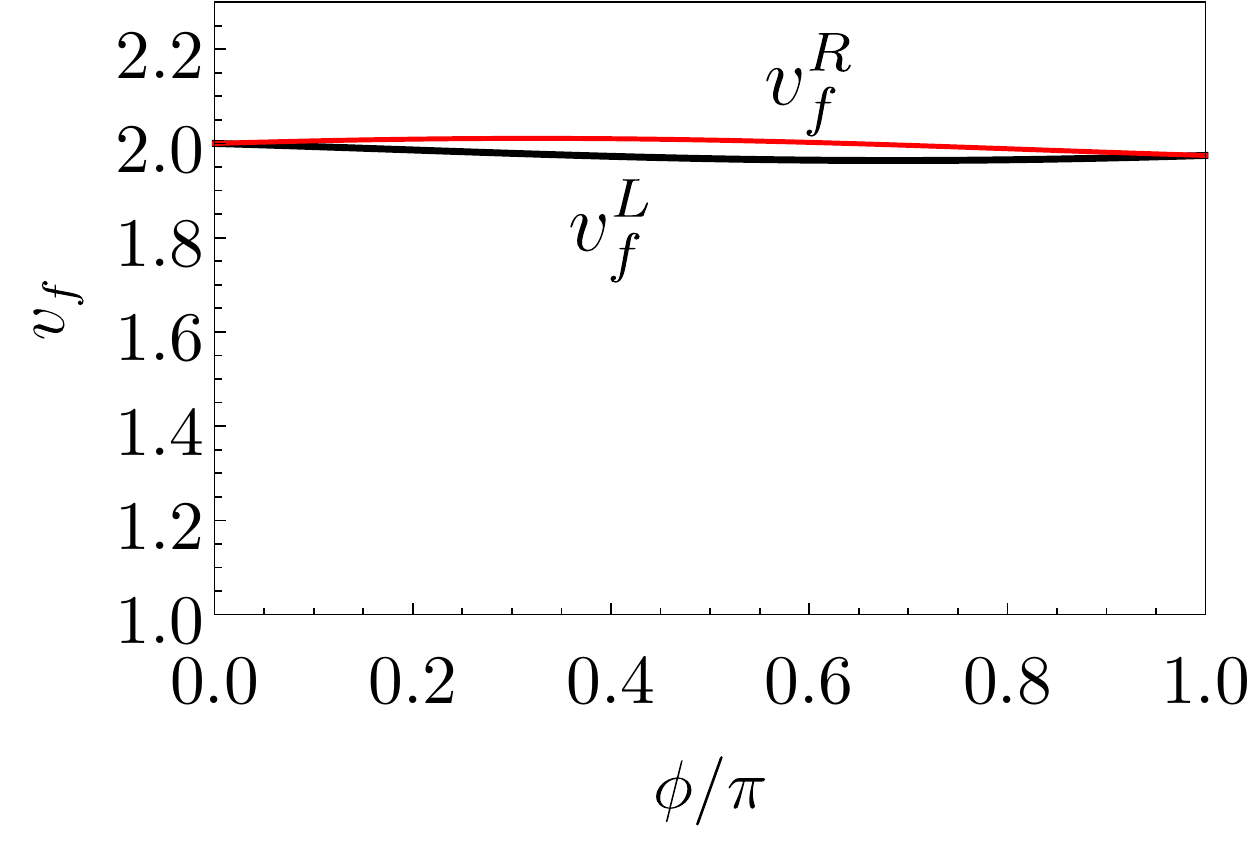}
	\includegraphics[width=0.23\columnwidth]{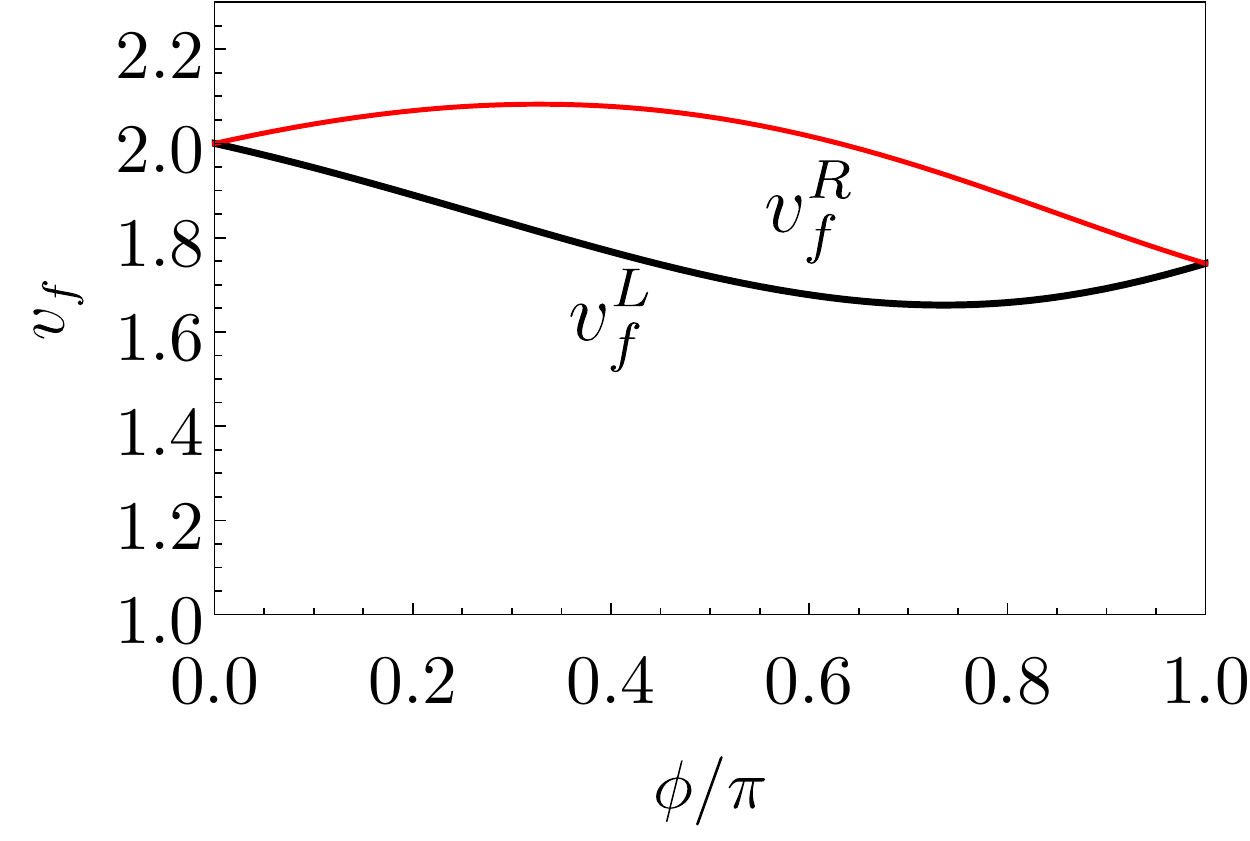}
	\includegraphics[width=0.23\columnwidth]{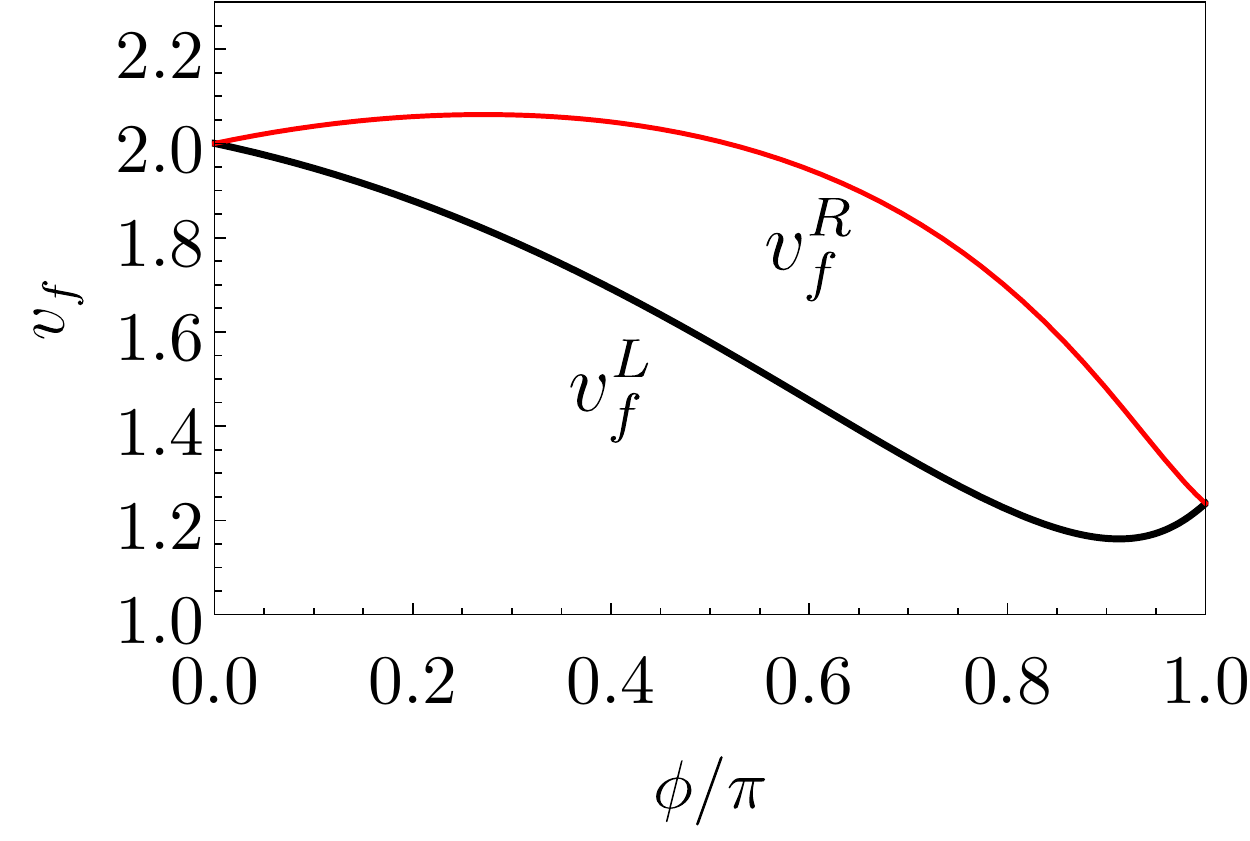}
	\includegraphics[width=0.23\columnwidth]{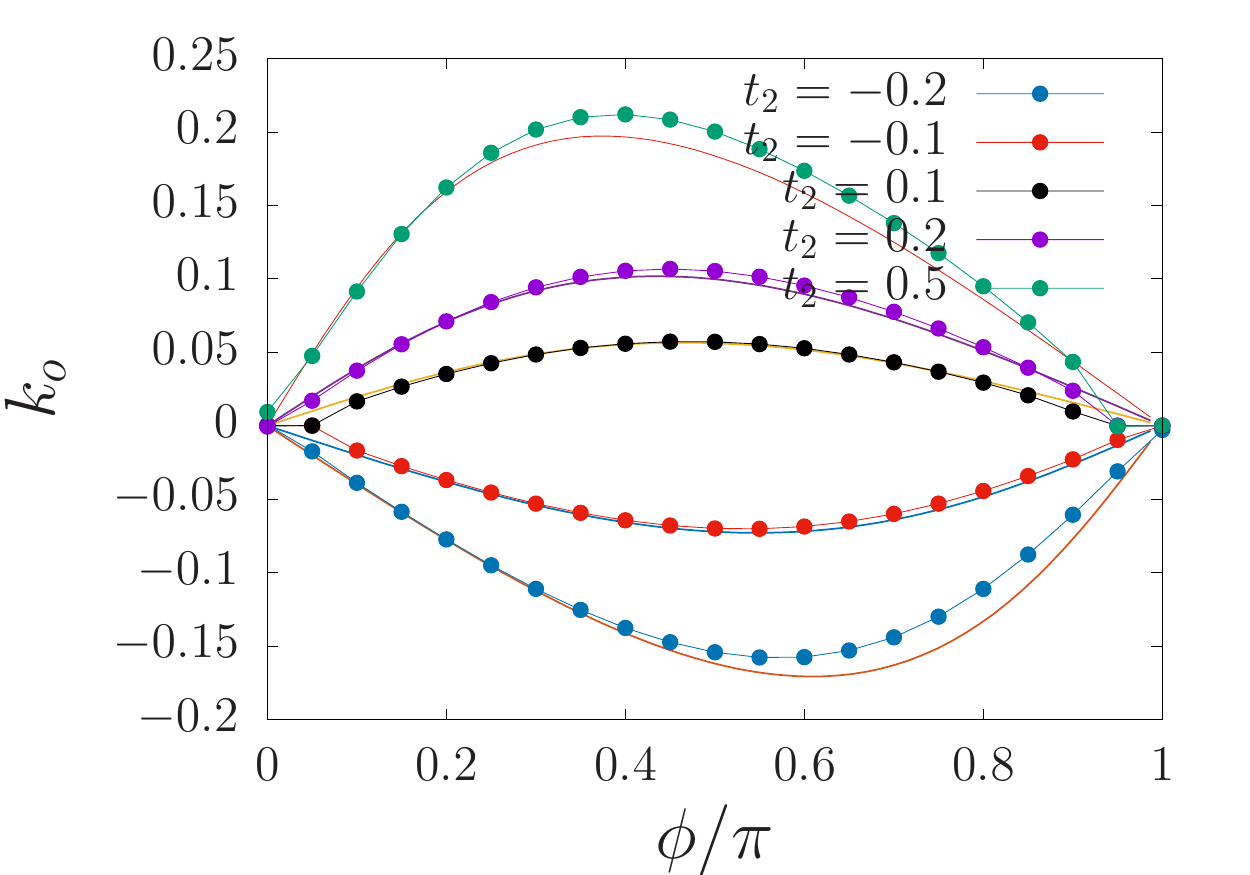}
	\caption{The right and left Fermi velocities ($v^R_f$ and $v^L_f$) as a function of $\phi$ for $t_2=-0.01$, $t_2=-0.1$  and $t_2=-0.3$. Notice that the right and left velocities are different.  The right most figure shows the comparison of shifted center of the Fermi sea $\equiv k_o$ from both HF and the DMRG study (see \eqn{fitcorr}) for various values of $t_2$ and as a function of $\phi$.}
	\label{diffvpar}
\end{figure}
\section{Details of the numerical DMRG calculations}
Here we provide various details of our DMRG calculations including the comparison with exact diagonalisation (ED) results for small systems.
\subsection{Determination of the phase diagram : Excitation gap, fidelity and central charge}
\subsubsection{Excitation gap}
$\Delta_L$ is the gap to single particle excitation defined as  
\beq
\Delta_L = E(L, N+1) + E(L,N-1) -2E(L,N)
\eeq
where $E(L, N)$ is the ground state energy for a system with $N$ fermions on $L$ sites. In the gapless regime $\Delta_L$ scales as $1/L$ (see \Fig{gapscaling}) and reaches $zero$ while in the gapped regime the value saturates to a finite value $\equiv \Delta_{\infty}$. Variation of $\Delta_{\infty}$ as a function  of $t_2$ for various values of $\phi$ is shown in \Fig{gapscaling} which is also the regime shown in Fig.~2 of the main text.
%%%%%%%%%%

\paragraph{\ul{BO phase and associated BKT transition} :}

\begin{figure}
	\centering
	\subfigure[]{
	\includegraphics[scale=0.425]{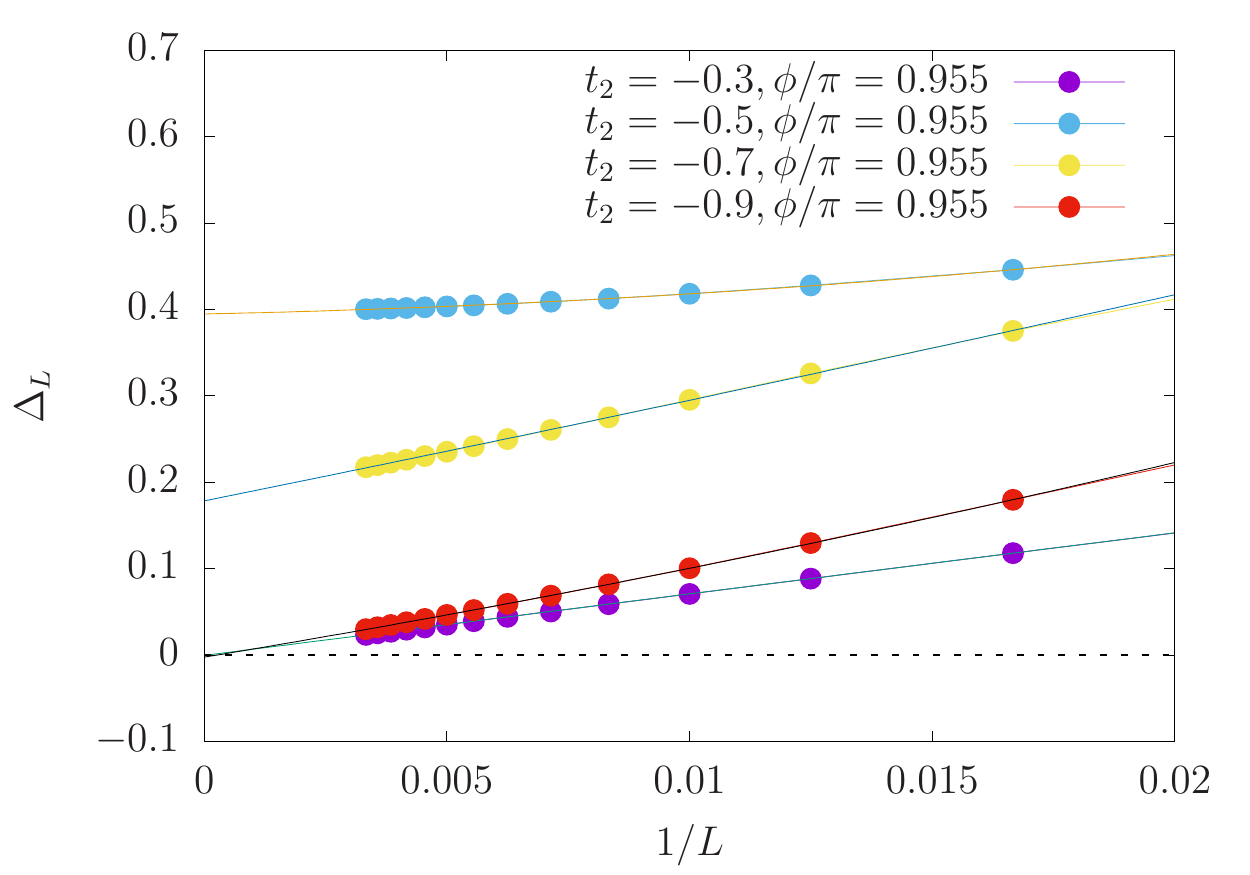}
	}
	\subfigure[]{
	\includegraphics[scale=0.425]{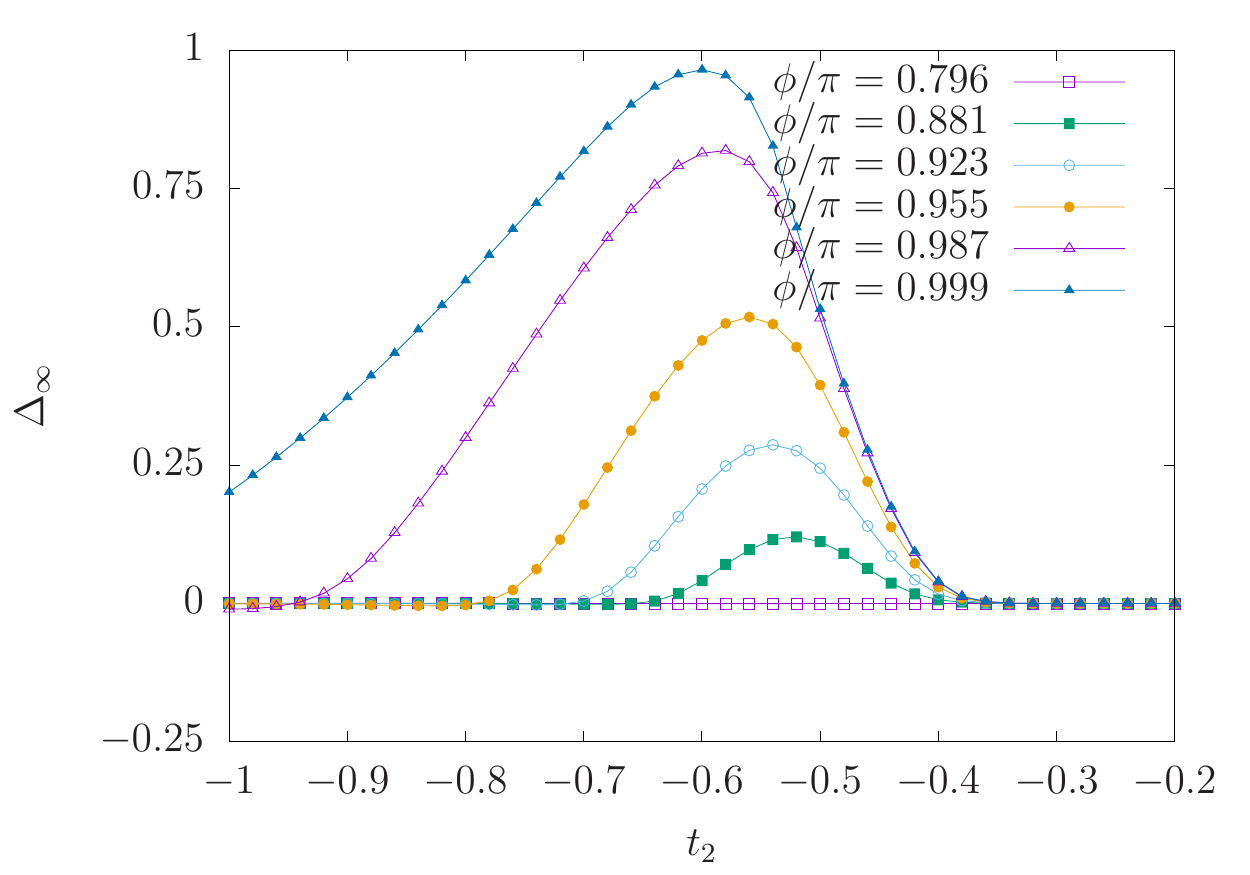} 
	}
	\subfigure[]{
	\includegraphics[scale=0.45]{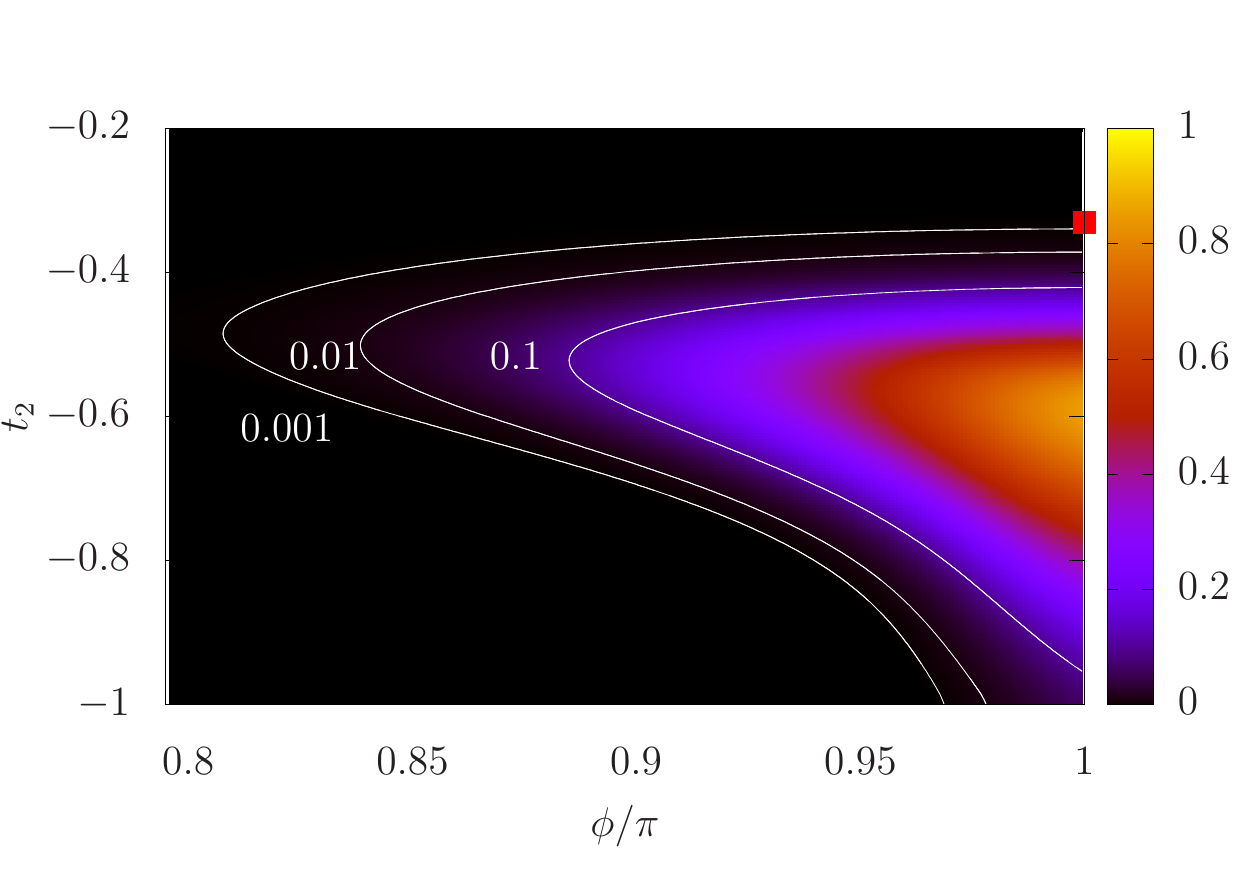} 
	}
	\caption{(a) Behavior of $\Delta_L$ at few parameters of $\{t_2, \phi\}$ as a function of $1/L$. For gapless phases $\Delta_L$ reaches $zero$ as a function of $1/L$ and saturates to a finite value ($\equiv \Delta_\infty$) for gapped regime. (b) Behavior of $\Delta_\infty$ in the gapped regime is shown as a function of $t_2$ for few values of $\phi$ and in the extended region (c).}
	\label{gapscaling}
\end{figure}

 The transition between the $c=1$ TLL and the BO phase is expected to be of BKT type (see main text). In order to pin the phase boundary, earlier works \cite{Dalmonte_PRB_2015, Carrasquilla_PRA_2013} have shown that behavior of correlation functions, fidelity etc.~may have significant errors near the transition. Instead, an alternate prescription of a data collapse of $\Delta_L$ with a scaling form of the correlation length provides a rather accurate estimate.  Briefly, the method entails that the variation of $\Delta_L L$ with $\log L - \frac{a}{\sqrt{V-V_c}}$ follows a universal behavior, where $V$ is a tuning parameter and $V_c$ is the critical value. Defining $x_L= \log L - \frac{a}{\sqrt{V-V_c}}$, values of $\Delta_L L$ and corresponding $x_L$s for various $L$s and $V$s near $V_c$ (in the gapped regime) can be fitted to a curve. Treating $a$ and $V_c$ as variational parameters, the least square fitting error is minimized to optimize $a,V_c$.  A representative estimation of the same is shown in \Fig{gapBKTScaling}. This procedure is used to for determining the gapless-gapped transition boundary as shown in the Fig.~1 of the main text. 

\begin{figure}
	\subfigure[]{
	\includegraphics[width=0.24\columnwidth]{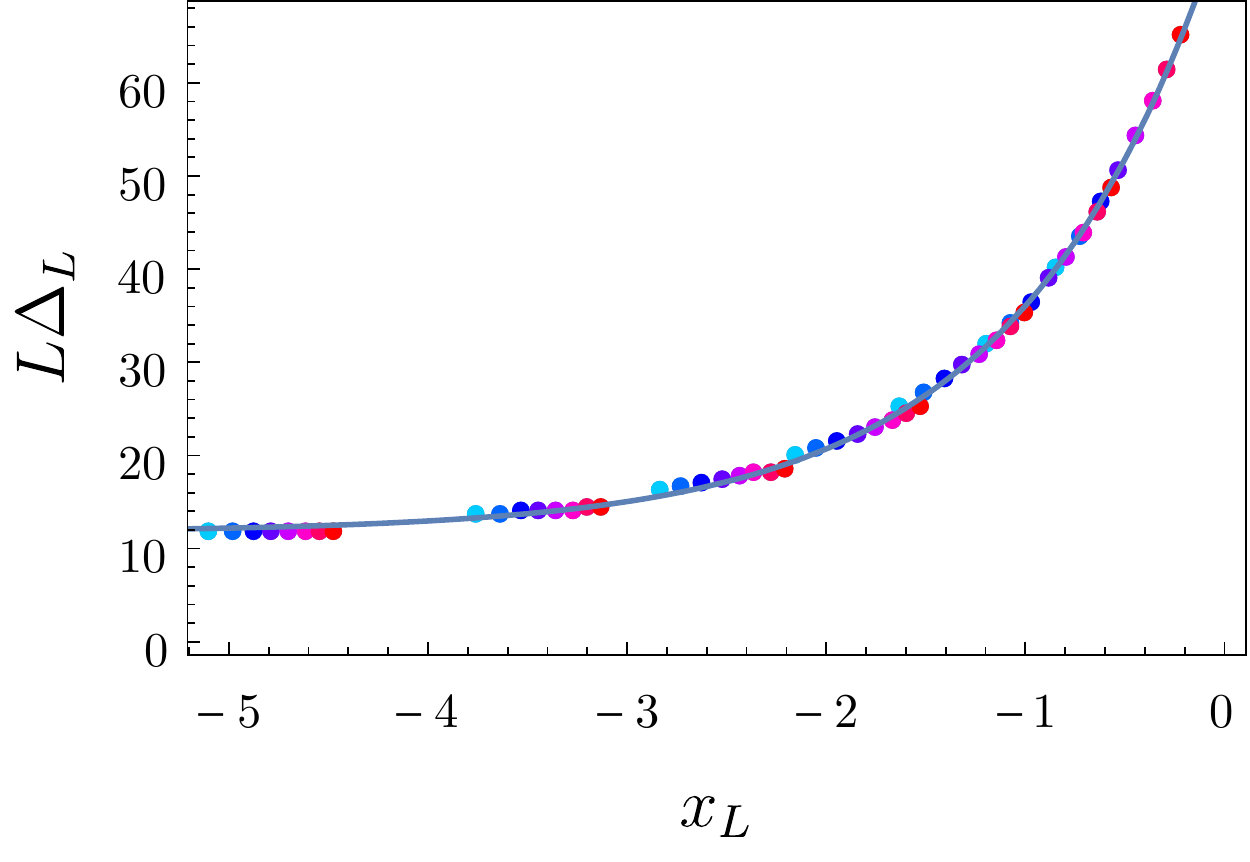}
	}
	\subfigure[]{
	\includegraphics[width=0.18\columnwidth]{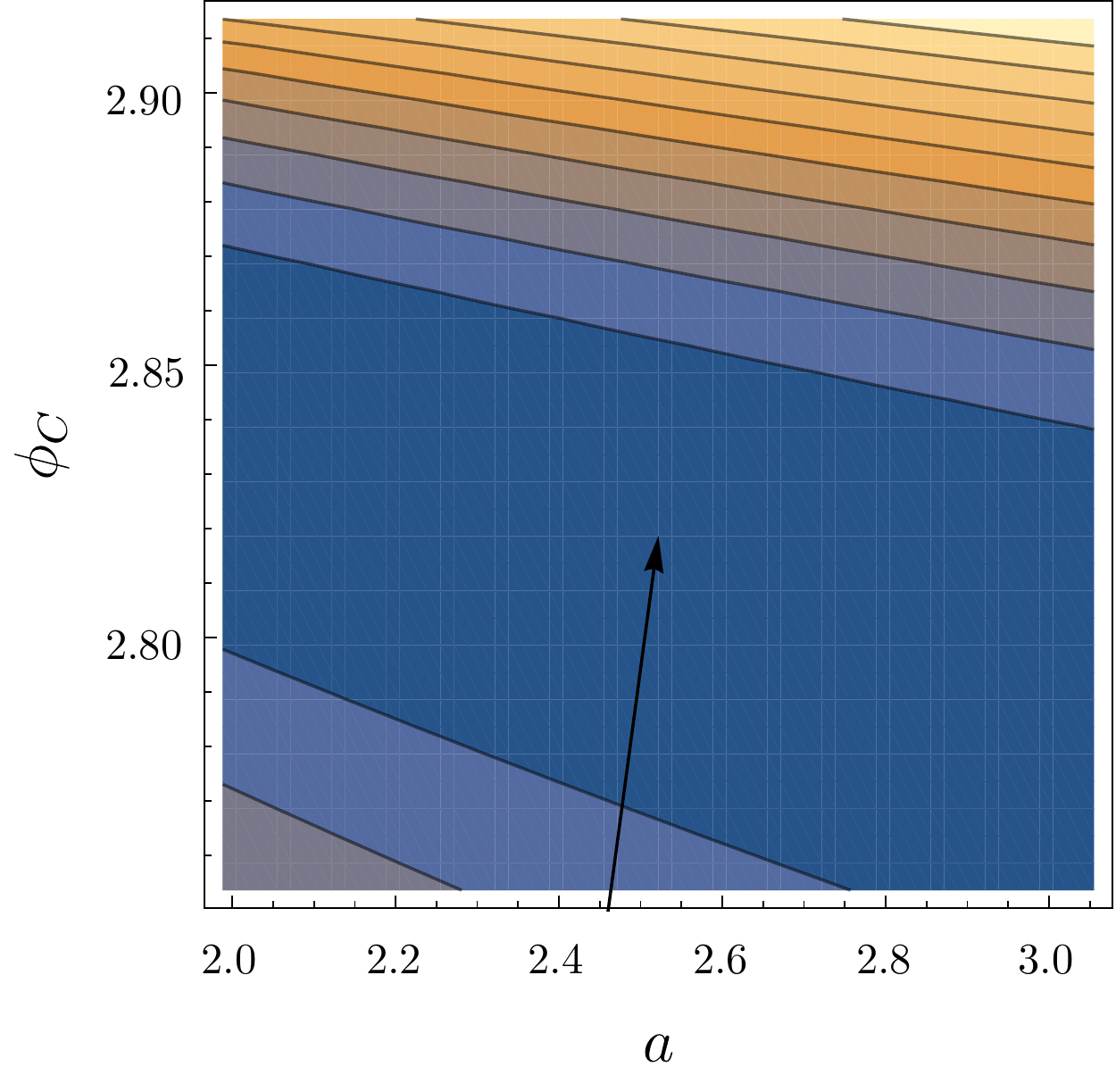} 
	}
	\subfigure[]{
	\includegraphics[width=0.24\columnwidth]{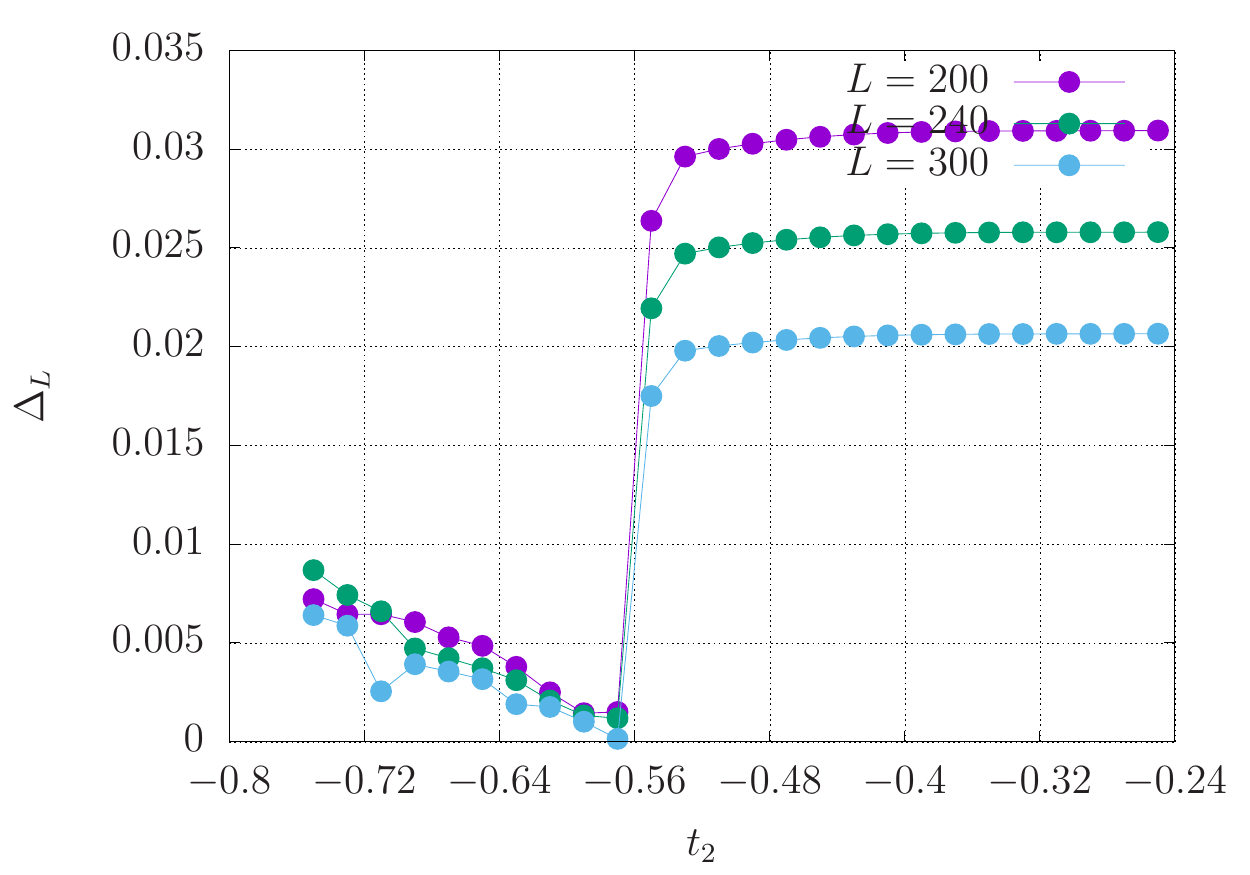}
	}
	\subfigure[]{
	\includegraphics[width=0.24\columnwidth]{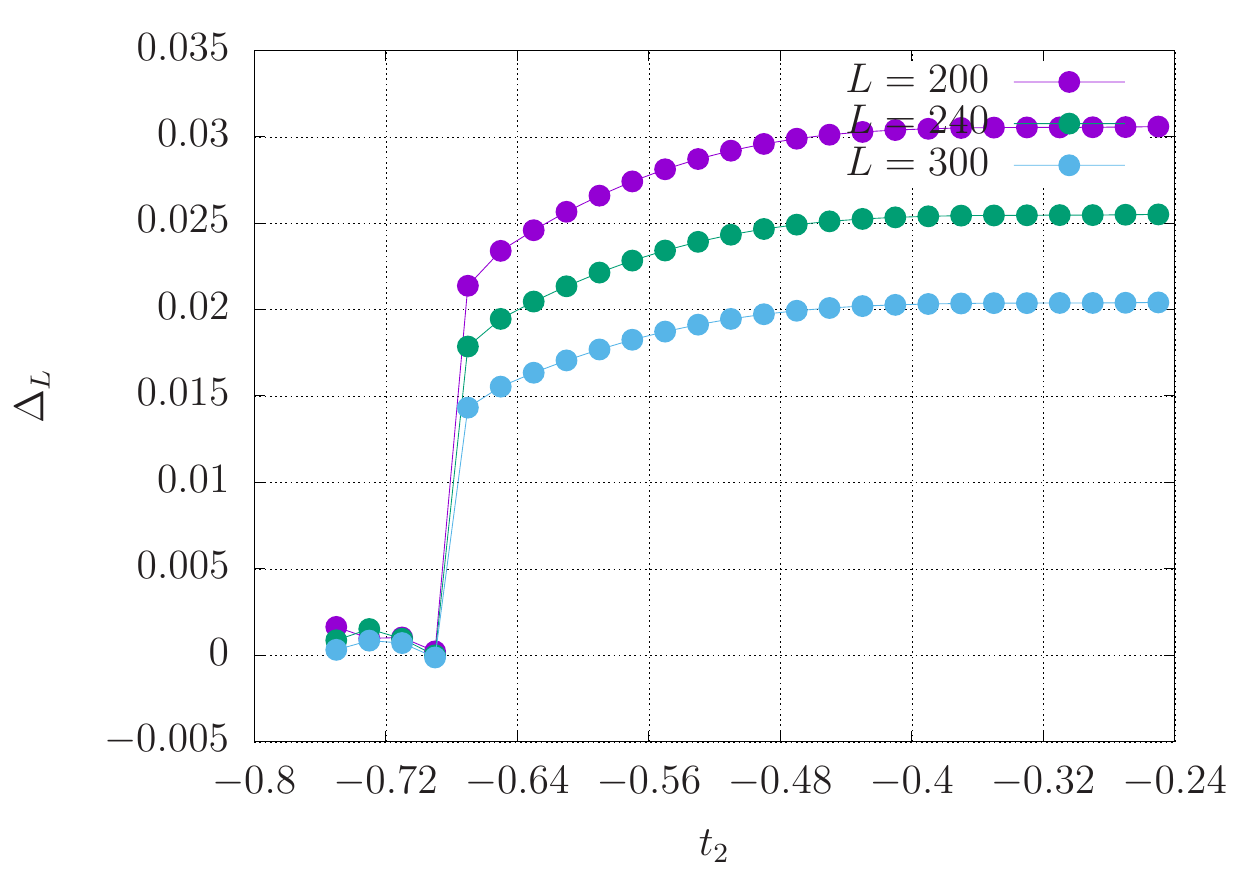} 
	}
	\caption{(a-b) Best obtained collapse of $\Delta_L L$ and $x_L = \log L - \frac{a}{\sqrt{\phi-\phi_c}}$ by variation of $a, \phi_c$. The estimation of $a, \phi_c$ is done by minimizing the fitting error. Here, this is done for the transition from gapless to gapped point as a function of $\phi$ for $t_2=-0.70$. (c-d) Behavior of $\Delta_L$ as a function of $t_2$ across the gapless-gapless transition for $\phi/\pi=0.28$ and $\phi/\pi=0.42$ for three different values of $L=200,240,300$. Notice that unlike the gapless-gapped transition, for any finite $L$, $\Delta_L$ shows a jump to $zero$ at the Lifshitz transition.}
	\label{gapBKTScaling}
\end{figure}

\paragraph{\ul{Behavior of excitation gap in the TLL phases and near the metal-metal transition} :}
Behavior of $\Delta_L$ as a function of $t_2$ across the gapless-gapless transition for $\phi/\pi=0.28$ and $\phi/\pi=0.42$ for three different values of $L=200,240,300$ is shown in \Fig{gapBKTScaling}. Note the rather abrupt jump in $\Delta_L$ to $zero$ unlike the gapless-gapped transition. This jump points out the boundary of the transition between the $c=1$ and $c=2$ phase.
\subsubsection{Fidelity}
Given two groundstate wavefunctions evaluated at parameters $\lambda$ and $\lambda + d\lambda$, the fidelity susceptibility is given by \cite{Gu_IJMB_2010} 
\beq
\chi_F (\lambda) = \lim_{d\lambda \rightarrow 0} \frac{-2\log(|\langle \psi(\lambda + d\lambda)| \psi(\lambda) \rangle |)}{(d\lambda)^2}.
\eeq
Signatures in $\chi_F$ signal phase transitions. \Fig{fidebondorder} shows the behavior of $\chi_F$ while going from gapless to gapped regime (a-b) and between gapless to gapless regime (c-d) for different system sizes. Note that $\chi_F$ behaves rather differently at the two kinds of transition.  
In order to corroborate our DMRG results we compare the results for small system sizes with exact diagonalization (ED) studies. Some representative figures are shown in \Fig{SMcomp}.
\begin{figure}
	\centering
	\subfigure[]{
	\includegraphics[scale=0.325]{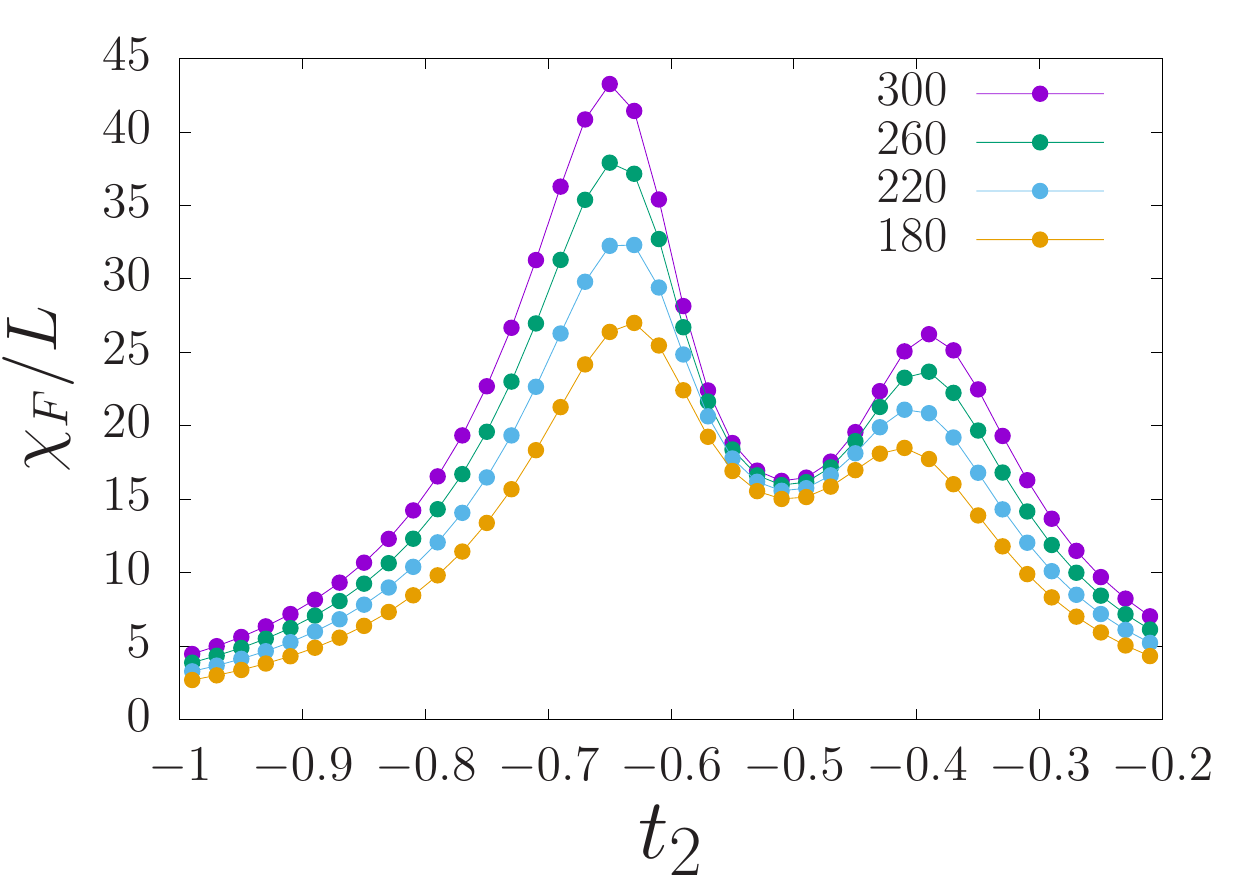} 
	}
	\subfigure[]{
	\includegraphics[scale=0.325]{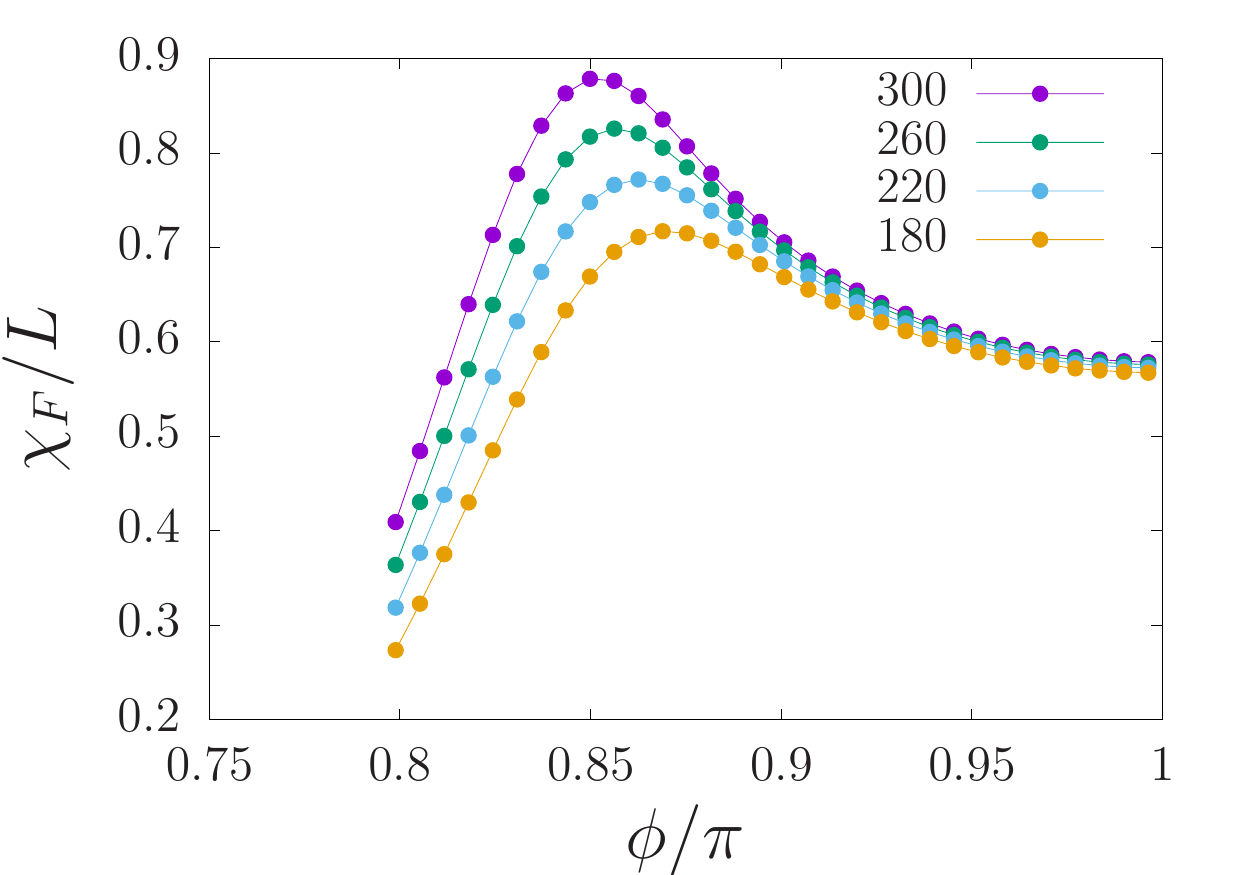}
	}
	\subfigure[]{
	\includegraphics[scale=0.325]{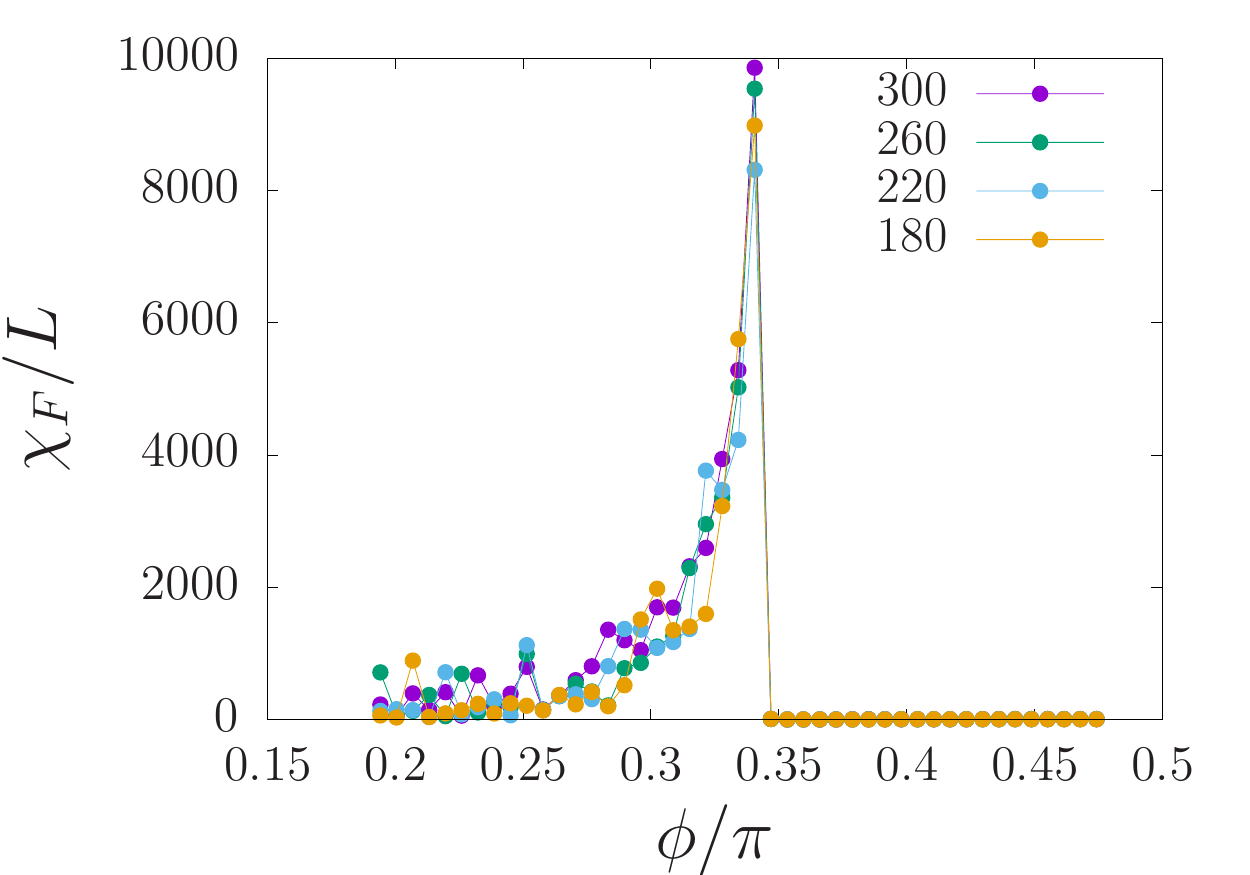} 
	}
	\subfigure[]{
	\includegraphics[scale=0.325]{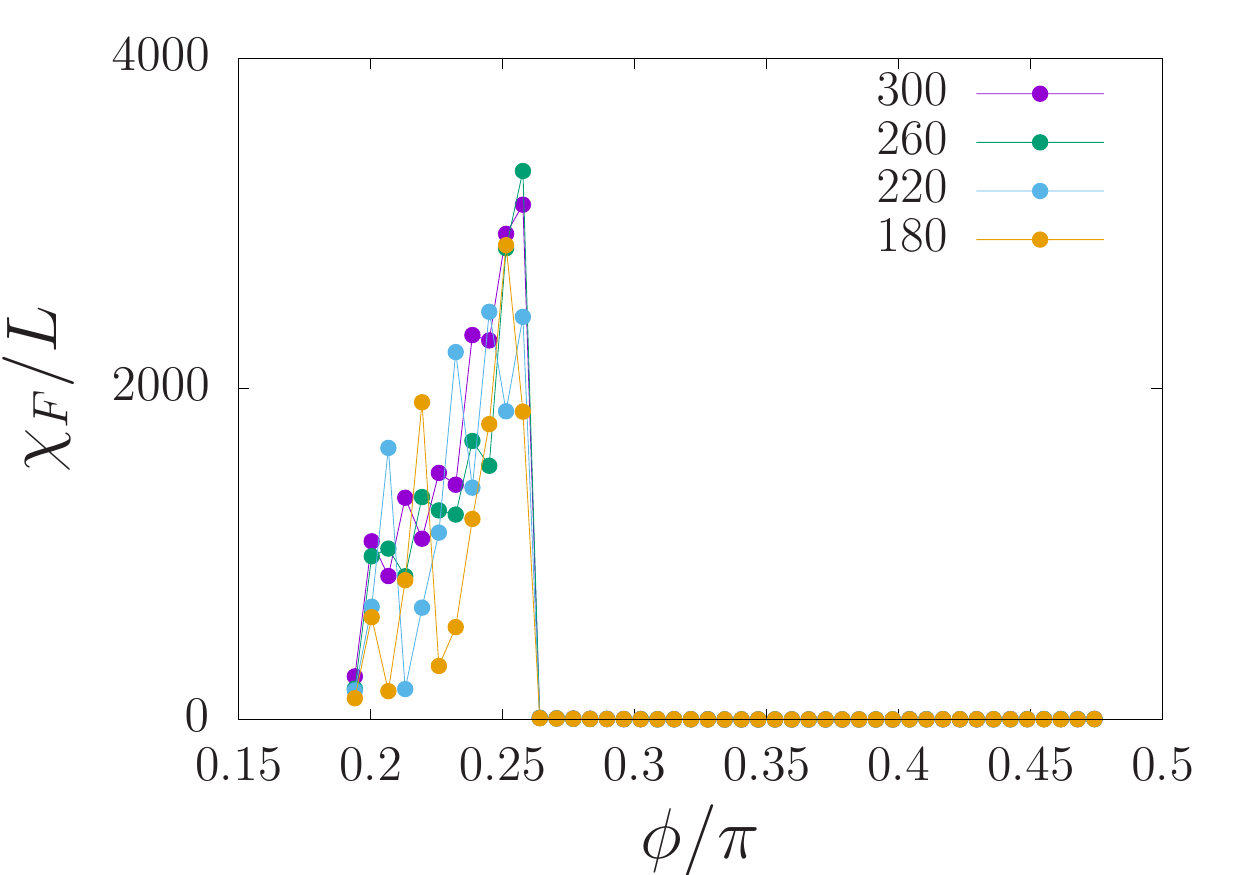}
	}
	\caption{Fidelity susceptibility $\chi_F$ : (a) as a function of $t_2$ at $\frac{\phi}{\pi}=0.8732$ and (b-d) as a function of $\phi/\pi$ for $t_2=-0.46, 1.00$ and $0.7$. The humps signal phase transitions. }
	\label{fidebondorder}
\end{figure}

\begin{figure}
	\centering
	\includegraphics[width=0.3\columnwidth]{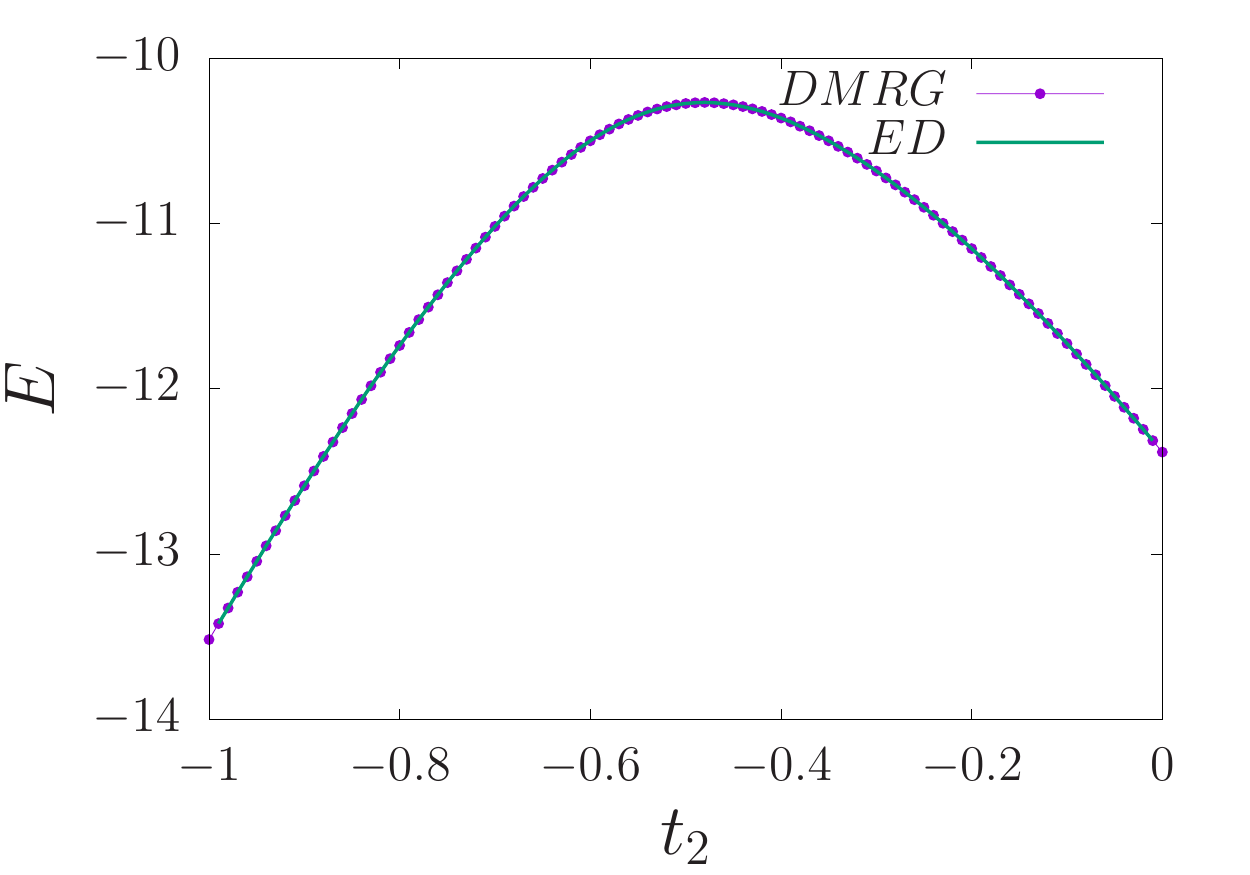} 
	\includegraphics[width=0.3\columnwidth]{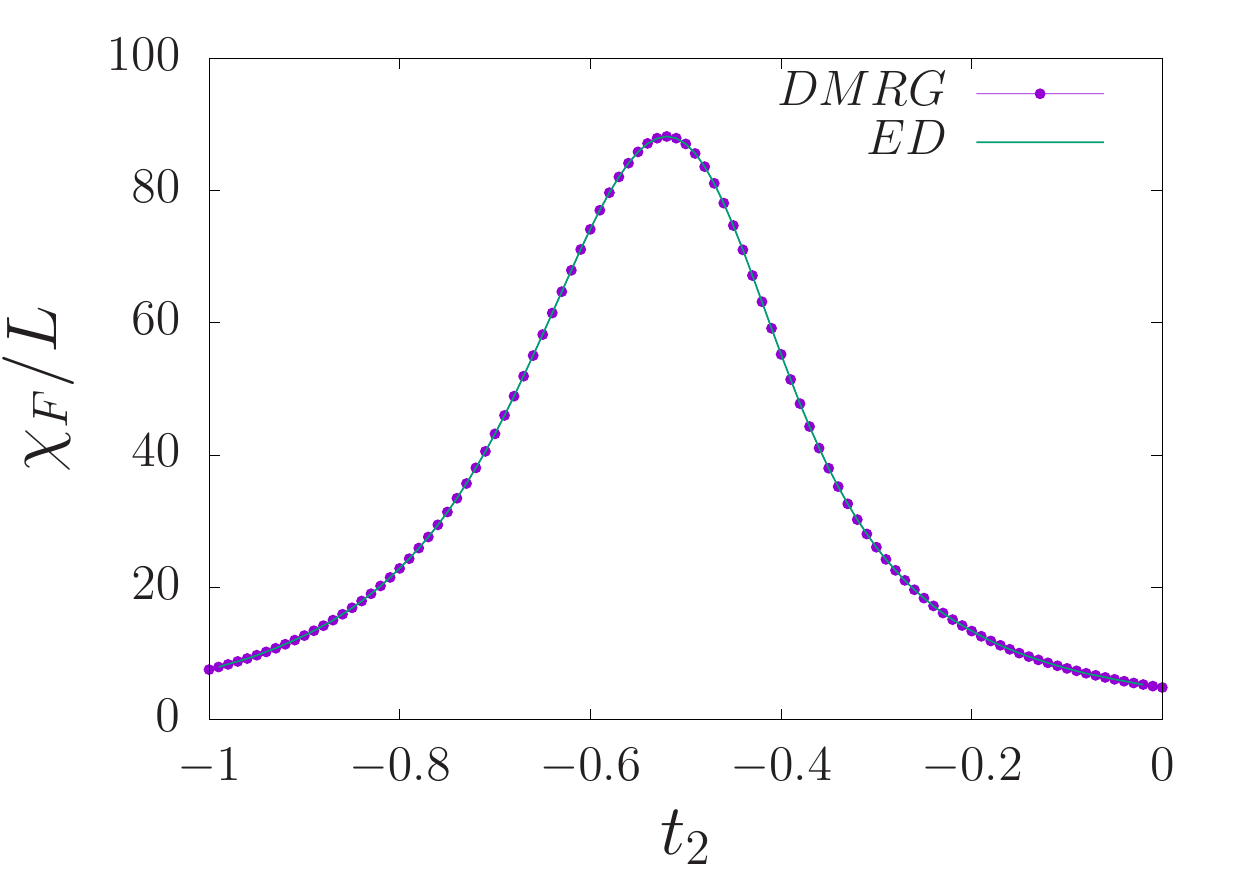}
	\caption{Ground state energy and fidelity susceptibility $\chi_F/L$ as a function of $t_2$  for $\phi/\pi=0.859$ for a system size $L=20$ under open-boundary condition, both using exact diagonalization and DMRG.}
	\label{SMcomp}
\end{figure}

\subsubsection{Entanglement Entropy and Central Charge}

In DMRG we work with open-boundary condition -- here it is known from Cardy-Calabrese formula, that entanglement entropy of a subsystem size $l$ as a function of $l/L$ can be fitted to the following form \cite{Calabrese_JPA_2009}
\beq
S(l) = \frac{c}{6} \ln [\frac{L}{\pi} \sin(\frac{\pi l}{L}) ] + \ldots
\eeq
to estimate $c$, the central charge. %\Fig{Entangle} shows fitting procedure for some parameter values. 

\paragraph{\ul{Central charge at the BKT transition} :} Behavior of central charge across the BKT transition is shown in \Fig{Entangle} for few parameters where central charge changes from $1$ to $zero$.

\begin{figure}
	\centering
	\subfigure[]{
	\includegraphics[scale=0.4]{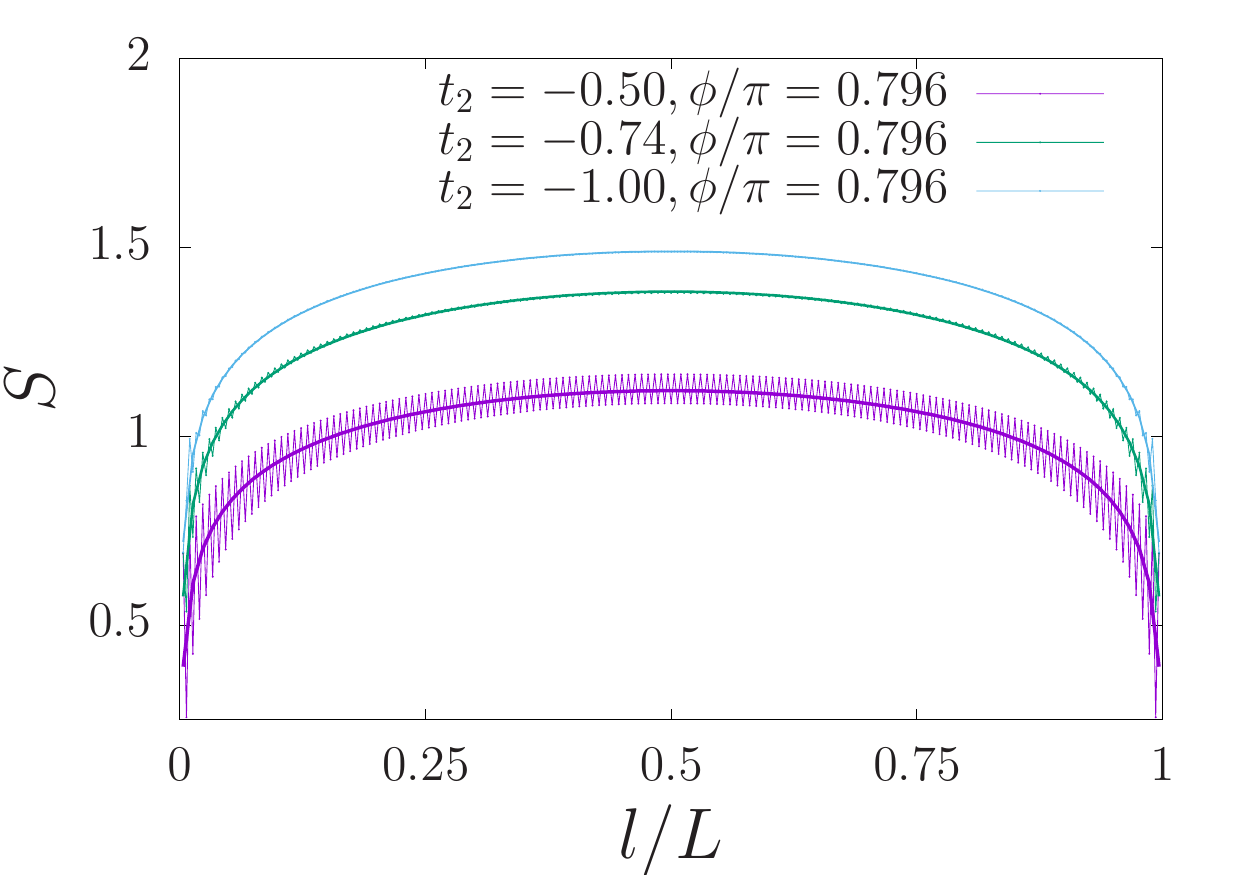} 
	}
	\subfigure[]{
	\includegraphics[scale=0.4]{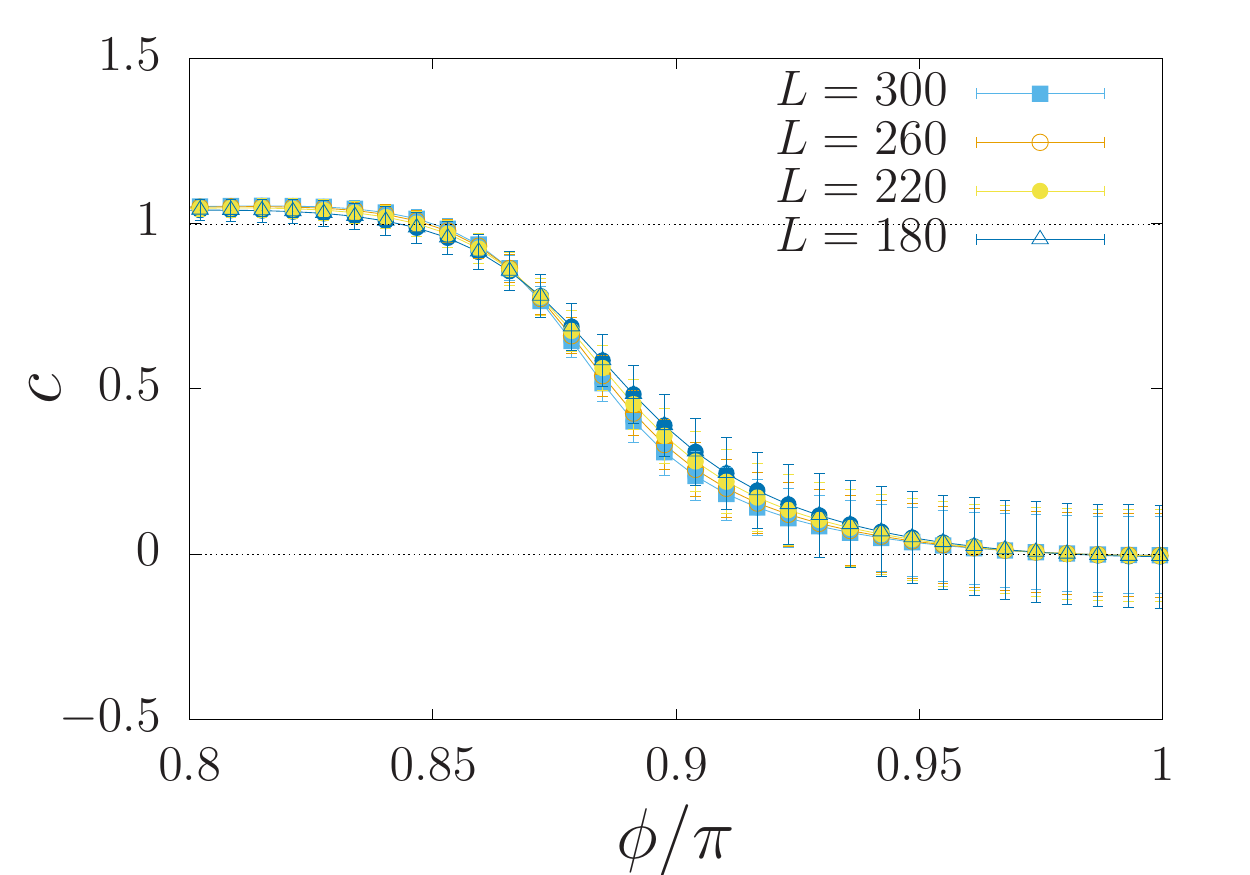}
	}
	\subfigure[]{
	\includegraphics[scale=0.4]{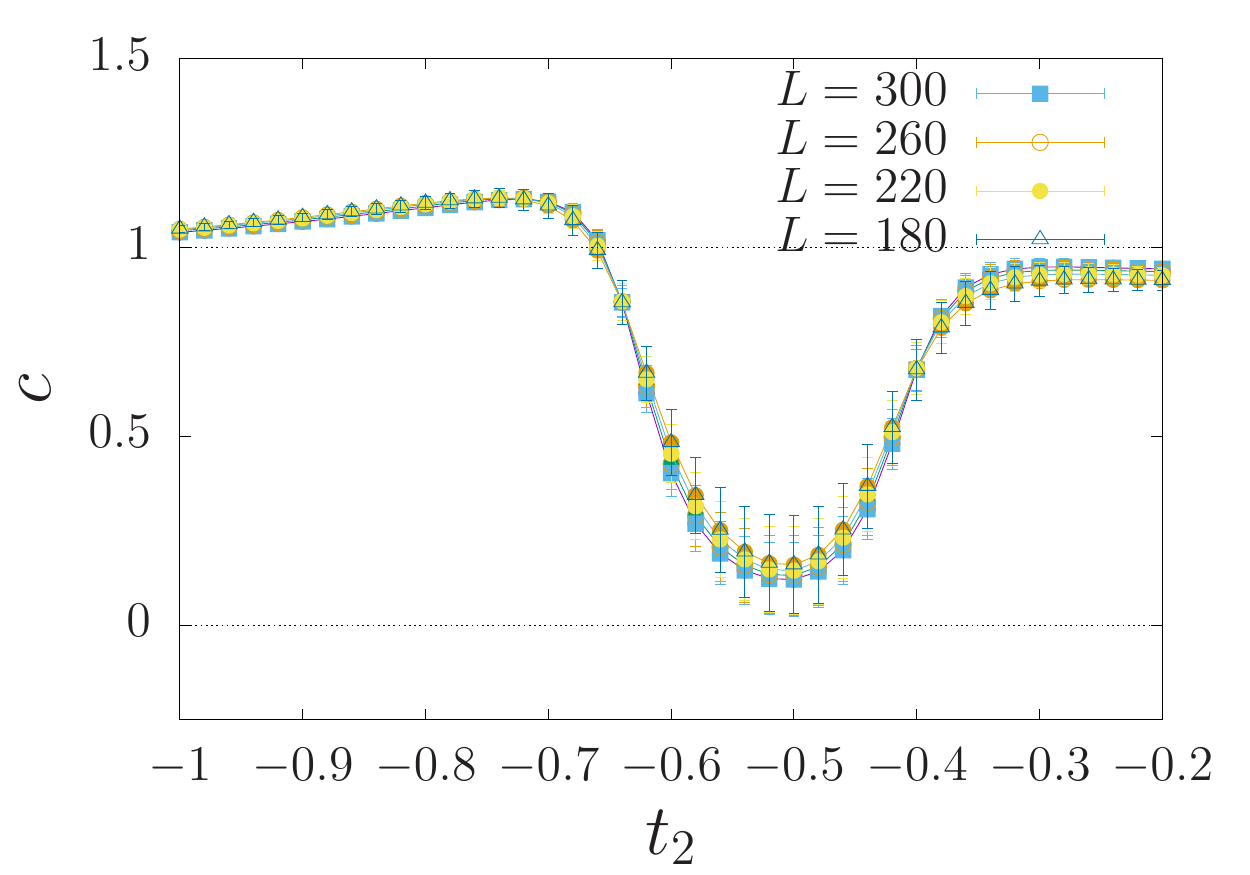}
	}
	\caption{(a) Entanglement entropy for three different parameters of $t_2,\phi$ as a function of bond length $l/L$ for $L=300$. The central charge($c$) is estimated using the Cardy-Calabrese formula.  Variation of $c$ for four different system sizes as a function of $\phi$ for $t_2=-0.60$ (b) and as a function of $t_2$ for $\phi/\pi=0.891$ $(c)$.} 
	\label{Entangle}
\end{figure}

\paragraph{\ul{Central charge at the gapless-gapless transition} :} Behavior of central charge transition at gapless-gapless transitions are shown in \Fig{centralgapless}. Note the abrupt change in $c$ in contrast to the smooth variation in $c$ for gapless-gapped transition.

\begin{figure}
	\centering
	\subfigure[]{
	\includegraphics[width=0.23\columnwidth]{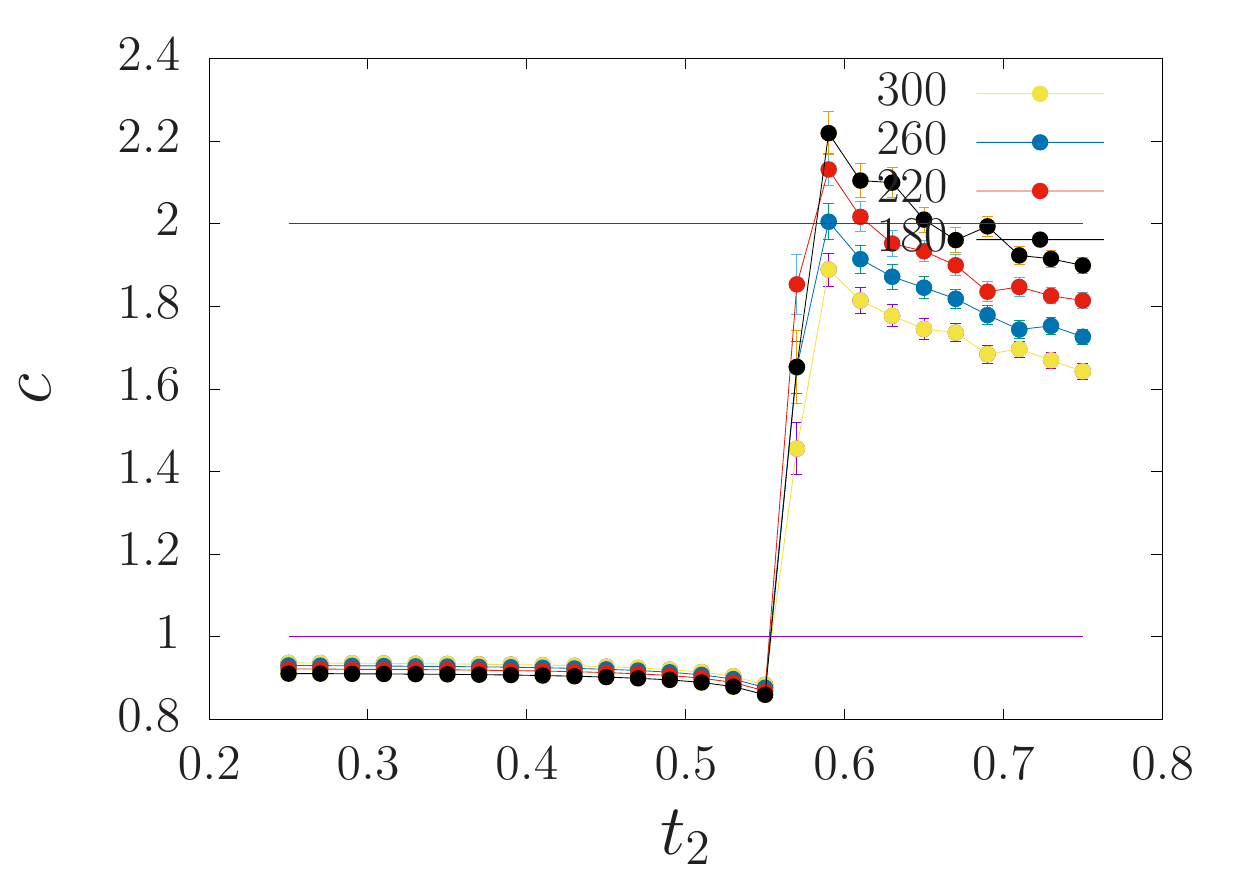}
	}
	\subfigure[]{
	\includegraphics[width=0.23\columnwidth]{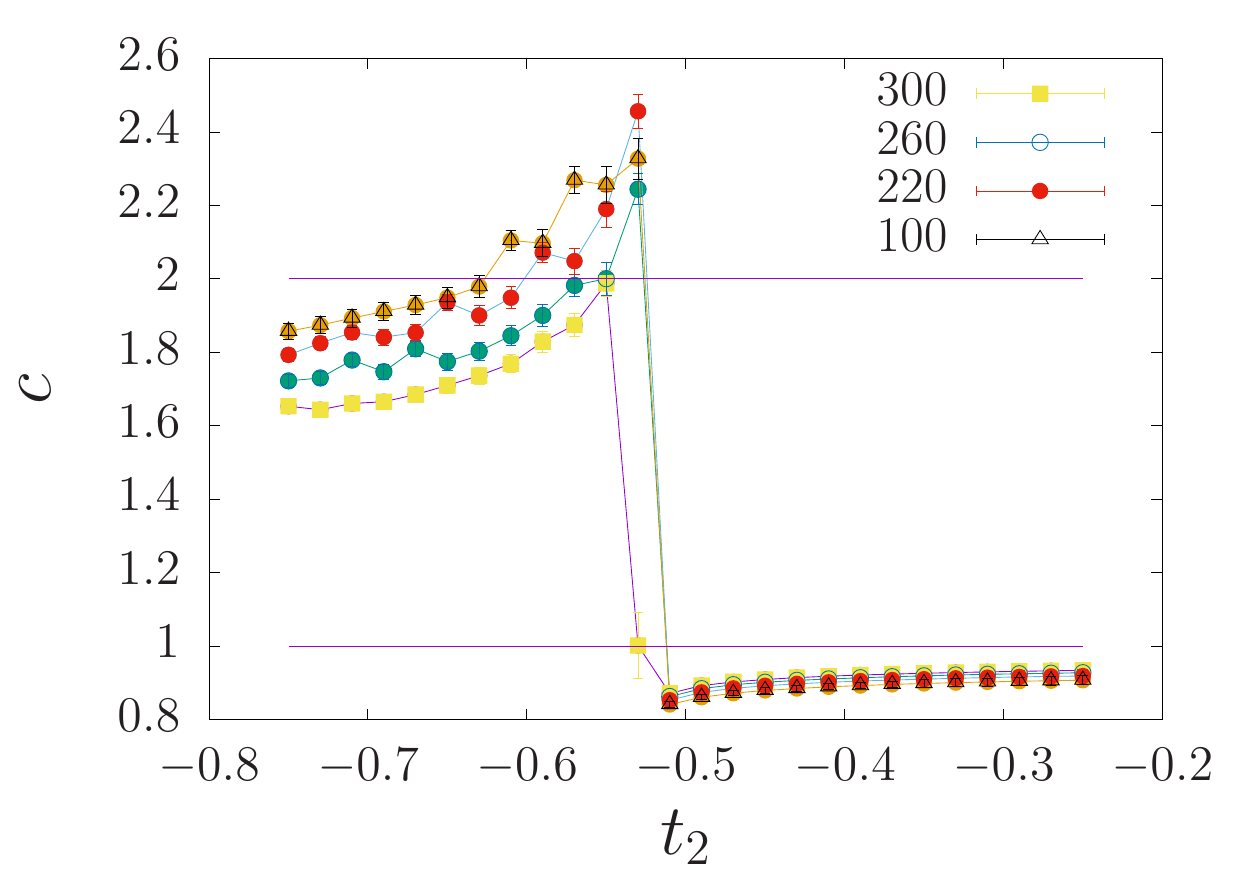}
	}
	\subfigure[]{
	\includegraphics[width=0.23\columnwidth]{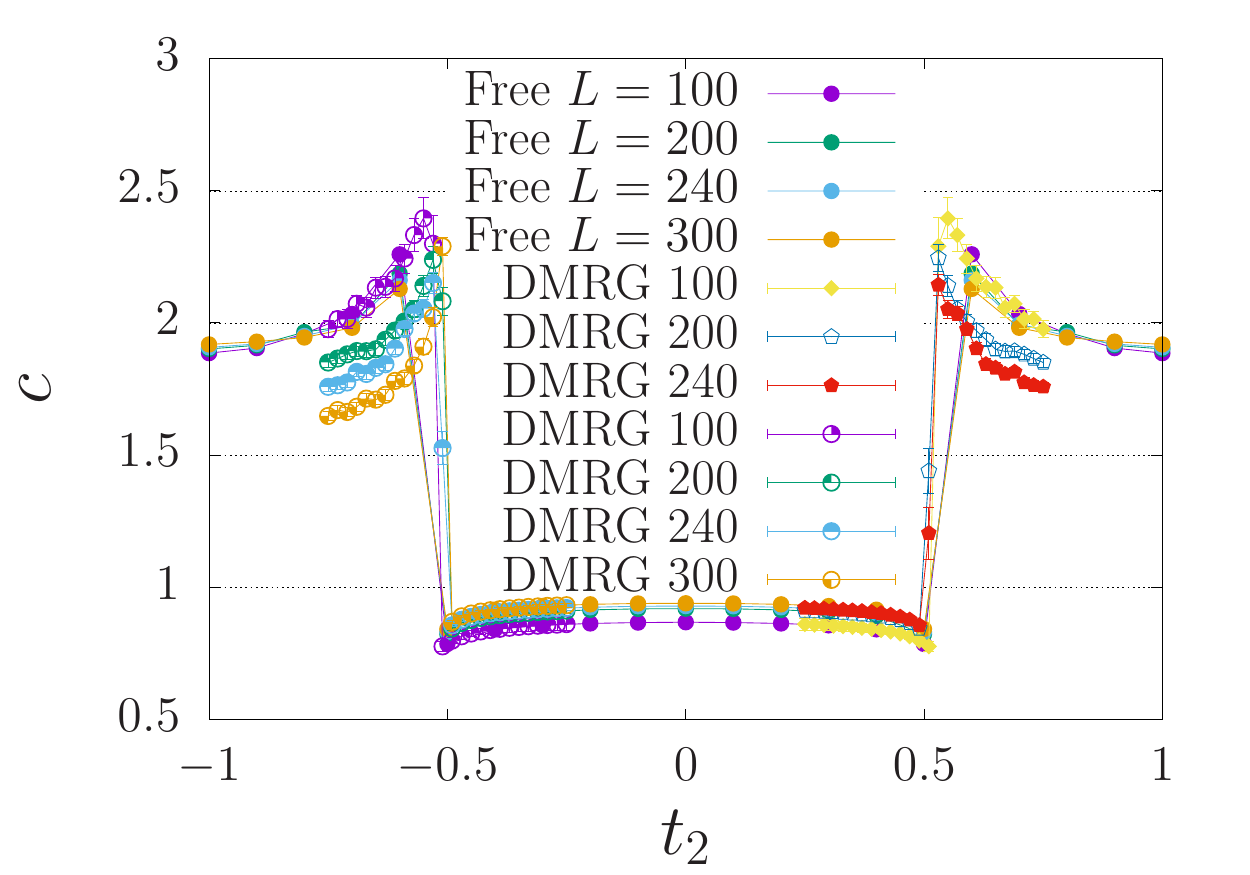} 
	}
	\subfigure[]{
	\includegraphics[width=0.23\columnwidth]{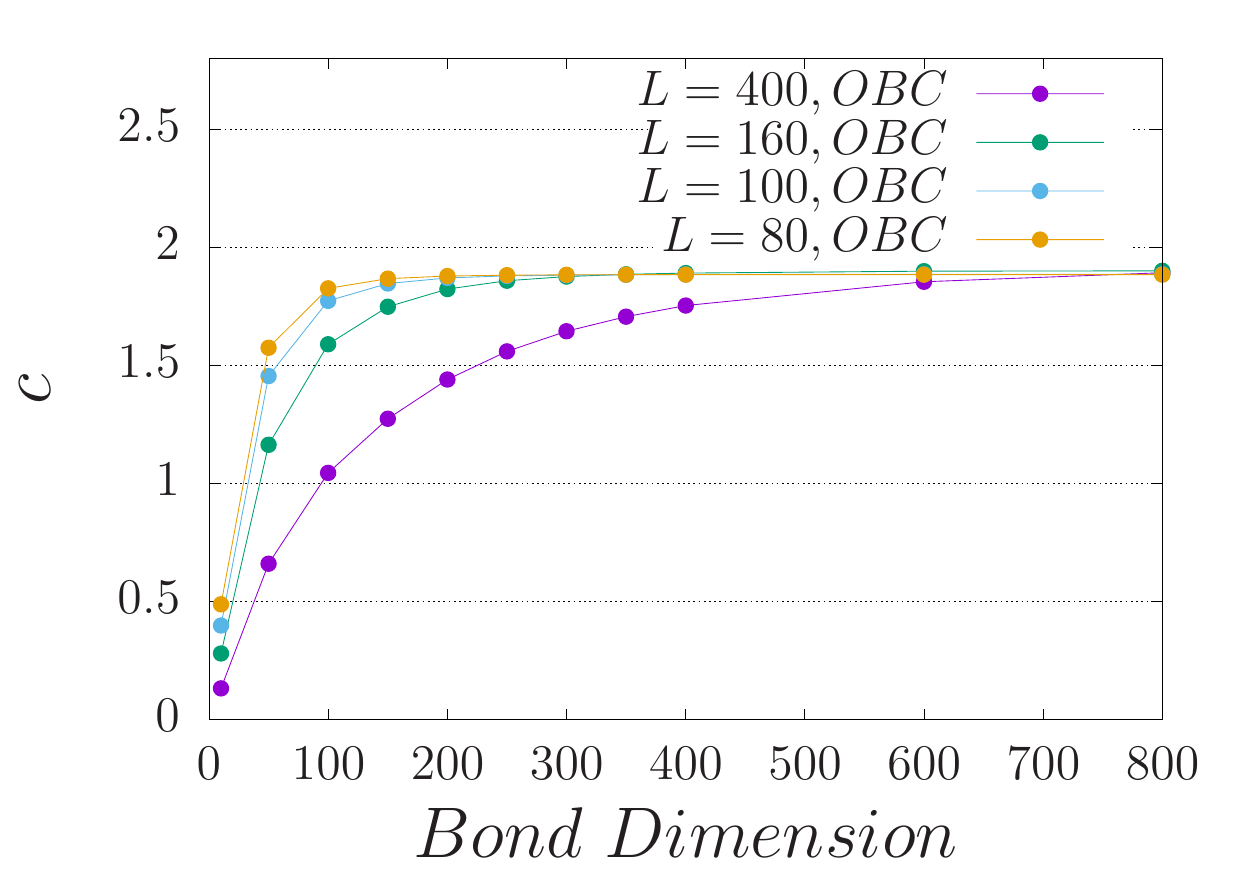}
	}
	\caption{The variation of central charge $c$ as a function of $t_2$ for $\phi/\pi =0.159$ (a) and $\phi/\pi=0.18$ (b) showing a gapless-gapless transition between the $c=1$ and $c=2$ phases. (c-d) Comparison of $c$ evaluated using DMRG and correlation matrix at the free fermionic limit ($\phi$=0) as a function of $t_2$ for different system sizes.  (d) shows dependence of $c$ evaluated at $t_2=-1, \phi=0$ for different system sizes and bond-dimension. Even at the free fermion limit, to capture the $c=2$ phase one needs to have a significantly large bond-dimension.}
	\label{centralgapless}
\end{figure}

The behavior of $c$ near $\phi=0$ limit can be understood by comparing the  results of DMRG with that obtained by calculating entanglement entropy using correlation matrix \cite{Chung_PRB_2001, Cheong_PRB_2004, Peschel_JPA_2003,Vidal_PRL_2003} ($C_{ij} = \langle c^\dagger_i c_j \rangle$ where the expectation is taken over the occupied states). The value of central charge calculated using this is shown in \Fig{centralgapless}.

\subsection{Characterisation of the phases}

\subsubsection{Fermionic occupation number in momentum space}

The fermion mode occupancy $\langle n(k)\rangle=\langle c_k^\dagger c_k\rangle$ is shown in \Fig{nofk}. While the similarity with the HF results is quite striking, we note that there is no jump discontinuity for the fermion occupation as this is a TLL. $\langle n(k) \rangle$ allows us to calculate ``moment of intertia" of the Fermi sea given by
 \beq
 I = \int dk \langle n(k) \rangle (\sin(k))^2
 \eeq
 the variation of which as a function of $t_2$ for different values of $\phi$ is shown in  \Fig{momentInertia}. While the gapless to gapless transition is characterized by a change in $I$, the gapless to gapped transition shows no such variation. 

\begin{figure}
	\includegraphics[width=0.245\columnwidth]{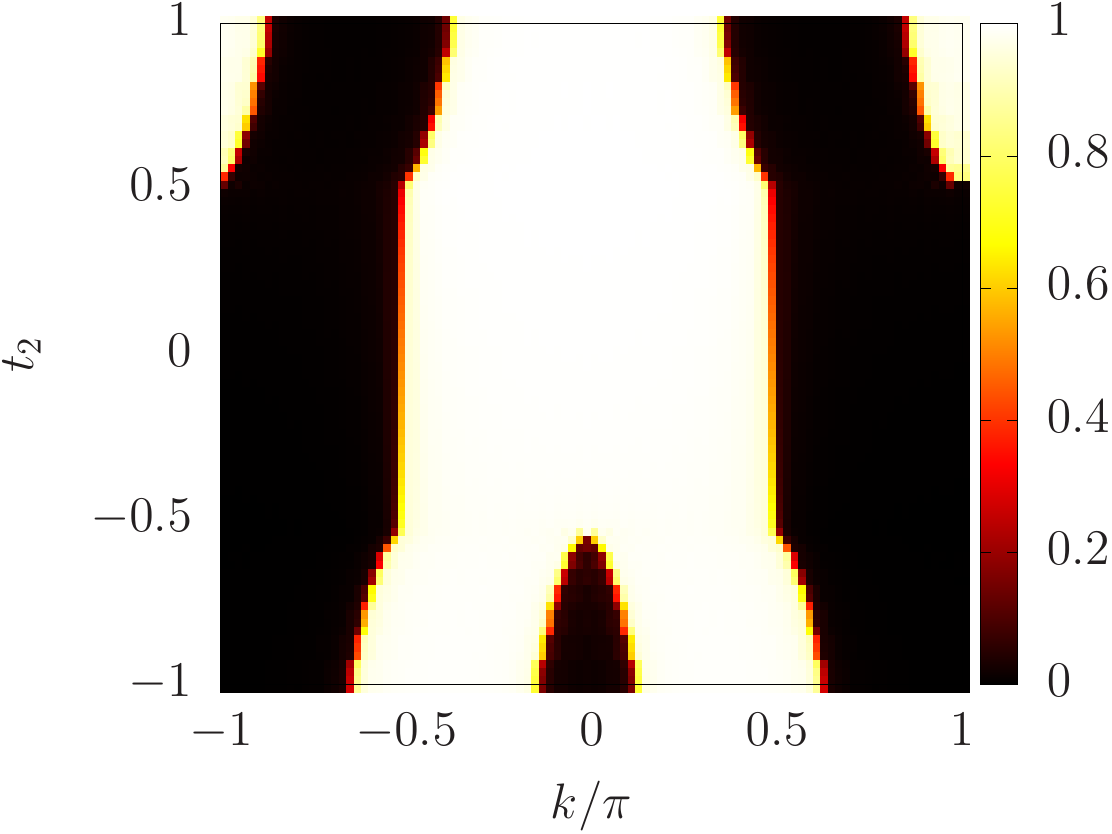}
	\includegraphics[width=0.245\columnwidth]{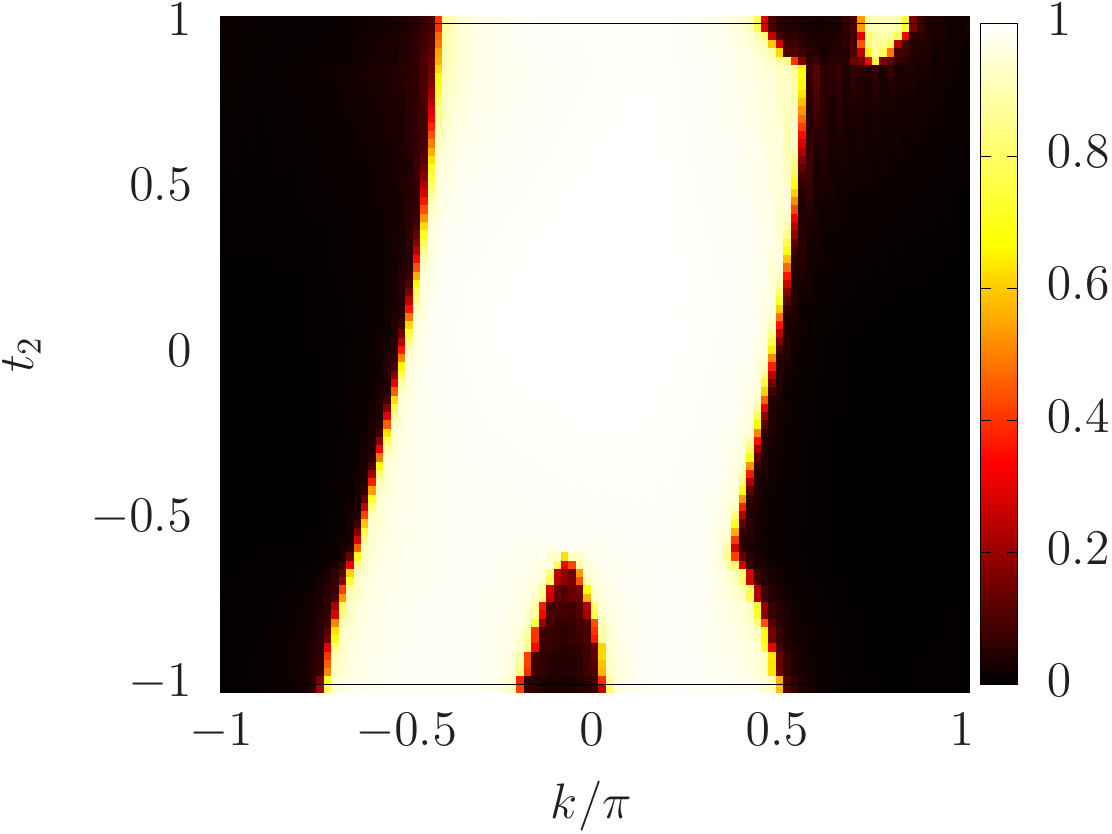}
	\includegraphics[width=0.245\columnwidth]{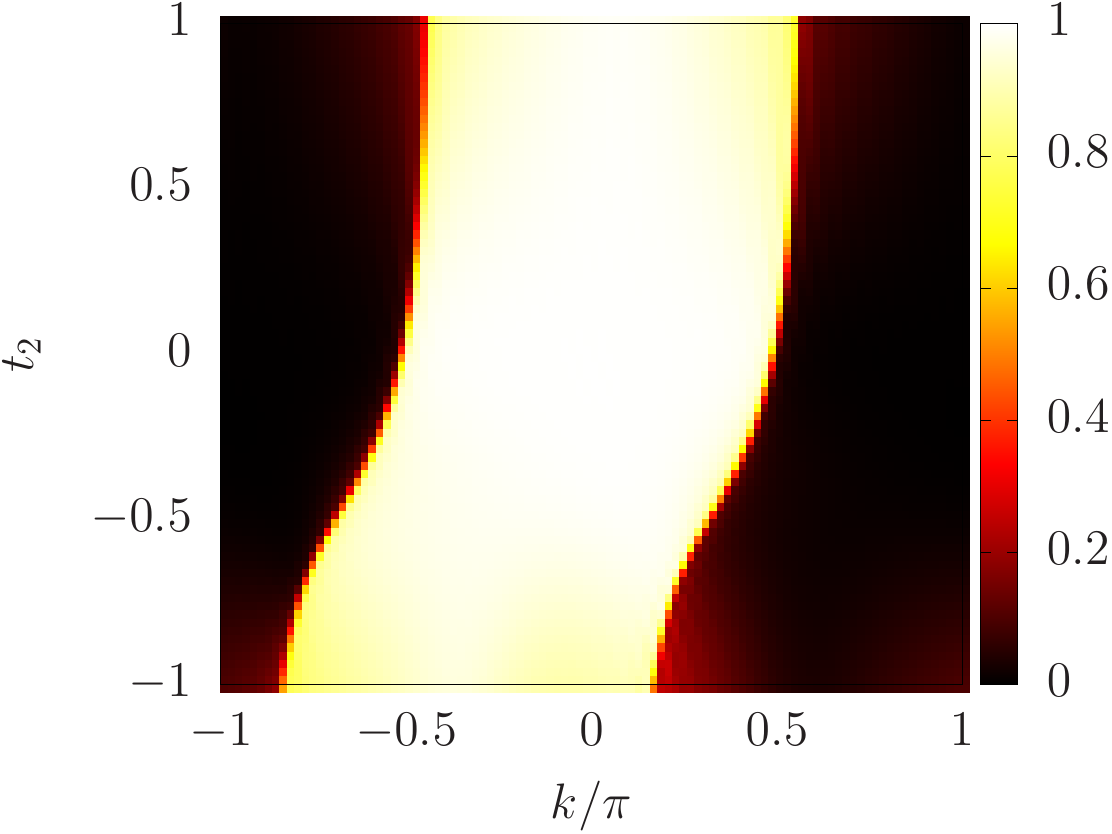}
	\includegraphics[width=0.245\columnwidth]{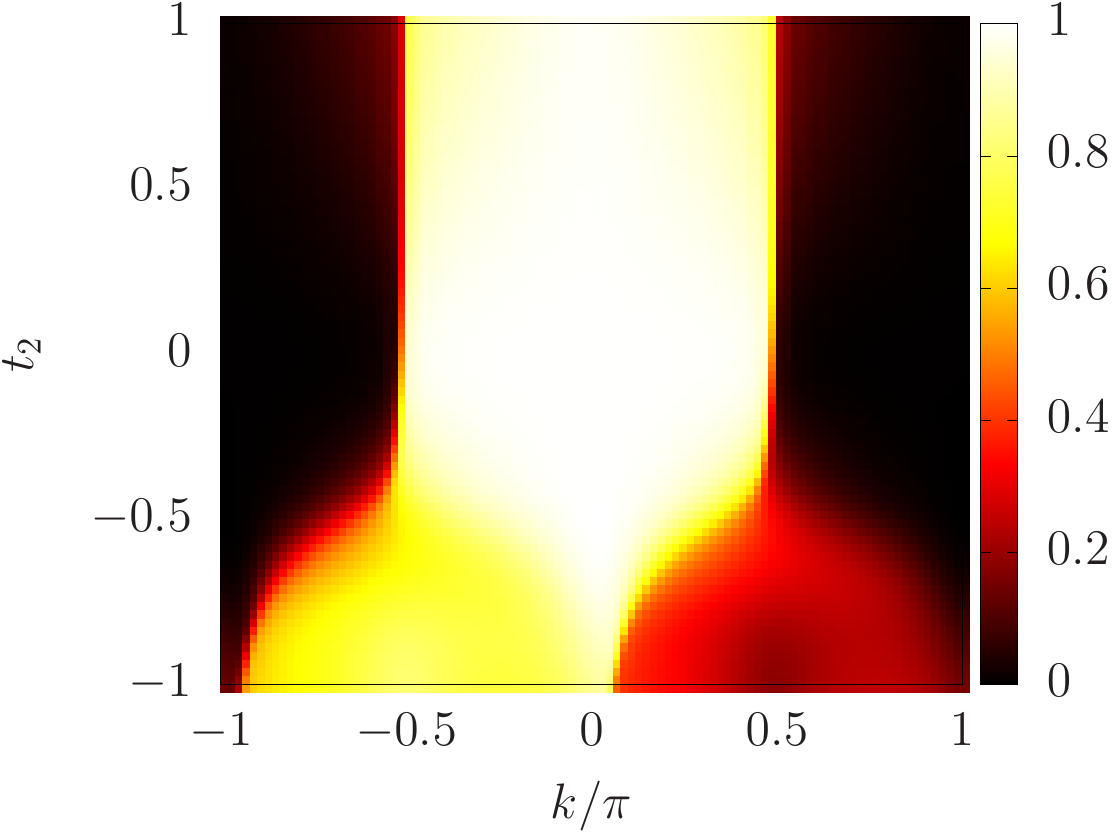}
	\caption{$\langle n(k) \rangle $ as a function of $k$ and $t_2$ for different values of $\phi = 0.0, 1.0, 2.0, 3.0$ via DMRG calculations.}
	\label{nofk}
\end{figure}

\begin{figure}
	\includegraphics[width=0.35\columnwidth]{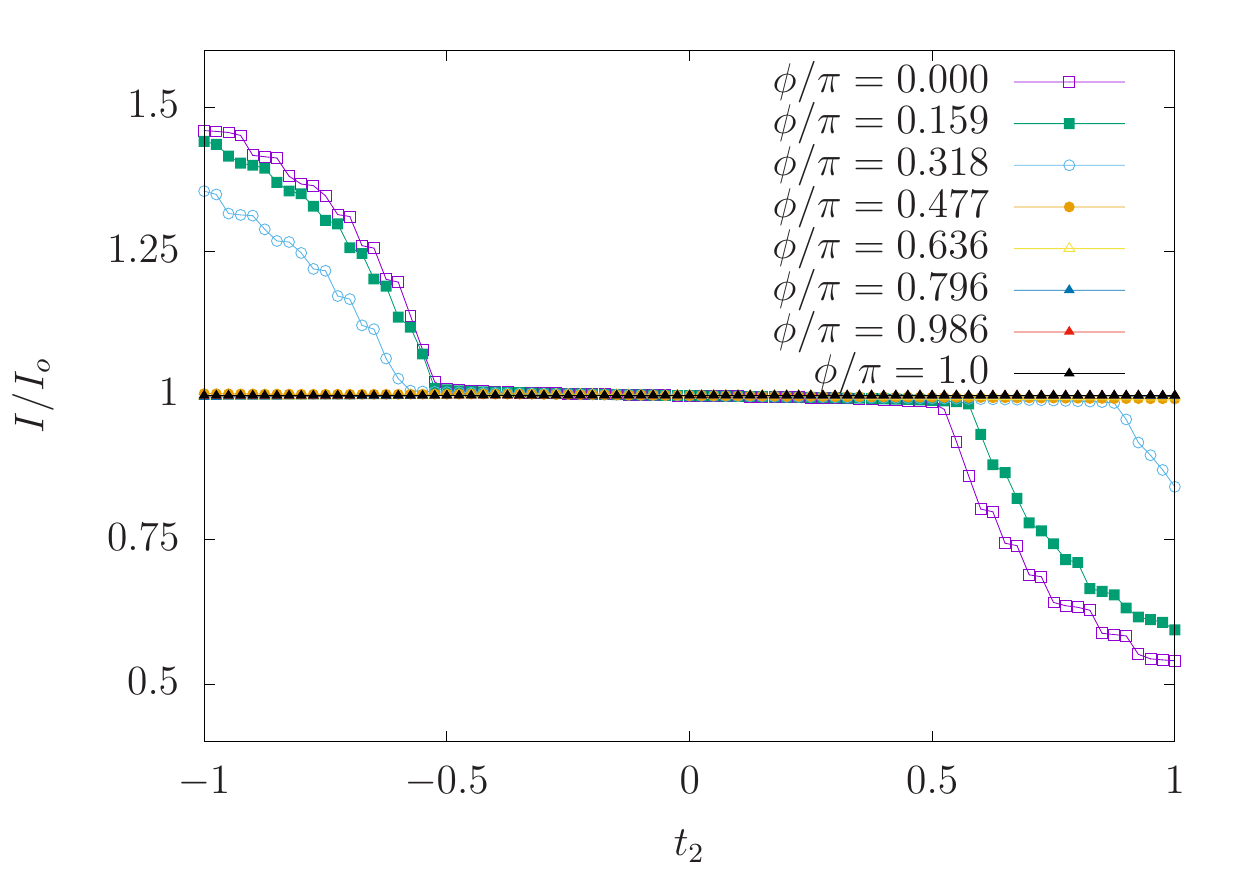}
	\caption{Variation of $I = \int dk \langle n(k) \rangle (\sin(k))^2 $ with $t_2$ for different values of $\phi$. $I_o$ is the corresponding value for a half-filled Fermi sea at $t_2=0$.}
	\label{momentInertia}
\end{figure}

\subsubsection{Bosonic occupation number in momentum space}

\begin{figure}
	\includegraphics[width=0.24\columnwidth]{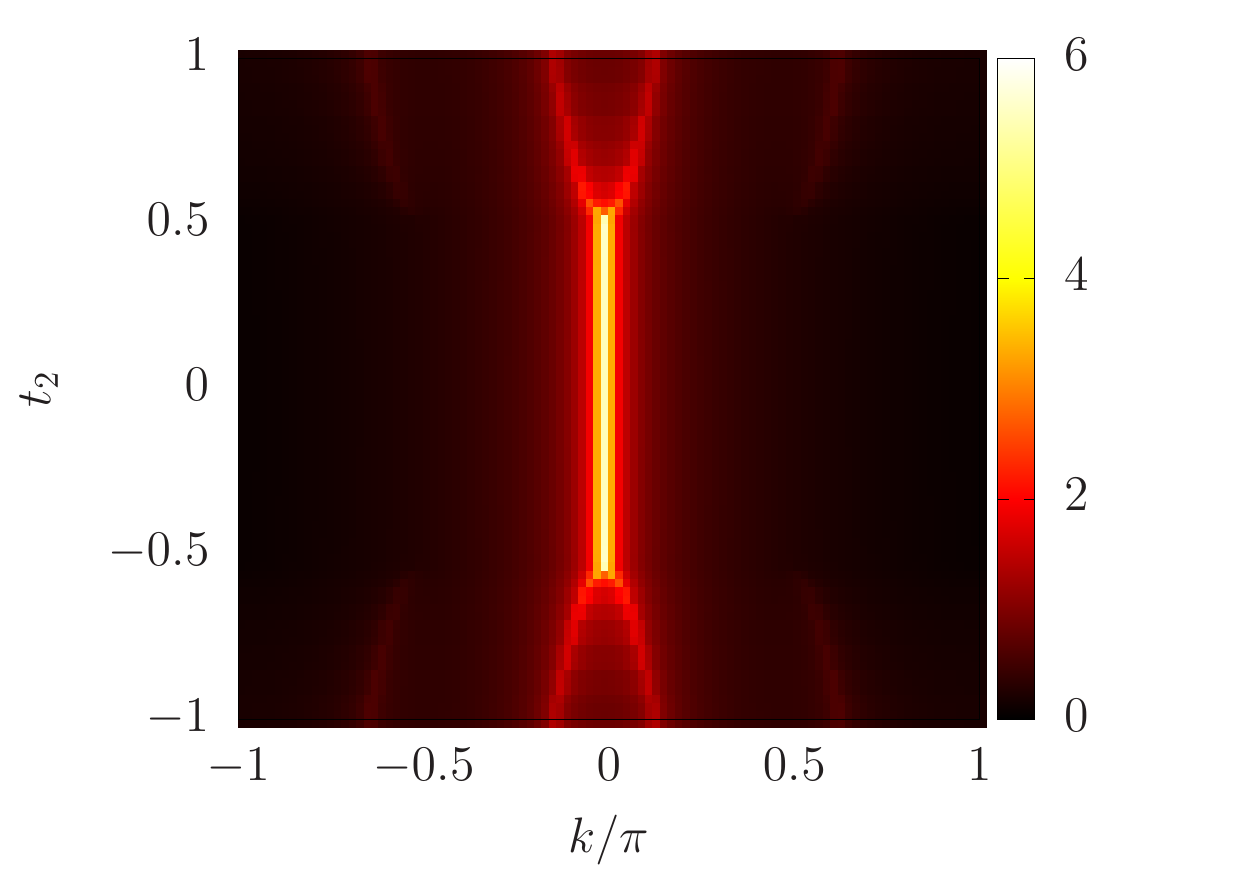}
	\includegraphics[width=0.24\columnwidth]{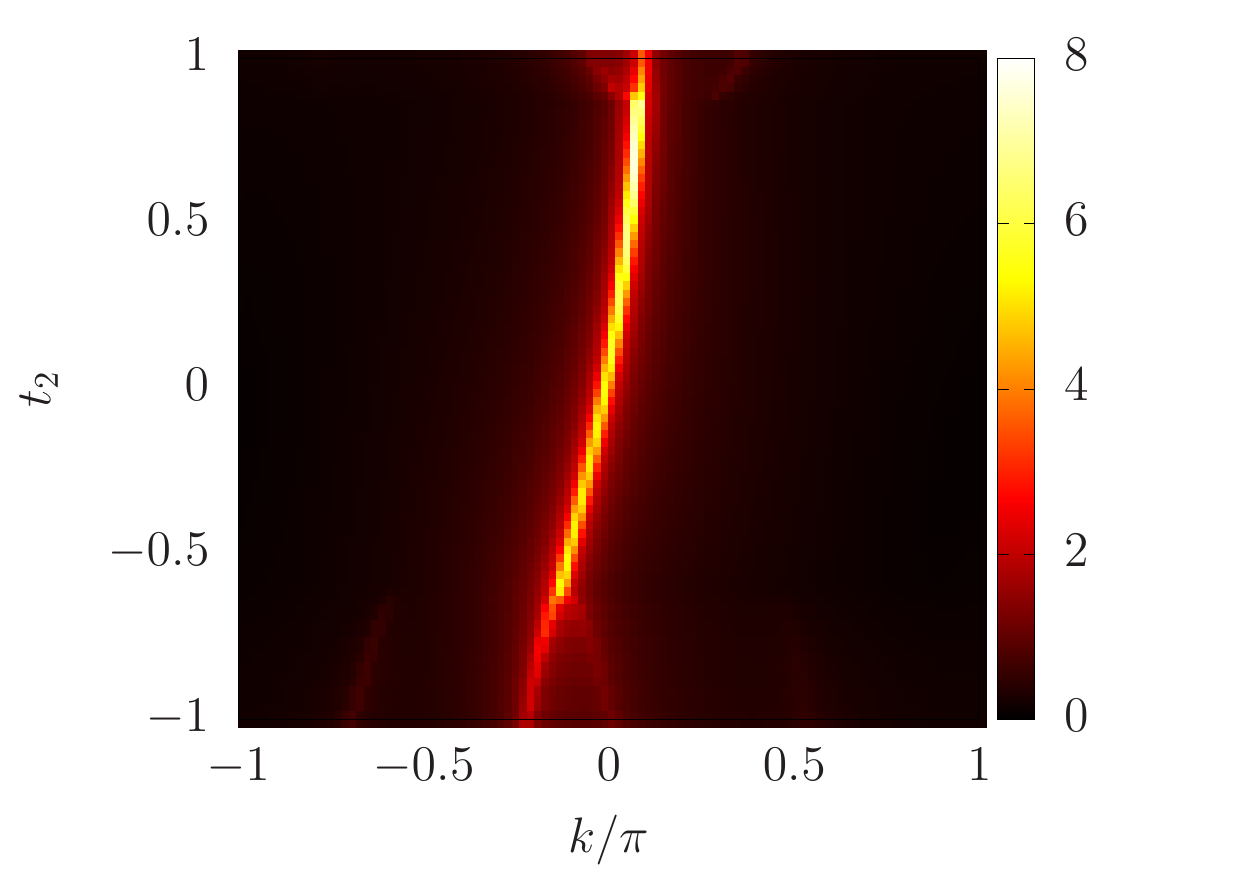}
	\includegraphics[width=0.24\columnwidth]{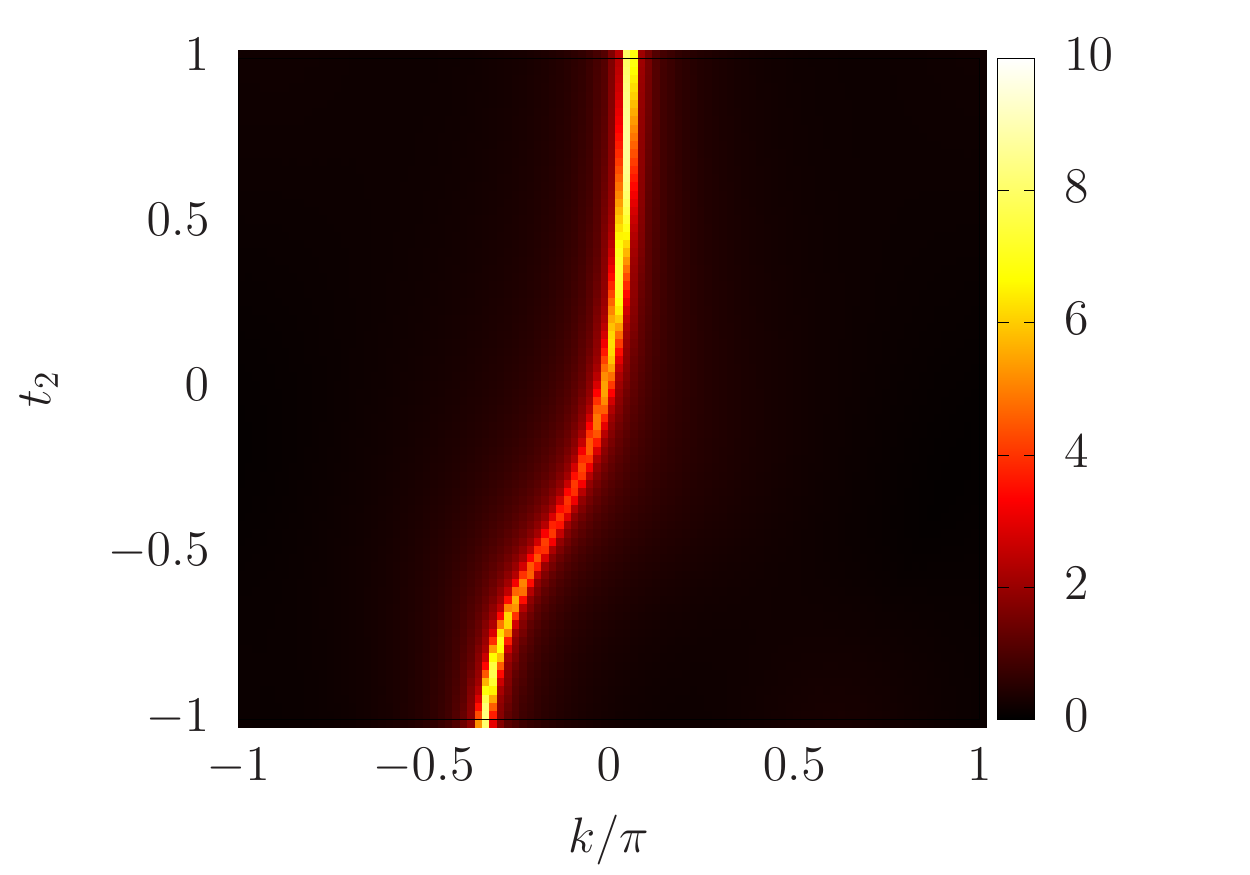}
	\includegraphics[width=0.24\columnwidth]{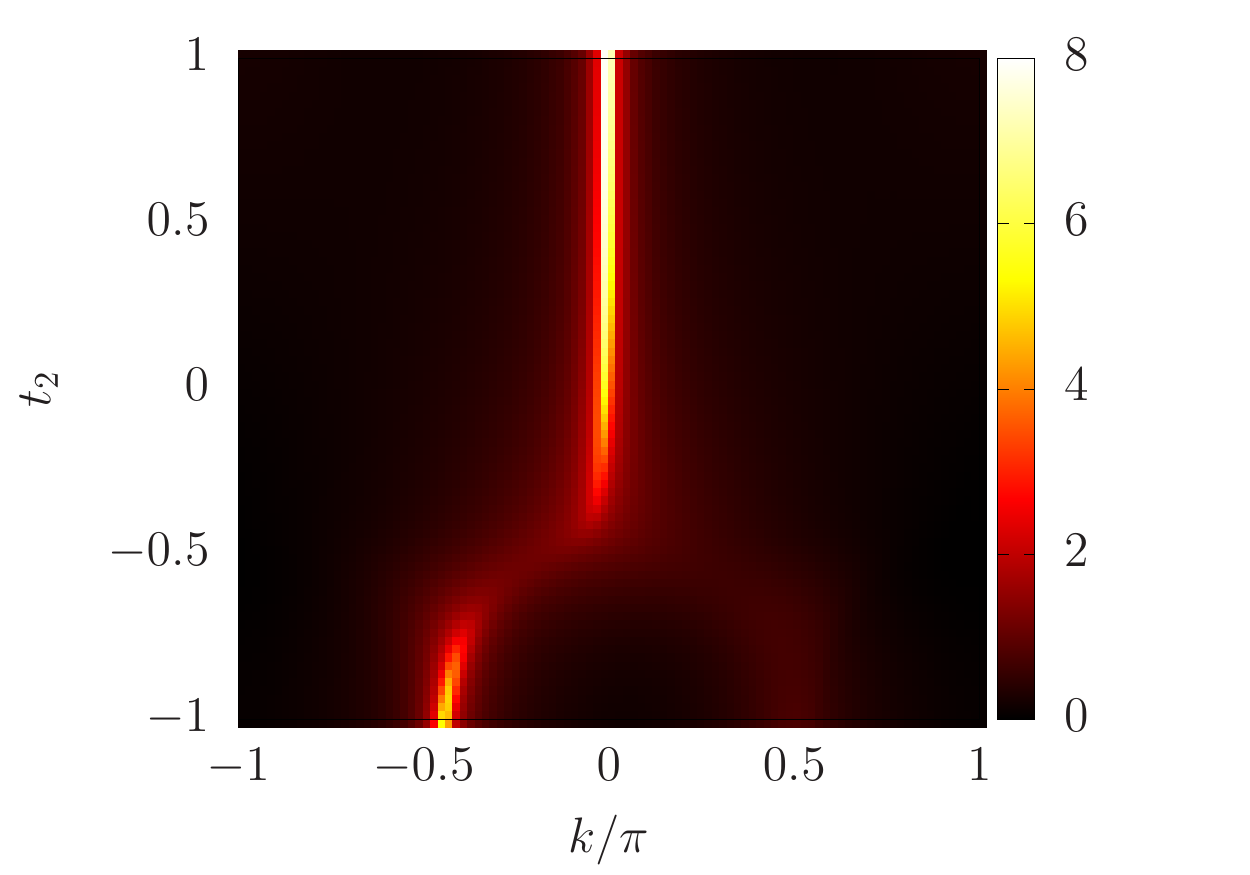}
	\caption{Bosonic $\langle n_b(k) \rangle$ as a function of $k$ and $t_2$ for different values of $\phi = 0.0, 1.0,2.0$ and $3.0$.}
	\label{nofkbos}
\end{figure}
While most of our discussion of the gapless phases has been in terms of TLL, a
quantity worth investigating is the expectation value of $\langle n^b_k \rangle \equiv \langle b^\dagger_k b_k \rangle$ which is expectation of each of the occupancy of the $k^{th}$ mode in bosonic language. \Fig{nofkbos} shows this for various values of $\phi$ and $t_2$ ($\langle n_b(k) \rangle = \langle b^\dagger(k) b(k)  \rangle = \frac{1}{L}\sum_{i,j} \exp{ik(i-j)} \langle S^+_i S^-_j  \rangle  $). Clearly the TLL phase is equivalently a power law superfluid as pointed in the main text. The shifted center of the Fermi sea in TLL manifests as again a shifted $k$ point where the superfluid $\langle n_b(k)\rangle $ peaks. 

\subsubsection{Characterisation of the $c=1$ TLL phase : Luttinger Parameters from DMRG study}
\label{SMLutt}

To characterise the $c=1$ TLL, it is particularly useful to calculate the fermion-fermion correlation function in this system. In the regime when the $\langle n(k) \rangle$ shows ``two" Fermi surfaces -- i.e. $c=1$ TLL 
\beq
C(r) = \frac{1}{L}\sum_i \langle c^\dagger_i c_{i+r} \rangle = \frac{1}{2\pi}  \int_{-\pi}^{\pi} dk \langle n(k) \rangle e^{-ikr}.
\eeq
For the case when $\phi=0$ and $ -0.5 \le t_2 \le 0.5$, such that there is a filled Fermi sea between $-\frac{\pi}{2} \le k \le \frac{\pi}{2}$, 
\beq
C(r) =  \frac{\sin(\pi r/2)}{\pi r}.
\label{corrNI1}
\eeq
For $ |t_2| \ge 0.5$,  $C(r)$ shows an oscillating behavior due to new Fermi wavevectors. In particular, for $t_2<-0.5$ we have two Fermi seas, one between $ k^{L+}_F \le k \le k^{R+}_F $ and the other between $ k^{L-}_F \le k \le k^{R-}_F $. In this regime
\beq
C(r)= \frac{2 \cos \left( \arccos (\frac{t_1}{2\sqrt{2}|t_2|})r \right)\sin(\frac{\pi}{4}r)}{\pi r}.
\label{corrNI2}
\eeq
\begin{figure}
	\includegraphics[width=0.31\columnwidth]{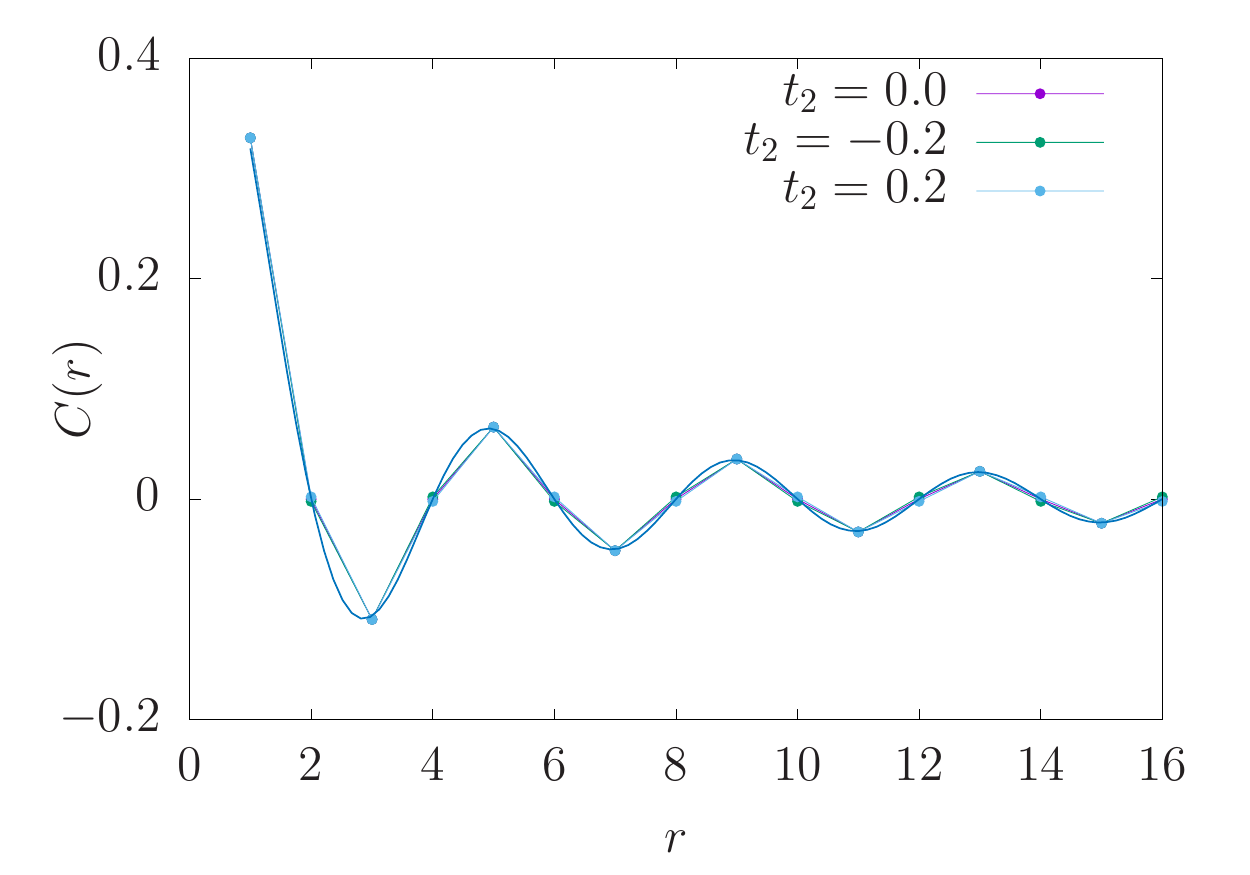}
	\includegraphics[width=0.31\columnwidth]{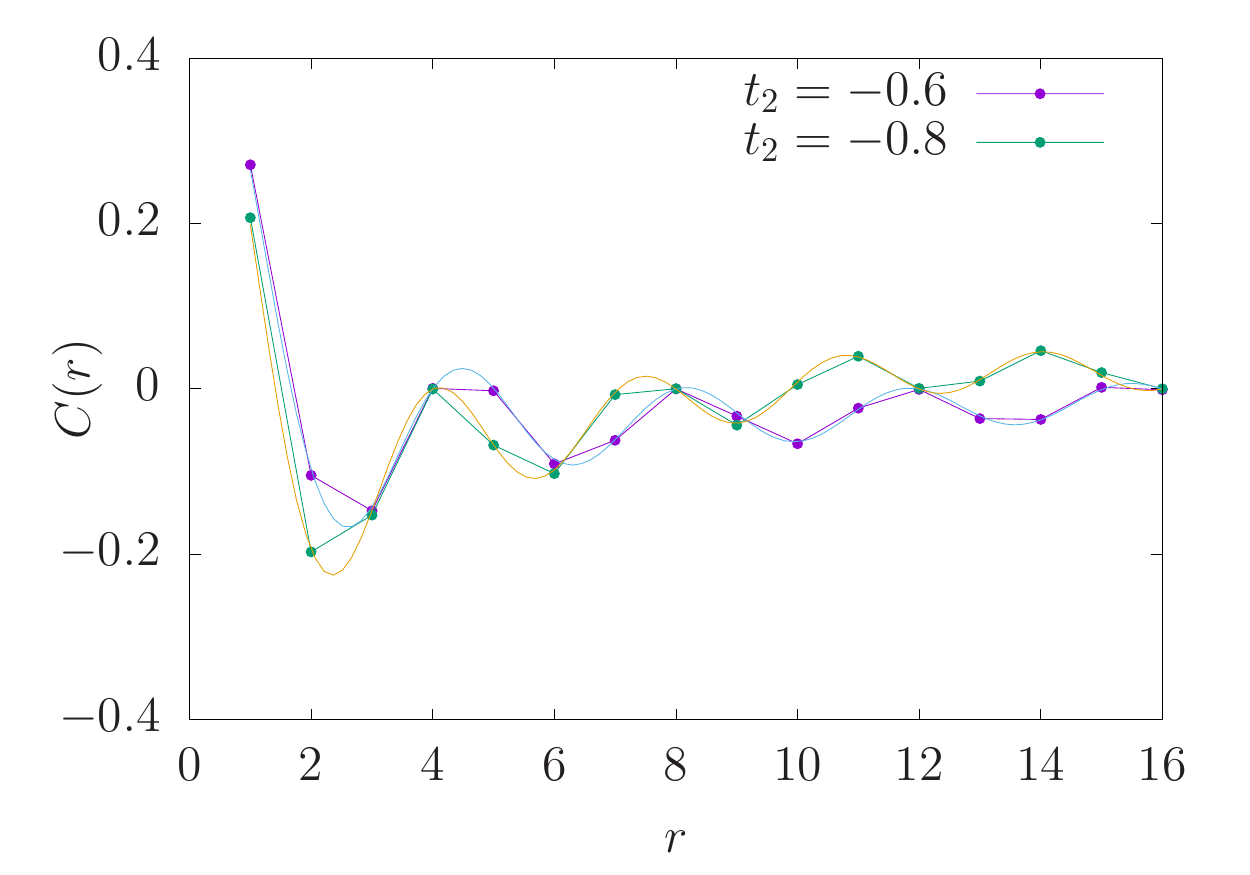}
	\includegraphics[width=0.31\columnwidth]{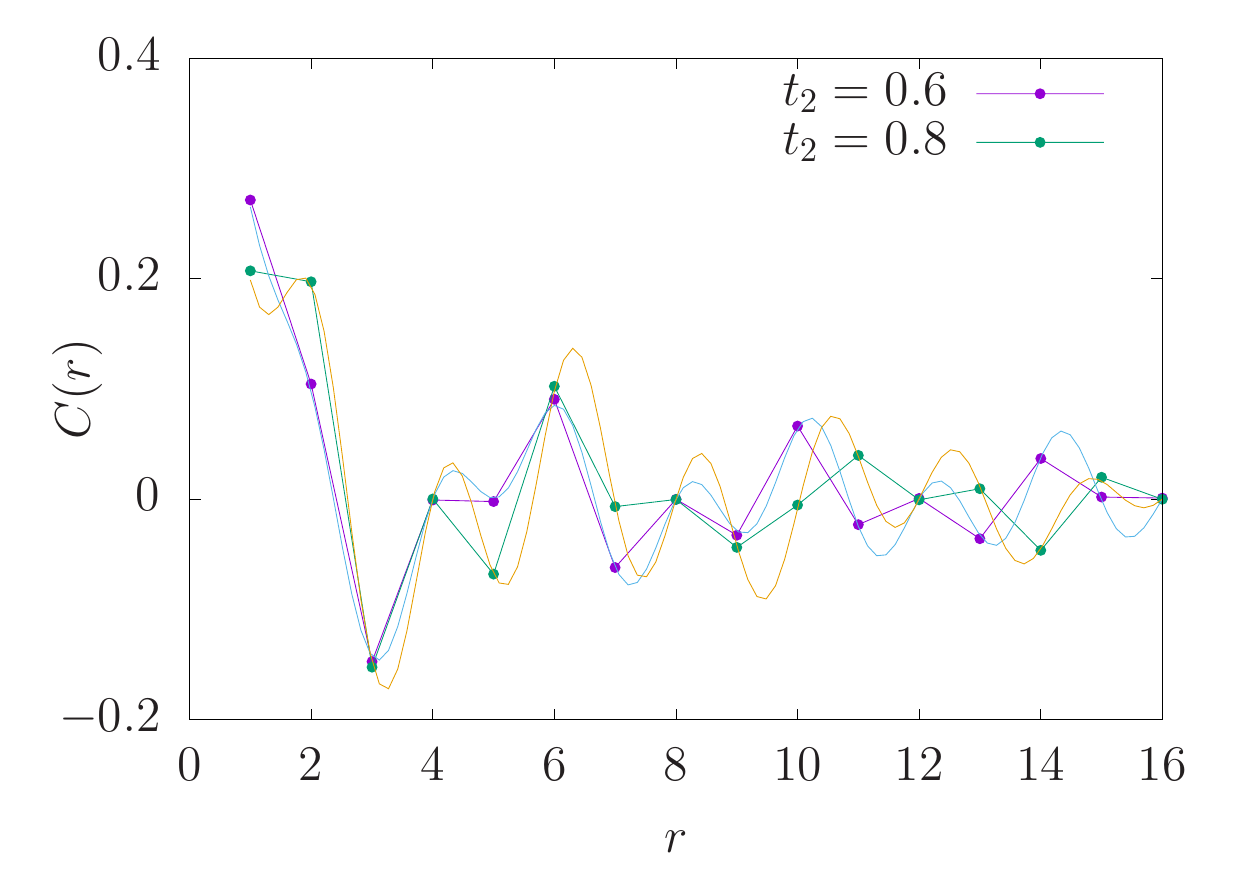}
	\caption{$C(r)$ has a function of $r$ for $\phi=0$ for different values of $t_2$. The numerical values from DMRG are the points. The continuous lines are the expressions shown near \eqn{corrNI2} for appropriate parameters.}
	\label{corrNI} 
\end{figure}
Similar calculations can be done for for $t_2>0.5$ (See \Fig{corrNI}). In presence of the interactions ($\phi\neq 0$)  the long wavelength scaling of the correlation function changes from $\sim\frac{1}{r}$ to $\frac{1}{r^\eta}$ where $\eta$ can be related to the Luttinger parameter \cite{Sachdev_book}. In $c=1$ TLL, which is the region of $-0.5<t_2<0.5$ and when $\phi \neq 0$, we fit $C(r)$ to a functional form 
\beq
C(r) \sim \frac{\cos(k_o x) \sin(\frac{\pi}{2}x)}{\pi x^\eta}
\label{fitcorr}
\eeq
where both $\eta$ and $k_o$ are fitting parameters. The intention is to capture the shift in the Fermi sea ($k_o$) and the Luttinger parameter ($K$)  where $K$ is related to $\eta$ via $\eta = \frac{1}{2}(K + \frac{1}{K})$. A comparison of $k_o$ obtained by above and the Hartree-Fock solution is shown in \Fig{diffvpar} for bench marking.

\section{Bosonization for the $c=1$ TLL}
\label{SMBoso}

The Hartree-Fock treatment as discussed above leads to a dispersion as schematically shown in \Fig{HFSchemFerm}(a). In general this {\it free} theory has two Fermi points centered about $k_o \neq 0$ and the Fermi velocities are not equal. A bosonization treatment \cite{Sachdev_book} in  terms of bosonic fields $\{\Theta, \Phi \}$ satisfying 
\beq
[\nabla \Phi(x), \Theta(y)] = [\nabla \Theta(x), \Phi(y)] = i \pi \delta (x-y)
\eeq
leads to the following free theory linearised about the HF ground state
\begin{align}
	\tilde{\mathcal{H}}_0=\frac{V_F}{2\pi}\int dx\left[(\partial_x\Phi)^2+(\partial_x\Theta)^2\right]+\frac{W}{2\pi}\int dx\left[\partial_x\Phi\partial_x\Theta+\partial_x\Theta\partial_x\Phi\right]
\end{align}
where $V_F=\frac{v_f^R+v_f^L}{2}$ and $W=\frac{v_f^R-v_f^L}{2}$.
 The interaction term is given by, 
\beq
\tilde{H}_{int}= -t_2 \sum_{k_1, k_2,k_3, k_4} \delta_{k_1+k_2-k_3-k_4}(e^{i\phi}-1) e^{i(-k_2+k_3+2k_4)} \left[:c^\dagger_{k_1}c_{k_4}::c^\dagger_{k_2}c_{k_3}: -:c^\dagger_{k_1}c_{k_3}::c^\dagger_{k_2}c_{k_4}:  \right] + h.c.
\eeq
where the normal ordering is done about the HF ground state. Identifying the slow modes, one finds two essential scattering contributions -- (i)~Forward scattering and (ii)~Umklapp scattering. Defining $\rho_R(q) =\sum_{q_1} : c^\dagger_{q_1 R} c_{q_1+q R}:$ and $\rho_R(x) = \sum_q e^{iqx} \rho_R(q)$, forward scattering contribution is
\beq
\tilde{H}^{Fow}_{int}= V \int dx \Big( (\rho_R + \rho_L )^2 - (\rho_R - \rho_L )^2  \Big) = V \int dx \left[(\partial_x\Phi)^2-(\partial_x\Theta)^2\right]
\eeq
where  $V= 4t_2(\cos(\phi+2k_o)-\cos(2k_o))$. For Umklapp process the contribution is
\bea
\tilde H_{int}^{Ump} %&=& \sum_i \Delta n_{i}  c^\dagger_{i-1} c_{i+1} + h.c. \\
&=& -\lambda \int dx \cos4\Phi
\eea where $\lambda=-2V$.
Gathering all the terms we get the bosonized Hamiltonian to be
\begin{align}
	H=\frac{v_f}{2\pi}\int dx\left[\frac{1}{K}(\partial_x\Phi)^2+K(\partial_x\Theta)^2\right]-\frac{W}{2\pi}\int dx\left[\partial_x\Phi\partial_x\Theta+\partial_x\Theta\partial_x\Phi\right]-\lambda\int dx\cos4\Phi
\end{align}
where the renormalised average Fermi-velocity is given by
$v_f=\sqrt{(V_F)^2-4\pi^2V^2}$
and the Luttinger parameter is given by
\begin{align}
	K=\sqrt{\frac{V_F-2\pi V}{V_F+2\pi V}}.
\end{align}
$V$ and $\lambda$
both go to zero at $\phi=0$ and hence at this point, $K=1$. Also, as expected, $W=0$ in this limit. This is nothing but the free fermions. To understand the effect of the other terms, we derive the corresponding real time action which is given by
\begin{align}
	\mathcal{S}=\frac{1}{2\pi v_f K}\int dt dx\left[\left(\partial_t\Phi+\frac{w}{2}\partial_x\Phi\right)^2-v_f^2(\partial_x\Phi)^2\right]+\lambda\int dt~dx\cos4\Phi
\end{align}
where $w=W/\Delta\tau$ where the limit is taken such that $w$ is constant. The effect of the ``boost" can then be gauged away \cite{Ray_AP_2017} after which we can wick rotate it to imaginary time to get the Euclidean action
\begin{align}
	\mathcal{S}_E=\frac{1}{2\pi v_f K}\int d\tau dx\left[\left(\partial_\tau\Phi\right)^2+v_f^2(\partial_x\Phi)^2\right]-\lambda\int d\tau~dx\cos4\Phi.
\end{align}
%%%%%%%%%%%%%%%%%%%%%%%%%%%%%%
\section{Details of the field theoretic calculations for the phase transitions}
\subsection{Field theory at the gapless-gapless transition}

Within HF, the gapless-gapless transition is a Lifshitz transition (see schematic \Fig{HFSchemFerm}(b)). The effective Hamiltonian here is given by
\bea
	{\cal H} &=&\frac{v_f}{2\pi}\int dx\left[\frac{1}{K}(\partial_x\Phi)^2+K(\partial_x\Theta)^2\right]-\frac{W}{2\pi}\int dx\left[\partial_x\Phi\partial_x\Theta+\partial_x\Theta\partial_x\Phi\right]-\lambda\int dx\cos4\Phi \notag\\ &+& g_1 \int dx [ \psi^\dagger_c(x) \psi_c(x) \partial_x \Phi] + g_2 \int dx   [ \psi^\dagger_c(x) \psi_c(x) \partial_x \Theta] +  \int dx \psi^\dagger_c [\frac{\partial^2_x}{2m*} - \mu ] \psi_c
\eea
where \bea
g_1 &=& -4t_2 \left( \{\cos(\phi+2k_c)-\cos(2k_c) \} - \{\cos(\phi+2k_o)-\cos(2k_o) \} \right) \\
g_2 &=& -8t_2 (\sin(k_o+k_c+\phi)-\sin(k_o+k_c)) 
%V &=& 4t_2\{ \cos(2k_o + \phi) - \cos(2k_o) \} 
\eea
By bosonising the left and right fermions we have been able to take into account their mutual interactions through the forward scattering channel by renormalizing the Luttinger parameter and the Fermi velocity as before. 

 Note that the microscopic symmetry of combination of parity($\mathcal{P}$) and time reversal($\mathcal{T}$) together, remains intact in this description (${\cal P}: \Phi \rightarrow -\Phi, \Theta \rightarrow \Theta$ and ${\cal T}: \Phi \rightarrow \Phi, \Theta \rightarrow -\Theta$). Note that at $\phi=0$, as is expected, $k_o=k_c=0$ and $g_1= g_2=0$ leading to the free fermionic description. Near $\phi=0$, $g_1 \sim -8 t_2 \phi \{k_o-k_c\}$ and $g_2 \sim -8t_2 \phi$ and $|(k_o -k_c)| \propto t_2 \phi$.   
 
The effective Euclidean action using the usual time-slicing is,
\bea
{\mathcal S} &=&  \frac{1}{2 v_f K \pi}\int dx d\tau \left[ (\dot{\Phi} - i\frac{W}{2} \partial_x \Phi + ig_2 \pi {\psi}^\dagger_c(x) {\psi_c}(x) )^2 + v^2_f (\partial_x \Phi)^2  \right]  + g_1 \int dx d\tau [ {\psi}^\dagger_c(x) \psi_c(x) \partial_x \Phi] \notag \\ &+&  \int dx \psi^\dagger_c [\frac{\partial^2_x}{2m^*} - \mu ] \psi_c  - \lambda \int dx d\tau \cos(4 \Phi) 
\eea
Redefining a boosted field such that $\dot{\Phi} \equiv \dot{\Phi} -i \frac{W}{2}\partial_x \Phi$, $\partial_x \Phi = \partial_x \Phi$ 
\bea
{\mathcal S} &=&  \frac{1}{2 v_f K \pi}\int dx d\tau \left[ (\dot{\Phi} + ig_2 \pi {\Psi}^\dagger_c(x) {\Psi_c}(x) )^2 + v^2_f (\partial_x \Phi)^2  \right]  + g_1 \int dx d\tau [ {\Psi}^\dagger_c(x) \Psi_c(x) \partial_x \Phi]  \notag \\ &+&  \int dx \psi^\dagger_c [\frac{\partial^2_x}{2m^*} - \mu ] \psi_c  - \lambda \int dx d\tau \cos(4 \Phi)
\eea

In the Fourier space the above action becomes  $\mathcal{S}=\mathcal{S}_\Phi+\mathcal{S}_c +\mathcal{S}_{int}$ where,
\begin{align}
\mathcal{S}_\Phi=\frac{1}{2\pi v_f K}\int dq~d\omega~\left[\omega^2+v_f^2 q^2\right]|\Phi(q,\omega)|^2
\end{align}
\begin{align}
\mathcal{S}_c=\int dq~d\omega \left[i\omega +\left(\frac{k^2}{2m^*}-\mu\right)\right]\psi_{c}^\dagger(k,\omega)\psi_c(-k,-\omega)
\end{align}
\begin{align}
\mathcal{S}_{int}=ig\int d\omega_1dq_1d\omega_2dq_2~\left[i\omega_1-\alpha q_1\right]\Phi(q_1,\omega_1)\psi_c^\dagger(q_2,\omega_2)\psi_c(q_1+q_2,\omega_1+\omega_2)
\end{align}
where $g=g_2/(v_f K)$ and $\alpha=g_1v_f K/g_2$. The short range four fermion term for the middle mode ($\psi_c$) are irrelevant at this critical point which is understandable due to the paucity of the phase space for such density-density scatterings. The scattering vertex is shown in \eqn{ScatVer}.
The chemical potential is always a relevant perturbation for the $\psi_c$ fermions at the $\mu=0$ fixed point. 
\beq
\includegraphics[width=0.3\columnwidth]{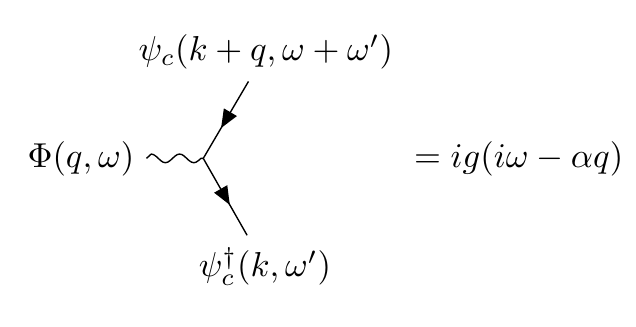}
%%\feynmandiagram [inline=(d.base), horizontal=d to b] {
%	a [particle=\(
%	{\psi_c(k+q,\omega+\omega')}
%	\)] -- [fermion]  
%	b  -- [fermion] c [particle=\(
%	{\psi^\dagger_c(k,\omega')}
%	\)],
%	b -- [boson] 
%	d [particle=\(
%	{\Phi(q,\omega)}
%	\)],};
%= i g (i \omega - \alpha q)
\label{ScatVer}
\eeq
A momentum-Shell RG scheme upto second order in perturbation theory in $g$ produces the following flow equation for $K$ at one-loop level (bubble diagram)
\begin{align}
	\frac{dK}{dl} = C \mu \left( \left(\frac{g_1K}{g_2} \right)^2   -1 \right)  
\end{align}
where $C \propto \frac{g_2^2}{2v_f}$. We find that near $\phi \rightarrow 0$ this term does not lead to a run-away flow signaling a stable phase within this approximation.	

\subsection{Transition at $\phi=\pi$ : XY Duality in $(1+1)$ dimensions.}

While most of our discussion of the gapless phases has been in terms of fermions, the $\phi=\pi$ line (and its vicinity) can be understood starting with hard-core bosons or spin-$1/2$s including the transition between the powerlaw SF and the BO phase \cite{sachdev2002quantum}. For universal properties such as the nature of the transition, we expect that it is sufficient to study the rotor Hamiltonian

\begin{align}
{\cal H}=-\sum_i\left[t_1\cos(\tilde\phi_i-\tilde\phi_{i+1})+t_2\cos(\tilde\phi_i-\tilde\phi_{i+2})\right]+U\sum_i(n_i-\bar n)^2
\label{eq_rotor}
\end{align}
 To understand the phase diagram of the above rotor model we first dualise the theory following Fisher and Lee \cite{baeriswyl2005strong}\label{key}. This is achieved by introducing the dual variables $(\phi'_{\bar{i}},\theta'_{\bar{i}})$ where the dual variables sit on the bonds of the original sites and the connection between the two sets of coordinates is $\bar{i}=i+1/2$. With this notation we now write down the mapping as
\begin{align}
e^{i\tilde\phi}=\prod_{\bar{j}<i}e^{i\theta'_{\bar{j}}},\quad\quad\quad\quad n_i=\phi'_{\bar{i}}-\phi'_{\bar{i}-1}
\end{align}
The dual algebra is given by $\left[e^{i\theta'_{\bar i}},\phi'_{\bar j}\right]=\delta_{\bar i\bar j}e^{i\theta'_{\bar i}}$. 
The eigenvalues of $\theta'_{\bar i}\in (0,2\pi]$ and $\phi'_{\bar i}\in \mathbb{Z}$. The rotor Hamiltonian of Eq. \ref{eq_rotor} becomes 
\begin{align}
\mathcal{H}=-\sum_{\bar i}\left[t_1\cos(\theta'_{\bar i})+t_2\cos(\theta'_{\bar i+1}+\theta'_{\bar i})\right]+U\sum_{\bar i}\left(\Delta_x\phi'_{\bar i}-\bar n\right)^2
\label{eq_dual_rotor}
\end{align}
where $\Delta_x\phi'_{\bar i}=\phi'_{\bar i+1}-\phi'_{\bar i}.$ The partition function corresponding to the Eq. \ref{eq_dual_rotor} is given by  $ Z=\int[\mathcal{D}\phi'][\mathcal{D}\theta']e^{-\mathcal{S}}$, 
 where, the Euclidean action on the discrete $(1+1)$ dimensional space-time lattice is given by
\begin{align}
\mathcal{S}=&-\sum_\tau\sum_{\bar i,\tau}\left[\tilde t_1\cos(\theta'_{\bar i,\tau})+\tilde t_2\cos(\theta'_{\bar i,\tau}+\theta'_{\bar i+1,\tau})\right]+\tilde U\sum_\tau\sum_{\bar i}\left[\Delta_x\phi'_{\bar i,\tau}-\bar n\right]^2+i\sum_\tau\sum_{\bar i}\theta'_{\bar i,\tau}\Delta_\tau\phi'_{\bar i,\tau}-\lambda\sum_\tau\sum_{\bar i}\cos(2\pi\phi'_{\bar i})
\end{align} 
where $\tilde t_1=\delta t_1$, $\tilde t_2=\delta t_2$ and $\tilde U=\delta U$ with $\delta$ being the time-step. The last term has been added with $\lambda>0$ as a soft potential promoting $\phi'$ to a real field \cite{baeriswyl2005strong}. These changes should keep the universal features of the phase and the phase transition intact. Implementing the scaling transformations  :$\phi\rightarrow 2\pi\phi$ and  $\theta\rightarrow\theta/2\pi$, the above action becomes
\begin{align}
\mathcal{S}=&-\sum_\tau\sum_{\bar i,\tau}\left[\tilde t_1\cos(2\pi\theta'_{\bar i,\tau})+\tilde t_2\cos(2\pi(\theta'_{\bar i,\tau}+\theta'_{\bar i+1,\tau}))\right]+\frac{\tilde U}{4\pi^2}\sum_\tau\sum_{\bar i}\left[\Delta_x\phi'_{\bar i,\tau}-2\pi\bar n\right]^2+i\sum_\tau\sum_{\bar i}\theta'_{\bar i,\tau}\Delta_\tau\phi'_{\bar i,\tau}-\lambda\sum_\tau\sum_{\bar i}\cos(\phi'_{\bar i})
\end{align}
Since, in presence of the soft potential, $\phi'$ is promoted to a real field, the conjugate $\theta'$ field is also no longer compact and hence we can expand the cosines to get
\begin{align}
\mathcal{S}'&
&=2\pi^2\sum_\tau\sum_{\bar i,\tau}\left[(\tilde t_1+2\tilde t_2)(\theta'_{\bar i,\tau})^2+2\tilde t_2\theta'_{\bar i,\tau}\theta'_{\bar i+1,\tau}\right]+\frac{\tilde U}{4\pi^2}\sum_\tau\sum_{\bar i}\left[\Delta_x\phi'_{\bar i,\tau}-2\pi\bar n\right]^2+i\sum_\tau\sum_{\bar i}\theta'_{\bar i,\tau}\Delta_\tau\phi'_{\bar i,\tau}-\lambda\sum_\tau\sum_{\bar i}\cos(\phi'_{\bar i})
\end{align}
Now, if we write the filling as $\bar n=\bar z+f $, where $\bar z$ is the integer part of the filling while $f$ is the fractional part, then the integer part of the filling can be removed by the  transformation $\phi'_{\bar i,\tau}\rightarrow\phi'_{\bar i,\tau}+2\pi\bar z\bar i$. We then define static background fields $\chi_{\bar i}$ such that $\Delta_x\chi_{\bar i}=f$ to get
\begin{align}
\mathcal{S}'&=2\pi^2\sum_\tau\sum_{\bar i,\tau}\left[(\tilde t_1+2\tilde t_2)(\theta'_{\bar i,\tau})^2+2\tilde t_2\theta'_{\bar i,\tau}\theta'_{\bar i+1,\tau}\right]+\frac{\tilde U}{4\pi^2}\sum_\tau\sum_{\bar i}\left[\Delta_x h_{\bar i,\tau}\right]^2+i\sum_\tau\sum_{\bar i}\theta'_{\bar i,\tau}\Delta_\tau h_{\bar i,\tau}-\lambda\sum_\tau\sum_{\bar i}\cos(h_{\bar i,\tau}+2\pi\chi_{\bar i})
\end{align}
where $h_i=\phi'_{\bar i,\tau}-2\pi\chi_{\bar i}$. Noting that the $\theta'$ field is not diagonal in real space due to the second term. Temporarily neglecting this term and integrating $\theta'$, we get
\begin{align}
 \mathcal{S}_h&=\frac{1}{8\pi^2(\tilde t_1+2\tilde t_2)}\sum_\tau\sum_{\bar i,\tau}\left(\Delta_\tau h_{\bar i,\tau}\right)^2+\frac{\tilde U}{4\pi^2}\sum_\tau\sum_{\bar i}\left(\Delta_x h_{\bar i,\tau}\right)^2-\lambda\sum_\tau\sum_{\bar i}\cos(h_{\bar i,\tau}+2\pi\chi_{\bar i})
\end{align}
Introducing $\chi(x)=\sum_{\bar i}f\bar i\delta(x-\bar i a)$, where $a$ is the lattice length-scale, we can take the space-time continuum limit of the above action to get
\begin{align}
\mathcal{S}_h=\frac{1}{2g}\int dx~d\tau~\left[\left(\partial_\tau h\right)^2+\left(\partial_x h\right)^2\right]-\lambda\int dx~d\tau~\cos\left(h+2\pi\chi\right)
\end{align}
where $g=4\pi^2\sqrt{(\tilde t_1+2\tilde t_2)/\tilde U}$ and we have scaled $x\rightarrow x/\sqrt{\tilde U(\tilde t_1+2\tilde t_2)}$.  For $\tilde U\ll t_{eff}(=\tilde t_1+2\tilde t_2)$, we have $g>1$ and the fluctuations of $h$ are soft. The pinning cosine term is irrelevant and we end up with a quadratic theory with powerlaw correlations which is nothing but the power-law SF.  On the other hand when $\tilde U>>t_{eff}(=\tilde t_1+2\tilde t_2)$, the fluctuations are energetically costly and the cosine term pins down the average value of $h$. Due to the background field $\chi$, this average value breaks translation symmetry and this is BO phase of the system. Without loss of generality we can take $\tilde t_1=1$ and find that any positive $\tilde t_2$ will make the effective hopping larger (within this approximation) and thereby stabilising SF, whereas for $\tilde t_2<0$, the effective hopping is reduced and the SF is destabilised. Neglecting the cross term as above, we note that the $g\rightarrow 0$, as $\tilde t_2\rightarrow -1/2$ which is the MG point. For $t_2/t_1<-0.5$, the above approximation breaks down as the cross term becomes singularly important. However, for $t_2/t_1>-0.5$, the cross term is only expected to renormalise various coefficients. To take these and also to explore the nature of the transition for the entire range of  $\phi$, we use the bosonized theory.
%\bibliography{ref1DCorr}

\iffalse

\end{document}